\documentclass[10pt]{article}

\usepackage{authblk}

\makeatletter
\newcommand{\subjclass}[2][]{%
  \let\@oldtitle\@title%
  \gdef\@title{\@oldtitle\footnotetext{#1 \emph{Mathematics subject classification (2010).} #2}}%
}
\newcommand{\keywords}[1]{%
  \let\@@oldtitle\@title%
  \gdef\@title{\@@oldtitle\footnotetext{\emph{Keywords:} #1.}}%
}
\makeatother

\title{The algebras of graph rewriting}

\author[1]{Nicolas Behr\thanks{email: nicolas.behr@ed.ac.uk (corresponding author)}}
\author[2]{Vincent Danos\thanks{email: danos@di.ens.fr}}
\author[1]{Ilias Garnier\thanks{email: igarnier@inf.ed.ac.uk}}
\author[3]{Tobias Heindel\thanks{email: tohe@di.ku.dk}}
\affil[1]{Laboratory for Foundations of Computer Science, School of Informatics, University of Edinburgh, Informatics Forum, 10 Crichton Street, Edinburgh, EH8 9AB, Scotland, UK}
\affil[2]{LFCS, CNRS \& \'{E}quipe Antique, D\'{e}partement d'Informatique de l'\'{E}cole Normale Sup\'{e}rieure Paris, 45 rue d'Ulm, 75230 Paris Cedex 05, France}
\affil[3]{Department of Computer Science, Datalogisk Institut (DIKU), K{\o}benhavns Universitet, Universitetsparken 5, 2100 K{\o}benhavn {\O}, Denmark}


\date{\today}

\usepackage{graphicx}
\usepackage[dvipsnames,table]{xcolor}
\usepackage{etoolbox}
\usepackage{booktabs}
\usepackage{amsmath,amssymb,amsfonts,amsthm}

\usepackage{bbold}
\usepackage{mathtools}
\usepackage{verbatim}
\usepackage{rotating}

\theoremstyle{plain} 
\newtheorem*{theorem*}{Theorem}
\newtheorem{thm}{Theorem}[section]
\newtheorem{prop}{Proposition}[section]
\newtheorem{cor}{Corollary}[section]
\newtheorem{lem}{Lemma}[section]
\newtheorem{defn}{Definition}[section]

\newcommand{\bC}{\ensuremath{\mathbb{C}}}

\newcommand{\bG}{\ensuremath{\mathbb{G}}}

\newcommand{\bK}{\ensuremath{\mathbb{K}}}

\newcommand{\bR}{\ensuremath{\mathbb{R}}}

\newcommand{\bZ}{\ensuremath{\mathbb{Z}}}

\newcommand{\fd}{\mathfrak{d}}
\newcommand{\fD}{\mathfrak{D}}

\newcommand{\fG}{\mathfrak{G}}

\newcommand{\cA}{\ensuremath{\mathcal{A}}}

\newcommand{\cD}{\ensuremath{\mathcal{D}}}
\newcommand{\cE}{\ensuremath{\mathcal{E}}}
\newcommand{\cF}{\ensuremath{\mathcal{F}}}
\newcommand{\cG}{\ensuremath{\mathcal{G}}}
\newcommand{\cH}{\ensuremath{\mathcal{H}}}
\newcommand{\cI}{\ensuremath{\mathcal{I}}}

\newcommand{\cL}{\ensuremath{\mathcal{L}}}
\newcommand{\cM}{\ensuremath{\mathcal{M}}}

\newcommand{\cO}{\ensuremath{\mathcal{O}}}
\newcommand{\cP}{\ensuremath{\mathcal{P}}}

\newcommand{\cR}{\ensuremath{\mathcal{R}}}

\newcommand{\cT}{\ensuremath{\mathcal{T}}}
\newcommand{\cU}{\ensuremath{\mathcal{U}}}
\newcommand{\cV}{\ensuremath{\mathcal{V}}}
\newcommand{\cW}{\ensuremath{\mathcal{W}}}

\renewcommand{\vec}[1]{\underline{#1}}

\newcommand{\rConv}{\smallsmile}

\newcommand{\comp}[3]{#1 \stackrel{#2}{\blacktriangleleft} #3}

\newcommand{\GRule}[2]{#1\rightharpoonup#2}
  
\newcommand{\isoc}[1]{%
   [#1]_{\cong}
}

\newcommand{\ntC}{\circledast}

\newcommand{\includetikz}[1]{\ensuremath{\vcenter{\hbox{\includegraphics{images/#1}}}}}

\usepackage{hyperref} 
 \hypersetup{
	colorlinks=true, 
	breaklinks=true, 
	linkcolor=green!20!black, 
	citecolor=green!20!black,
	menucolor=green!20!black, 
	urlcolor=green!20!black
	}

\begin{document}

\maketitle

\begin{abstract}
The concept of diagrammatic combinatorial Hopf algebras in the form introduced for describing the Heisenberg-Weyl algebra in~\cite{blasiak2010combinatorial} is extended to the case of so-called rule diagrams that present graph rewriting rules and their composites. The resulting rule diagram algebra may then be suitably restricted in four different ways to what we call the rule algebras, which are non-commutative, unital associative algebras that implement the algebra of compositions of graph rewriting rules.  Notably, our framework reveals that there exist two more types of graph rewriting systems than previously known in the literature, and we present an analysis of the structure of the rule algebras as well as a form of Poincar\'{e}-Birkhoff-Witt theorem for the rule diagram algebra. Our work lays the foundation for a fundamentally new way of analyzing graph transformation systems, and embeds this very important concept from theoretical computer science firmly into the realm of mathematical combinatorics and statistical physics.
\end{abstract}

\tableofcontents

\section{Introduction}

The aim of this work is to build a unified framework for the analysis
of a broad class of stochastic processes from the combinatorics,
computer science and physics perspectives. The processes we consider
are based on \emph{graph rewriting}, a model of computation that
allows to frugally express complex behaviors by specifying local
transformations of graphical structures by the use of \emph{rules}. We
put forward a philosophy that transition systems should be described
algebraically and that in particular graph rewriting systems have an
underlying combinatorial algebra that we call the \emph{rule algebra}.\\

Graph rewriting is Turing-complete, hence sufficiently expressive to
encode any other model of computation. It is applied with considerable success
in modeling of biological systems (especially protein interaction
networks, see~\cite{danos2007rule}). In fact, many models in statistical
physics, theoretical chemistry and combinatorics can be seen as particular cases of graph rewriting systems. Most notably, the \emph{Heisenberg-Weyl algebra}, the combinatorial
algebra underlying chemical reaction systems and the theory of the
harmonic oscillator, is a particularly simple case of a graph rewriting algebra.\\

In this paper, the focus is on the construction of the
combinatorial Hopf algebra that allows to construct the actual rule
algebras. Representation theoretic aspects and the stochastic
mechanics framework are described in \cite{bdg2016}. For the benefit
of the interdisciplinary audience, in the remainder of this section we
will provide a brief overview of the various connections to previous
existing work at the end of this section.

\subsection{Overview of the paper}

Our new formulation relies on expressing graph rewriting rules as
so-called \emph{irreducible rule diagrams}, and their compositions in
terms of a diagrammatic form of compositions of rule
diagrams, followed by an operation we call reduction. Notably, while the diagram algebraic construction will be
universal, it is only via the definitions of so-called reduction
morphisms that we will be able to define the \emph{algebras of graph
rewriting}, i.e.\ the \emph{rule algebras}.  Two of the conceivable
variants of reduction result in the two main variants of graph
rewriting, namely \emph{\underline{D}ouble
\underline{P}ush\underline{o}ut} ($DPO$)
~\cite{DBLP:conf/focs/EhrigPS73,DBLP:conf/gg/1997handbook} and
\emph{\underline{S}ingle \underline{P}ush\underline{o}ut}
($SPO$)~\cite{lowe1993algebraic,DBLP:conf/gg/EhrigHKLRWC97}
rewriting. However, as we will present in detail, there exist two
further natural variants of reduction, resulting in two entirely
\emph{new forms of graph rewriting}. Additional benefits of the novel
framework include a clear separation between the \emph{algebras} of
graph rewriting (hitherto only considered indirectly in as far as
individual contributions to sequential compositions of two rules were
concerned) and their \emph{representations}, which implement the
action of rewriting rules on graphs. Via this clear distinction, we
gain a substantial advantage over the traditional framework (in
addition perhaps to a more intuitive formulation): we may for the
first time in this line of research directly study the structure and
combinatorics of graph rewriting algebras!\\

Our main hypothesis for the construction of the general rule algebras
has been that the Heisenberg-Weyl algebra should form a special case of generic
graph rewriting. Since in all four types of rewriting the rewriting of
discrete graphs takes the same form, we took the additional hypothesis
as a constructive guideline that also the general diagrammatic
construction must be fitted into a \emph{combinatorial Hopf algebra}
framework. Remarkably, as presented in Section~\ref{sec:RDhopf}, these
two hypotheses together with a natural Ansatz for the ``irreducible''
rule diagrams (in the form of the graphs of linear rules) were
sufficient to derive a consistent construction of the so-called
\emph{rule diagram algebra}.\\

The most complicated aspect in the derivation of the rule algebra
framework proved to be the question of how the previously known two
main variants of graph rewriting,
$DPO$~\cite{DBLP:conf/focs/EhrigPS73,DBLP:conf/gg/1997handbook} and
$SPO$ rewriting~\cite{lowe1993algebraic,DBLP:conf/gg/EhrigHKLRWC97},
could be implemented. A first clue towards a mathematical Ansatz was
yet again provided by Blasiak et al.'s construction of the HW algebra
as a reduction of the HW diagram
algebra~\cite{blasiak2010combinatorial}: employing our novel
definition of diagrams representing HW algebra elements and their
compositions, it became evident that this reduction consists in
assigning to a given composite diagram its normal form. For general
graph rewriting and the corresponding rule diagrams that include
rewriting of graphs with edges however, a naive Ansatz for such a
reduction to normal forms proved to be problematic: experts in the
field of graph rewriting might be familiar with the idea of ``dangling
edges'', an artifact that can arise when trying to compose rewriting
rules inconsistently. Our second main result of this paper consists
thus in the aforementioned observation that one may define \emph{four
different natural strategies} to define reductions of rule diagrams!
It is only in the definitions of these four variants of reduction
morphism that we obtain four different rule algebras, two of which
reproduce the traditional $DPO$ and $SPO$ rewriting frameworks.\\

The second part of this paper is devoted to the study of the structure
of the algebras of rule diagrams and that of the rule algebras. In
Section~\ref{sec:LieUnAlg}, we study the universal enveloping and Lie
algebraic structures of the rule diagram algebra, and demonstrate that
our presentation of the algebra in terms of rule diagrams and
compositions thereof is naturally isomorphic to the respective
universal enveloping algebra, an important consistency check for our
hypothesis of combinatorial Hopf algebra construction. In
Section~\ref{sec:HW}, we demonstrate how the results of Blasiak et
al.~\cite{blasiak2010combinatorial} on the HW and HW diagram algebras
are compatible with our more general framework, and how the
traditionally known combinatorial formulae in this context follow
efficiently.\\

Our third main result then consists in the structure theory of the
rule algebras as presented in Section~\ref{sec:structureTheory}. There
are three natural ``generator'' subalgebras definable for all four
types of rule algebras (independently on the type of the rule algebra)
that may serve as an Ansatz for the generators of the rule
algebras. The vertex-only subalgebra not only subsumes the well-known
HW algebra, but is in fact a realization of the so-called Boson
algebra. We find remarkably simple commutation relations for the three
``generator''' subalgebras (aka the vertex, the loop and the edge
subalgebras), yet find that apart from the vertex algebra the other
subalgebras do not admit a simple presentation as monoid algebras over
words in the generators modulo the commutation relations. This failure
precisely signals the necessity of \emph{new approaches to
combinatorics} for the analysis of graph rewriting systems, and some
first ideas towards such concepts are presented. In the second part of
Section~\ref{sec:structureTheory}, we will then discuss a second type
of subalgebras that is particularly interesting in view of
applications, the so-called ``structural'' subalgebras. The most
remarkable such subalgebras is a subalgebra that may be seen as a
\emph{direct generalization of the HW algebra}, but we also find the
subalgebras of graph observables in a number of variants.\\

Our new framework closes one very important gap in the literature: how
are the theoretical computer science, the mathematical combinatorics,
the chemical reaction systems and the statistical physics descriptions
of probabilisitic systems with the Markov property related? While we
leave it to~\cite{bdg2016,bd2016} and future work for the practical
answers to these questions, it is the core framework presented in this
paper that makes an entirely novel way of approaching these research
questions possible. We took great care in the present paper to make
the material accessible to a broader audience, and made an attempt to
render it self-contained (including an extended Appendix A on
background material for combinatorial Hopf algebras).\\

The reader mostly interested in the core results of the framework
might wish to focus on Sections~\ref{sec:RDpart}, \ref{sec:HW}
and~\ref{sec:structureTheory}, yet we would hope that in particular
the mathematical combinatorists might take interest also in the
mathematical details of the framework as presented in 
Sections~\ref{sec:RDhopf} and~\ref{sec:structureTheory}.

\subsection{Relations to previous work}

\paragraph{Combinatorics}

Consider the following very simple graph rewriting system, consisting
of the rule for creating a vertex ($\GRule{\emptyset}{\bullet}$) and the one for deleting a vertex ($\GRule{\bullet}{\emptyset}$). This
heuristic description is quite reminiscent of the interpretation of
the generators of the Heisenberg-Weyl algebra as creation and
annihilation of indistinguishable particles
\footnote{More precisely,
  in the canonical representation of the HW algebra, one chooses a
  basis of \emph{number vectors} $|n\rangle$ (each of which in
  physical terms presents a pure state of exactly $n$
  indistinguishable particles), and defines the representation via the
  two equations
  \begin{align*}
    a^{\dag}|n\rangle:= |n+1\rangle\,,\quad
    a|n\rangle:=\delta_{n,0}0|0\rangle+(1-\delta_{n,0})n|n-1\rangle\,,
  \end{align*}%
where $a^{\dag}$ and $a$ are the representations of the creation and
of the annihilation operator, respectively.}.
The deeper combinatorial
meaning of this algebra was clarified by Blasiak, Duchamp, Solomon, Horzela and Penson in their seminal work on the combinatorial Hopf algebra construction of the
Heisenberg-Weyl algebra~\cite{blasiak2010combinatorial}. At the core of their framework is a
diagrammatic description of the generators as well as a form of
diagram composition operation. In our new interpretation, this
diagrammatic composition is seen as a combinatorial description of the
possible sequential applications of the respective graph rewriting steps.\\

As we present in this paper, this construction can be extended to
the more general combinatorial Hopf algebra of generic graph rewriting rules,
which we call the rule diagram algebra. This opens a new perspective
for the study of graph rewriting systems by techniques well established
in the combinatorics literature, such as formal power series, generating
functions and related concepts (see e.g.~\cite{blasiak2011combinatorial}).

\paragraph{Computer science}

\emph{Traces are to be taken seriously} -- this is the slogan of
concurrency theory, the subfield of computer science concerned with
endowing models of computation with adequate notions of trajectories (called
\emph{traces} in that literature) that capture particular
behavioural properties such as causal relationships and conflicts.
Arguably, the most successful approaches provide models which are diagrammatic in essence: confer the work on concurrent game models for Petri nets (aka chemical reaction
systems) \cite{games2016}  or for the Pi Calculus~\cite{eberhart2015intensionally}.\\

This work is an initial step towards a theory relating these processes
and their diagrammatic trace models to realizations of such processes
as elements of the relevant algebras -- especially generalizations of the Heisenberg-Weyl
algebra. We give a new perspective on both matters through the definition of
\emph{rule diagrams}, a notion of behavioral traces that collectively form a
Hopf algebra. We foresee a great potential for future cross-fertilization,
starting with the study of \emph{trace compression}~\cite{danos2012graphs}
in relation with our notion of diagram reduction. Conversely, the foundational
work on diagrammatic models evoked above will provide inspiration for the
definition the diagram algebras of more structured processes such as
the Pi calculus.

\paragraph{Statistical physics and theoretical chemistry}

Doi's work on the second quantization of the description of chemical reaction systems~\cite{doi1976second}
relies on the implementation of chemical reactions as representations of
elements of the Heisenberg-Weyl algebra acting on probability distributions
over the species count state space. Moreover, the use of diagrams reminiscent of Feynman diagrams by Baez and Biamonte~\cite{baez2012course} to interpret chemical reactions is also reminiscent of the rule diagrams used in this paper. Quite remarkably, at the intersection of techniques from combinatorics and statistical physics, it is possible to derive closed-form solutions for the time evolution of species count probability distributions for certain types of chemical reaction systems~\cite{jahnke2007solving}, which elicits the hope that similarly powerful results may be obtainable via the analogues of these techniques for stochastic graph rewriting systems.\\

More abstractly, the physical concept of \emph{worldlines} of particles
is reflected to a certain extent in the syntax of the rule diagrams
through causality constraints bearing on vertices and edges. A given
rule diagram represents the ``time evolution'' of vertices and edges
through the course of sequential applications of rewriting steps.
In this way, one might indeed interpret rule diagrams as some form of analogues of
Feynman diagrams for modeling interactions in particle physics.

\section*{Acknowledgements} 

The authors acknowledge support by the \emph{Advanced ERC grant RULE/320823}. NB and IG would like to thank the Antique team at the D\'{e}partement d'Informatique of the ENS Paris for hospitality, where part of this work was carried out. TH would like to acknowledge support by the \emph{Marie Curie action IEF PEOPLE 628877 RUBYX}.

\section{Core concept: the rule diagram algebra}\label{sec:RDpart}

The aim of a \emph{diagram algebra} construction is to be able to enumerate all combinatorial possibilities of compositions of operations from a given associative algebra of interest, and to provide a \emph{reduction morphism} that allows to interpret each possible diagrammatic composite as an element of said associative algebra, thereby not only reproducing the original associative algebra composition in a fashion amenable to combinatorial studies, but also providing deeper mathematical insights into the structure of the associative algebra. Moreover, just as the Heisenberg-Weyl algebra possesses an essentially unique generic representation (in the form of the bosonic Fock space construction), more generally a diagram algebra approach may be employed to derive such an essentially unique representation also for a more general class of combinatorial algebras. The representation-theoretic aspect will be one of the main topics of the third paper in our current series~\cite{bd2016}.\\

In the present paper, we will introduce the construction of the diagram algebra for the associative algebra of graph transformation systems, which we call the \emph{rule diagram algebra}. %
From this rule diagram algebra, we obtain the four types of rule algebras, i.e.\ the algebras of the four types of graph rewriting, via the operation of \emph{reduction} (to be defined shortly). We will refer to this realization from hereon as \emph{diagrammatic realization of the rule algebra} (as opposed to a category-theoretical realization of the rule algebras that we plan to present in future work). Notably, two of the four types of graph rewriting defined via this new framework had been hitherto unknown in the literature. %
Moreover we will be able to extract important aspects of the combinatorics of graph rewriting from the rule diagram construction by demonstrating that it can be endowed with the structure of a \emph{combinatorial Hopf algebra}. To the best of our knowledge, both the universal aspects of the construction of a diagram algebra as well as specific constructions of the rule diagram and rule algebras are novel and original results.\\

Our construction may be summarized as follows: we will first define a presentation of \emph{graph transformation rules} in terms of so-called \emph{irreducible rule diagrams}. The crucial step then consists in lifting the \emph{set} of isomorphism classes of such diagrams to  the $\bK$-\emph{vector space} of isomorphism classes of irreducible rule diagrams, denoted $Irr(\cD)$ (where $\bK$ is a field such as $\bR$ or $\bC$). Next, the \emph{algebras of graph transformations} aka \emph{rule algebras} $\cR_T$ of the four different types $T$ ($T\in\{DPO,SPO_A,SPO_B,SPO_{AB}\}$)  are implemented via first embedding the irreducible rule diagrams into a space of generic rule diagrams $\cD$ via an \emph{insertion morphism} $\psi:Irr(\cD)\rightarrow \cD$, and defining a \emph{diagrammatic composition operation} on this space of generic rule diagrams $\cD$, thereby endowing it with the structure of a $\bK$-algebra, $\cD\equiv (\cD,+,\cdot,*_{\cD})$. We will then provide  \emph{reduction morphisms} $\bar{\varphi}_T:\cD\rightarrow Irr(\cD)$ that are shown to be algebra homomorphisms, thereby defining a notion of composition operation $*_{\cR_T}$ on $\cR_T\equiv(Irr(\cD),+,\cdot,*_{\cR_T})$. The precise definition of the reduction morphisms is in the form of the composition of a number of auxiliary morphisms, as presented below:
\begin{equation}\label{eq:outline}
\includetikz{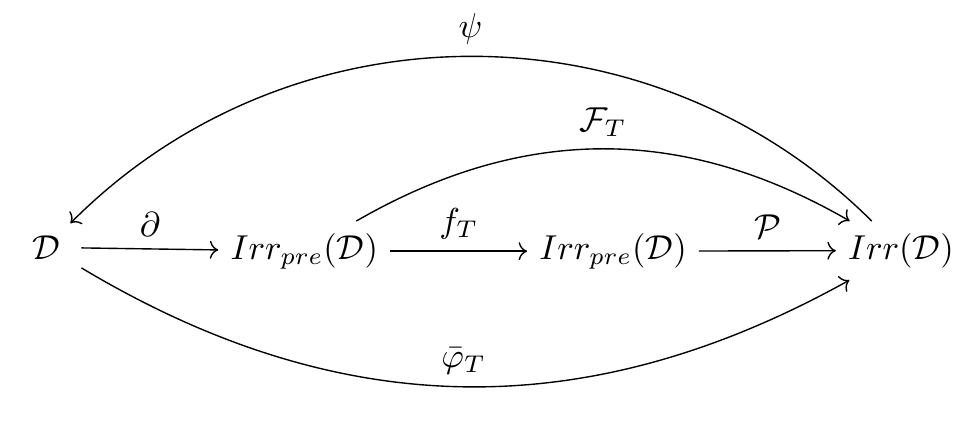}
\end{equation}
Remarkably, the rule algebras derived from the two known main variants of graph rewriting, namely \emph{double pushout (DPO)} and \emph{single pushout (SPO)} rewriting\footnote{To be slightly more precise, our variant of $SPO$ graph rewriting aka $SPO_A$ rewriting is in fact equivalent to the sesqui-pushout rewriting construction of~\cite{corradini2006sesqui} for the special case of injective matches and linear rules (cf.\ loc.\ cit.~Proposition 13).}, plus two hitherto more variants can be all encoded in \emph{one} universal construction, based on a single construction of a \emph{diagram algebra}! In fact, as we will present in the following, the four variants differ only in the definitions of the respective ``fixing morphisms'' $\cF_T$, and the existence of the two novel variants of graph rewriting was discovered in analyzing the definitions of $\cF_T$ for $T\in\{DPO,SPO_A\}$.\\

Let us begin by recalling some necessary definitions and concepts from the theory of relations, graph theory and from the theory of graph rewriting, thereby in particular also fixing notations and conventions for the rest of the paper. 

\subsection{Preliminaries on relations}

Relations will play a key role in all of our constructions, whence let us briefly recall some notations and conventions from the standard literature (cf. e.g.~\cite{brink2012relational}):
\begin{defn}
	A \emph{(binary) relation} $R$ of two finite sets $A$ and $B$ is a subset of $A\times B$. We denote the \emph{set of relations} between any two finite sets $A$ and $B$ by $Rel(A,B)$. Some standard definitions for relations that we will use are the \emph{domain} $dom(R)$ of a relation $R$,
\begin{equation}
	dom(R):=\{ a\in A\mid \exists b\in B: (a,b)\in R\}\,,
\end{equation}
and the \emph{image} $im(R)$,
\begin{equation}
	im(R):=\{ b\in B\mid \exists a\in A: (a,b)\in R\}\,.
\end{equation}
The \emph{converse} $R^{\rConv}\in Rel(B,A)$ of a relation $R\in Rel(A,B)$ is defined as
\begin{equation}
	R^{\rConv}:=\{ (b,a)\in B\times A\mid (a,b)\in R\}\}\,.
\end{equation}
The \emph{complement} $\bar{R}\in Rel(A,B)$ of a relation $R\in Rel(A,B)$ is defined as
\begin{equation}
	\bar{R}:= (A\times B)\setminus R\,.
\end{equation}
Interesting operations on relations include their \emph{union}, \emph{intersection} and \emph{inclusion}, all inherited from the respective operations on the underlying sets of pairs.\\

We denote the set of \emph{one-to-one relations} between two finite sets $A$ and $B$ by $Rel_{11}(A,B)$, where a relation $R\in Rel(A,B)$ is called one-to-one if and only if\footnote{Whence we may associate to each injective partial map between to finite sets a one-to-one relation.}
\begin{equation}
	(a_1,b_1),(a_2,b_2)\in Rel(A,B):
	(a_1,b_1)\neq(a_2,b_2)\Rightarrow
	(a_1\neq a_2\land b_1\neq b_2)\,.
\end{equation}
As an important special case, we define the \emph{identity relation} $id_A\in Rel_{11}(A,A)$ as
\begin{equation}
	id_A:=\{ (a,a)\mid a\in A\}\,,
\end{equation}
which coincidentally constitutes an example of a function on $A$.
\end{defn}
We will make extensive use of the notion of \emph{composition} of relations:
\begin{defn}
	Let $R\in Rel(A,B)$ and $S\in Rel(B,C)$ be relations of finite sets. Then we may define the \emph{composition} of the two relations, denoted by\footnote{The variant employed in this paper appears to be the version common in mathematics textbooks, since it is compatible with the notation for compositions of functions, which constitutes as special case of the composition of relations. However, in logic and computer science textbooks, the alternative notation $R;S:=S\circ R$ seems to be more common.}
 $S\circ R$:
 \begin{equation}
 \begin{aligned}
 	Rel(A,C)&\ni S\circ R:=\{
 	(a,c)\in A\times C\mid \exists b\in B, (a,b)\in R\land (b,c)\in S\}\,.	
 \end{aligned}
 \end{equation}
 \end{defn}
 The concepts and notations introduced thus far allow to express special properties rather succinctly (\cite{kahl2011collagories}, p.~300, Def.~2.1.6):
\begin{defn} A relation $R\in Rel(A,B)$ is 
\begin{itemize}
	\item \emph{univalent} iff $R\circ R^{\smallsmile}\subseteq id_B$,
	\item \emph{total} iff $id_A \subseteq R^{\smallsmile}\circ R$,
	\item \emph{injective} iff $R^{\smallsmile}\circ R\subseteq id_A$,
	\item \emph{surjective} iff $id_B\subseteq R\circ R^{\smallsmile}$,
	\item a \emph{mapping} iff it is univalent and total,
	\item \emph{bijective} iff it is injective and surjective.
	\end{itemize}
This identifies one-to-one relations as univalent injective relations.
\end{defn}
A useful corollary is the following:
\begin{cor}
The composition of two one-to-one relations $R\in Rel_{11}(A,B)$ and $S\in Rel_{11}(B,C)$ is a one-to-one relation, i.e.\ $S\circ R\in Rel_{11}(A,C)$.
\begin{proof}
	By assumption and by the definition of one-to-one relations, we have that
	\begin{align*}
		R&\circ R^{\smallsmile}\subseteq id_B\,,\; R^{\smallsmile}\circ R\subseteq id_A\\
		S&\circ S^{\smallsmile}\subseteq id_C\,,\; S^{\smallsmile}\circ S\subseteq id_B\,,
	\end{align*}
	and moreover evidently
	\[
		(S\circ R)^{\smallsmile}=R^{\smallsmile}\circ S^{\smallsmile}\,.
	\]
	The proof follows via a short computation:
	\begin{align*}
		(S\circ R)\circ ((S\circ R)^{\smallsmile})&=
		S\circ R\circ 
		R^{\smallsmile}\circ
		S^{\smallsmile}
		\subseteq S\circ id_B \circ S^{\smallsmile}\subseteq id_C\\
		((S\circ R)^{\smallsmile})\circ (S\circ R)&=
		R^{\smallsmile}\circ
		S^{\smallsmile}\circ 
		S\circ R\subseteq R^{\smallsmile}\circ id_B \circ R\subseteq id_A\,.
	\end{align*}
\end{proof}	
\end{cor}
We will need the following special property of relations $R\in Rel(A,A)$ in the following:
\begin{defn}
A relation $R\in Rel(A,A)$ is called \emph{cyclic} if and only if there exists a finite sequence of elements $a_1,\dotsc, a_k$ with $k>1$ such that $(a_j,a_{j+1})\in R$ for all $j=1,\dotsc,k-1$ and $(a_k,a_1)\in R$. Otherwise, it is called \emph{acyclic}.
\end{defn}
As a final piece of elementary definitions, let us introduce a useful notation from the theory of Kleene algebras:
\begin{defn}
The \emph{Kleene star}~\cite{kozen1994completeness} ${}^{*}$ of the composition operation $\circ$ for a relation $R\in Rel(A,A)$ of a finite set $A$ is defined as\footnote{It is clear that since any relation $R\in Rel(A,A)$ is by definition a relation of a finite set $A$, and whence a subset of the finite set $A\times A$, also $R^{*}$ will be a finite relation by construction}
\begin{equation}
	R^{*}:=id_A\cup \bigcup_{n=1}^{\infty} R^{\circ\:n}\,.
\end{equation}
More intuitively, the Kleene star operation maps a relation $R\in Rel(A,A)$ to its reflexive transitive closure.
\end{defn}

\subsection{Graph theory preliminaries}

For the sake of concreteness and simplicity, we will mainly focus in this paper on \emph{directed multigraphs} (without colors), albeit we will briefly comment in Section~\ref{sec:otherGraphs} on more generic graphs:
\begin{defn}[Directed multigraphs]
A \emph{finite directed multigraph} $G\equiv(V,E,s,t)$ is given in terms of the following pieces of data:
\begin{itemize}
	\item A (finite) \emph{set of vertices} $V$;
	\item a (finite) \emph{set of edges} $E$;
	\item (total) \emph{source and target maps} $s:E\rightarrow V$ and $t:E\rightarrow V$.
\end{itemize}	
In particular, the definitions ensure that each edge $e\in E$ has precisely one source and one target vertex, albeit we allow for the case of these vertices to coincide, thus allowing for the occurrence of \emph{self-loops}. We will denote the \emph{set of all finite directed multigraphs} by $\fG$. As a convention, we will refer to elements of $\fG$ as ``graphs'' from hereon unless explicitly mentioned otherwise.
\end{defn}
An important concept for our work is the notion of graph morphisms, which in the current framework may be defined succinctly as follows:
\begin{defn}
	A \emph{graph morphism} $f:\GRule{G}{H}$ between two graphs $G$ and $H$ is a pair of relations $f\equiv(f_V,f_E)$ such that the mapping of vertices and edges are compatible:
	\begin{equation}
	  \includetikz{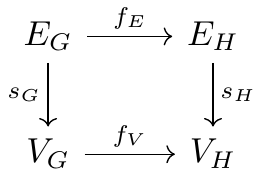}\,,\quad 
     \includetikz{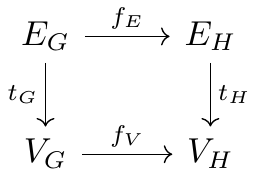}\,.
	\end{equation}
These properties may be alternatively expressed as
\begin{equation}
	(e,e')\in f_E \Rightarrow \{(s_G(e),s_H(e')),(t_G(e),t_H(e'))\}\sqsubseteq f_V\,.
\end{equation}
A graph morphism is called \emph{injective}, \emph{surjective} or \emph{bijective} if the vertex and edge morphisms have the respective properties. The set of bijections aka isomorphisms from a graph $G$ to a graph $G'$ will be denoted $Iso(G,G')$. We will frequently use the term \emph{linear rules} for injective partial graph morphisms for brevity, and refer to the set of linear rules from a graph $G$ to a graph $G'$ as $Lin(G,G')$.
\end{defn}

For the purpose of graph rewriting, we will be interested to work with isomorphism classes of graphs:
\begin{defn}
Let $G\in\bG$ be a graph. Then we denote by $[G]_{\cong}$ the \emph{isomorphism class} of $G$, defined as
\begin{equation}
[G]_{\cong}:=\{ G'\mid \exists f\in Iso(G.G'): G' =f(G)\}\,.
\end{equation}
The \emph{set of isomorphism classes} of finite directed multigraphs will be denoted $\fG_{\cong}$.
\end{defn}

An example for an injective partial map aka linear rule (with the map itself reconstructible from the labels provided, i.e.\ a vertex that is in the image of $t$ carries the same label as the vertex in the domain, but decorated with a $'$ accent, and analogously for the edges) is provided below:
\begin{equation}\label{eq:ruleEx2}
\bullet_1\xrightarrow[a]{}\bullet_2\xrightarrow[b]{}\bullet_3\quad \stackrel{t}{\rightharpoonup}
\quad\bullet_{2'}\xrightarrow[b']{} \bullet_{3'}\xrightarrow[c']{}\bullet_{4'}
\end{equation}

It will prove necessary to consider linear rules as maps of \emph{isomorphism classes} of graphs, a concept which is made concrete via the following definition:
\begin{defn}
An \emph{isomorphism of linear rules} $r_i:\GRule{I_i}{O_i}$ ($i=1,2$) is defined as a pair of graph isomorphisms $i\equiv(i_I,i_O)\in Iso(I_1,I_2)\times Iso(O_1,O_2)$, which induces the transformation $r_1\equiv (r^{V_1},r^{E_1})\mapsto r_2\equiv(r^{V_2},r^{E_2})$ according to
\begin{equation}
\includetikz{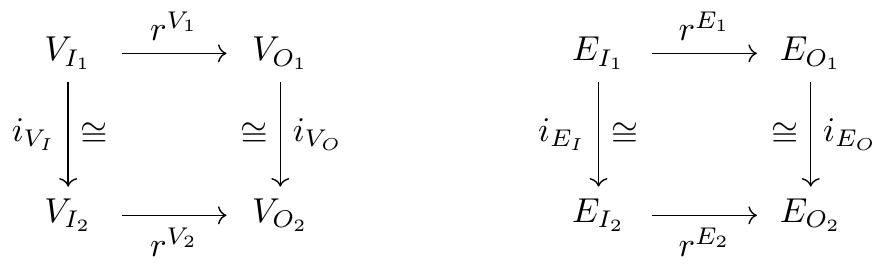}\,.
\end{equation}
Here, the notations $V_G$ and $E_G$ refer to the vertex and edge sets of a given graph $G$, respectively, and analogously for the various components of the morphisms. We will denote the set of \emph{isomorphism classes of linear rules between two  graphs\footnote{According to the definition, we are thus more precisely considering the isomorphism class of a linear rule between two isomorphism classes of graphs $I$ and $O$ for each representative of $Lin(I,O)$.} $I$ and $O$ by $Lin(I,O)$}.
\end{defn}
As a convention, we will always imply by ``a linear rule'' its isomorphism class, of which a given explicit realization via a choice e.g.\ of labelling is only a representative.\\

A standard operation from the theory of graphs is the notion of forming the superposition aka disjoint union of graphs, which in our generalized context amounts to the following:
\begin{defn}
Let $G_i\equiv(V_i,E_i,s_i,t_i)\in \fG$ ($i=1,2$) be two graphs. Then we define their \emph{superposition} as
\begin{equation}
\begin{aligned}
	G_1\uplus G_2&:=(V_1\uplus V_2,E_1\uplus E_2,s_1\uplus s_2,t_1\uplus t_2)\,,
	\end{aligned}
\end{equation}	
where the notation $\uplus$ stands for the disjoint union of sets. This concept naturally extends to disjoint unions of isomorphism classes of graphs and of  linear rules.
\end{defn}

With these preparations, we will distinguish amongst all possible linear rules the following cases:
\begin{defn}[Connected linear rules]\label{def:connRules}
A linear rule $r:\GRule{I}{O}$ is called \emph{connected} if and only if it cannot be written as a disjoint union of two rules.
\end{defn}
More intuitively, a rule $r:\GRule{I}{O}$ is connected if its graph (obtained by drawing both $I$ and $O$ as well as the relation $r$) is connected.

\subsection{Rule diagrams}

One of the main concepts of the present paper is the notion of rule diagrams:
\begin{defn}[Rule diagrams]
A \emph{rule diagram} 
\[
\fd\equiv(I,O,r,m,s,t)
\]
consists of the following pieces of data: 
\begin{equation}
\begin{aligned}
	I&\equiv \uplus_{i=1}^n I_i\,,\; O\equiv \uplus_{i=1}^n O_i\\\
	r&\equiv\left(\uplus_{i=1}^n r_i\right)\in Rel_{11}(I,O)\,,\quad m\equiv \left(\biguplus_{\stackrel{i,j=1}{i<j}}^n 
	\{m_{i,j}\}\right)\in Rel_{11}(O,I)\\
	s&:= \biguplus_{i=1}^n (s_{I_i}\uplus s_{O_i})\,,\; 
	t:= \biguplus_{i=1}^n (t_{I_i}\uplus t_{O_i})\,,
\end{aligned}
\end{equation}
Here, $r:\GRule{I}{O}$ is a collection of \emph{connected linear rules} (cf.\ Definition~\ref{def:connRules}) $r_i:\GRule{I_i}{O_i}$ (with $1\leq i\leq n$ for some $n\in \bZ_{>0}$), while $m:\GRule{O}{I}$ denotes a collection of \emph{matches}
\begin{equation}
	m_{i,j}\equiv I_i\stackrel{m_{ij}}{\leftharpoonup}O_j
	\equiv\{m_{i,j}^V,m_{i,j}^E\}\in Rel_{11}(V^{O_j},V^{I_i})\uplus Rel_{11}(E^{O_j},E^{I_i})\,,
\end{equation}
where $n>0$ is the (finite) \emph{number of constituents}\footnote{More precisely, we explicitly allow for the cases where we have $r_i=\emptyset$ and either $I_i=\emptyset$ or $O_i=\emptyset$, thus being able to realize also rules that have disconnected left or right hand side graph connected components that are not mapped by any part of the partial map $r$.}. We also define as a special case the \emph{empty rule diagram}
\begin{equation}
	\fd_{\emptyset}:=(\emptyset,\emptyset,\emptyset,\emptyset,\emptyset,\emptyset)\,.
\end{equation}
The rule diagrams must fulfill the following \emph{consistency condition}: if we define the \emph{worldline relations} of the rule diagram as
\begin{equation}
\begin{aligned}
		\omega_r&:=r\circ(m\circ r)^{*}\in Rel_{11}(V^I,V^O)\uplus Rel_{11}( E^I,E^O)\\
	\omega_m&:=m\circ(r\circ m)^{*}\in Rel_{11}(V^O,V^I)\uplus Rel_{11}( E^O,E^I)\,,
\end{aligned}
\end{equation}
then we require the \emph{delayed edge morphism condition} to be fulfilled, namely that
\begin{equation}
\begin{aligned}
	&(e,e')\in \omega_m\cap (E^O\times E^I)\\
	&\quad \xRightarrow{!}\quad (s(e),s(e'))\in \omega_m\cap V^O\times V^I\land (t(e),t(e'))\in \omega_m\cap V^O\times V^I\,.
\end{aligned}
\end{equation}
\end{defn}
An first example of a rule diagram is provided in~\eqref{eq:ruleEx2asRD}, while~\eqref{eq:rdIllustration2} illustrates the delayed edge morphism condition.\\

Since in our framework linear rules are \emph{isomorphism classes} of linear rules, we need to also define a consistent concept of \emph{isomorphisms of rule diagrams}:
\begin{defn}
Let $\fd\equiv(I,O,r,m,s,t)$ be a rule diagram as before. Then each pair of graph isomorphisms $i\equiv(i_I,i_O)\in Iso(I,I')\times Iso(O,O')$ of the input and output graphs $I$ and $O$ is defined to induce an \emph{isomorphism of rules diagrams} according to
\begin{equation}
	\begin{aligned}
		(I,O,r,m,s,t)&\xrightarrow[\cong]{i} (I',O',r',m',s',t')\\
		r'&:= i_O\circ r\circ i^{-1}_I\,,\quad m':=i_I\circ m\circ i^{-1}_O\\
		s'&:=i_I\circ s\circ i_I^{-1}\uplus i_O\circ s\circ i_O^{-1}\\
		t'&:=i_I\circ t\circ i_I^{-1}\uplus i_O\circ t\circ i_O^{-1}\,.
	\end{aligned}
\end{equation}
We denote the \emph{set of isomorphism classes of rule diagrams} by $\fD$, and the \emph{$\bK$-vector space of isomorphism classes of rule diagrams} by 
\begin{equation}
	\cD:=span_{\bK}(\fD)\equiv(\cD,+,\cdot)\,.
\end{equation}
\end{defn}

Let us pause for a moment to focus on a notational convention that is made throughout this paper in order to emphasize a very important distinction:
\begin{defn}
	In this paper, we will reserve \emph{fraktur symbols} for \emph{sets of objects}, while \emph{calligraphic letters} are reserved for \emph{$\bK$-vector spaces} (and more sophisticated structures thereover) constructed over a given set. More precisely, if $\frak{A}$ is a set, then we denote by
	\begin{equation}
		\cA:=span_{\bK}(\frak{A})\equiv(\cA,+.\cdot)
	\end{equation}
	the $\bK$-vector space of finite linear combinations of the basis vectors associated to the elements of the set $\frak{A}$. This entails that if we introduce an arbitrary, but fixed unique indexing scheme $\cI$ on the set $\frak{A}$,
	\begin{equation}
		\frak{A}=\biguplus_{i\in \cI}\{\frak{a}_i\}\,,
	\end{equation}
	every element of the $\bK$-vector space $\cA$ may be expressed uniquely as finite linear combination
	\begin{equation}
		\cA\ni a=\sum_{i\in \cI} \kappa_i\delta(\frak{a}_i)\,,
	\end{equation}
	with $\delta(\frak{a}_i)\in \cA$ the \emph{basis vector associated to the element $\frak{a}_i\in \frak{A}$}, and
	where finiteness entails that at most a finite number of the coefficients $\kappa_i\in \bK$ are nonzero.\\
	
	\underline{\textbf{Convention:}} In a slight abuse of notations, we will write
	\begin{equation}
	\frak{a}_i\in \cA	
	\end{equation}
	to denote the \emph{basis vector} $\delta(\frak{a}_i)$  associated to the element $\frak{a}_i$ (as opposed to $\frak{a}_i\in\frak{A}$ if we are explicitly referring to the element of the \emph{set} $\frak{A}$).\\
	
	For example, in these notations the operations of the $\bK$-vector space, the \emph{vector addition} $+$ and \emph{multiplication by scalars} $\cdot$, are defined such as to fulfill the following standard consistency condition (aka \emph{multilinearity}):
	\begin{equation}
		\forall a,b\in \cA,\; \alpha,\beta\in \bK:\quad \alpha\cdot a+\beta\cdot b
		=\sum_{i\in \cI} (\alpha \kappa_{a_i}+\beta \kappa_{b_i})\frak{a}_i\,.
	\end{equation}
\end{defn}

In the following, it will often prove crucial to explicitly distinguish between individual elements $\fd\in\fD$ of the \textbf{\emph{set}} of isomorphism classes of rule diagrams and generic elements $d\in \cD$ of the \textbf{\emph{vector space}} of isomorphism classes of rule diagrams, which are in general linear combinations of rule diagrams. We thus fix the following convention:
\begin{defn}
In a slight abuse of notations, we will write $\fd\in \fD$ for the element of the \emph{set} of linear rules $\fD$, as well as $\fd\in \cD$ for the associated basis vector $\delta(\fd)\in \cD$ of the \emph{vector space} of linear rules. In order to further economize the language, we will refer to rule diagrams $\fd\equiv \delta(\fd)\in \cD$ as  \textbf{\emph{basis diagrams}}.
\end{defn}

The moniker rule \emph{diagrams} is motivated by the following intuition: consider for example the linear rule presented as an example in~\eqref{eq:ruleEx2}. We may visualize it as a rule diagram in the form
\begin{equation}\label{eq:ruleEx2asRD}
\left(\bullet_1\xrightarrow[a]{}\bullet_2\xrightarrow[b]{}\bullet_3\quad \stackrel{t}{\rightharpoonup}
\quad\bullet_{2'}\xrightarrow[b']{} \bullet_{3'}\xrightarrow[c']{}\bullet_{4'}\right)
\quad \widehat{=}\quad
\includetikz{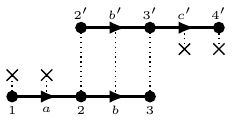}
\end{equation} 
We may thus interpret a rule diagram as a sort of ``time evolution'' of a set of graphs (along the ``optimistic arrow of time'' which points vertically upwards). In the example, the vertex labelled $1$ and the edge labelled $a$ are ``deleted'' (as indicated by the symbol $\times$ which marks the ``time of deletion''), while the vertices $2$ and $3$ as well as the edge $b$ are preserved (indicated by connecting dotted lines), with finally the vertex $3'$ and $c'$ being ``created'' (indicated by a symbol $\times$ marking the ``time of creation''). A much more thorough prescription of how to construct the actual diagrams as well as an interpretation of the dotted lines in the pictures as worldlines will be presented in Section~\ref{sec:viz}.\\

Amongst the possible generic rule diagrams, we discern two special classes:
\begin{defn}
The \emph{set of isomorphism classes of atomic rule diagrams} $\frak{Atom}(\fD)$ is defined as
\begin{equation}
	\frak{Atom}(\fD):=\{ \fd\in \fD\mid \fd\equiv(I,O,r,\emptyset,s,t)\}\,,
\end{equation}	
where $r:\GRule{I}{O}$ is a \emph{single connected rule}; the corresponding $\bK$-vector space is denoted by 
\begin{equation}
	Atom(\cD):=span_{\bK}(\frak{Atom}(\fD))\,.
\end{equation}
Moreover, we define the \emph{set of isomorphism classes of irreducible rule diagrams} $\frak{Irr}(\fD)$ as
\begin{equation}
\begin{aligned}
		\frak{Irr}(\fD)&:=\bigg\{\fd\in \fD\bigg\vert 
	\fd=\left(\biguplus_{m=1}^n \fd_{i_m}\right)\equiv(\uplus_{m=1}^n I_{i_m},\uplus_{m=1}^n O_{i_m},\uplus_{m=1}^n r_{i_m},\emptyset,s,t)\\
	&\qquad \qquad \qquad \fd_{i_m}\in \frak{Atom}(\fD)\,,\; {i_m}\in \cI\,, \; 0\leq n<\infty\bigg\}\,,
\end{aligned}
\end{equation}
where we consequently require the partial homomorphisms $r_i:\GRule{I_i}{O_i}$ to be connected linear rules; the corresponding $\bK$-vector space is defined as
\begin{equation}
	Irr(\cD):=span_{\bK}(\frak{Irr}(\fD))\,.
\end{equation}
\end{defn}
We evidently have that
\begin{equation}
	\frak{Atom}(\fD)\subsetneq \frak{Irr}(\fD)\subsetneq \fD\,,
\end{equation}
and moreover the inclusion of vector spaces
\begin{equation}
	Atom(\cD)\subsetneq Irr(\cD)\subsetneq \cD\,.
\end{equation}
Note that we consider the \emph{empty rule diagram}
\begin{equation}
	d_{\emptyset}\equiv\fd_{\emptyset}\equiv(\emptyset,\emptyset,\emptyset,\emptyset,\emptyset,\emptyset)
\end{equation}
as an element of all three vector spaces.\\

As mentioned in the introduction, the irreducible rule diagrams play an important role in our framework since they directly present superpositions of connected rules, and whence generic graph transformation rules. Our aim will now consist in equipping the $\bK$-vector space $Irr(\cD)$ with a notion of \emph{composition}, or rather with four different types of composition that implement the four possible types of compositions of graph transformation rules.

\subsection{Definition of the rule diagram algebra}

In the following, $Irr(\cD)$ will play two different roles -- it will evidently feature as a subvector space of $\cD$, but it will also serve as the vector space underlying the four types of rule algebras $\cR_T$ that will be defined in this subsection.\\

The starting point of our rule algebra constructions will consist in the definition of an \emph{insertion morphism} that simply acts as the identity on any irreducible rule diagram:

\begin{defn}\label{def:insertionMorph}
The \emph{insertion morphism} is defined on pure irreducible rule diagrams as
\begin{equation}
	\psi:\frak{Irr}(\fD)\hookrightarrow \fD: \fd\mapsto \fd\,,
\end{equation}	
extended to the a vector space homomorphism $\psi: Irr(\cD)\hookrightarrow \cD$ via multilinearity.
\end{defn}

We furthermore need to define the notion of interfaces of rule diagrams:
\begin{defn}
	The \emph{input interface} $\cI(\fd)$ and the \emph{output interface} $\cO(\fd)$ of a rule diagram $\fd\equiv(I,O,r,m,s,t)$ are defined as
	\begin{equation}
		\cI(\fd):= I\setminus codom(m)\,,\quad \cO(\fd):=O\setminus dom(m)\,.
	\end{equation}
\end{defn}
It is worthwhile emphasizing that in general the interfaces $\cI(\fd)$ and $\cO(\fd)$ of a rule diagram $\fd\in\fD$ are \emph{not} necessarily graphs, a fact which will play a prominent role in the sequel.\\

With these ingredients, we may now define the notion of \emph{composition} of rule diagrams:
\begin{defn}
Let 
\[
\fD\ni \fd_X\equiv (I_X,O_X,r_X,m_X,s_X,t_X)
\]
with $X\in \{A,B\}$ be two rule diagrams. Then the \emph{composition of $\fd_A$ with $\fd_B$ along the match $m_{AB}\in \cM_{\fd_A}(\fd_B)$}, denoted $\comp{\fd_A}{m_{AB}}{\fd_B}$, is defined as follows:
\begin{equation}
	\begin{aligned}
		&\comp{\fd_A}{m_{AB}}{\fd_B}\\
		&\quad :=(I_A\uplus I_B,O_A\uplus O_B,r_A\uplus r_B,m_A\uplus m_{AB}\uplus m_B,s_A\uplus s_B,t_A\uplus t_B)\,.
	\end{aligned}
\end{equation}
Here, the \emph{set of matches} $\cM_{\fd_A}(\fd_B)$ is defined as
\begin{equation}
	\cM_{\fd_A}(\fd_B):= \bigg\{
m_{AB}\in Rel_{11}(O_B ,I_A)\bigg\vert \comp{\fd_A}{m_{AB}}{\fd_B}\in \fD\bigg\}\,.
\end{equation}
\end{defn}
In other words, the \emph{set of matches} $\cM_{\fd_A}(\fd_B)$ consists of all those one-to-one relations $m_{AB}$ that render $\comp{\fd_A}{m_{AB}}{\fd_B}$ a consistent (pure) rule diagram. It may be constructed explicitly by fixing representatives of the isomorphism classes of $\fd_A$ and $\fd_B$, followed by listing the possible one-to-one relations for the given representatives, from which the consistent matches may be read off according to the consistency conditions for rule diagrams. Evidently, \emph{matches must respect orientations}, in the sense that directed edges can only be matched in the correct orientation. Note that an equivalent characterization of the set of matches $\cM_{\fd_A}(\fd_B)$ between two rule diagrams is given by
\begin{equation}
	\cM_{\fd_A}(\fd_B)=Inj(\cO(\fd_B),\cI(\fd_A))\,,
\end{equation}
whence a match between two rule diagrams $\fd_A$ and $\fd_B$ is nothing but an injective partial graph morphism from the output interface of $\fd_B$ to the input interface of $\fd_A$. A proof of this nontrivial statement may be found in~\cite{bdg2016}, Proposition~18.\\

The next step consists in defining a composition operation on $\cD$:
\begin{defn}
The \emph{composition} $*_{\cD}$ on the vector space $\cD$ is defined as follows: we first define a composition of basis diagrams $\fd_X\in \cD$:
\begin{equation}
	\begin{aligned}
		\fd_A*_{\cD}\fd_B&:=
		\sum_{m_{AB}\in \cM_{\fd_A}(\fd_B)} \comp{\fd_A}{m_{AB}}{\fd_B}\,.
	\end{aligned}
\end{equation} 	
We then extend this definition to the whole of $\cD$ via a bilinear extension of the composition operation,
\begin{equation}
\begin{aligned}
*_{\cD}:\cD\times\cD&\rightarrow \cD:\\
(d_A,d_B)\equiv \left(\sum_{i\in\cI}\kappa_{A_i}\fd_i, \sum_{j\in\cI}\kappa_{B_j}\fd_j\right)
&\mapsto 
d_A*_{\cD}d_B:= \sum_{i,j\in \cI} \kappa_{A_i}\kappa_{B_j} \fd_i *_{\cD} \fd_j\,.
\end{aligned}
\end{equation}
\end{defn}

As a special case, it follows from our definitions that the composition of the \emph{empty rule diagram} $d_{\emptyset}$ with any other rule diagram $d\in \cD$ fulfills
		\begin{equation}
			d*_{\cD}d_{\emptyset}\equiv d_{\emptyset}*_{\cD}d:=d\quad \forall d\in \cD\,.
		\end{equation}
This observation together with the definition of the composition operation leads to the following definition which is central to this work:
\begin{defn}
Endowing the $\bK$-vector space $\cD\equiv (\cD,+,\cdot)$ of isomorphism classes of rule diagrams with the composition operation $*_{\cD}$ results in a \emph{$\bK$-algebra} $\cD\equiv(\cD,+,\cdot,*_{\cD})$ which we call the \textbf{\emph{rule diagram algebra}}.
\end{defn}

We are now in a position to formulate the first central result of this paper:
\begin{thm}\label{thm:RDassoc}
	The rule diagram algebra $\cD\equiv(\cD,+,\cdot,*_{\cD})$ is an \emph{associative unital $\bK$-algebra}, with unit under composition $d_{\emptyset}$ (the empty rule diagram).
	\begin{proof}
		It is a straightforward exercise to demonstrate that the isomorphisms of two basis diagrams $\fd_A,\fd_B\in \fD$ are compatible with the isomorphisms of any given composite rule diagram $\comp{\fd_A}{m_{AB}}{\fd_B}$ according to our definitions of isomorphisms of rule diagrams, in the sense that any choice of representatives for $\fd_A$ and $\fd_B$ leads to the same set of matches and composite rule diagrams up to isomorphisms. Moreover, associativity follows directly from the definition of the composition operation, since both sequences of binary compositions of three pure rule diagrams $\fd_A,\fd_B,\fd_C\in \fD$ lead to the triple composite rule diagrams of the form
		\begin{equation}
			\begin{aligned}
			&(I_A\uplus I_B\uplus I_C,
			O_A\uplus O_B\uplus O_C,r_A\uplus r_B\uplus r_C, \\
			&\qquad m_A\uplus m_{B}\uplus m_C\uplus m_{AB}\uplus m_{AC}\uplus m_{BC},s_A\uplus s_B\uplus s_C,t_A\uplus t_B\uplus t_C)\,,
			\end{aligned}
		\end{equation}
		with the matches $m_{AB}$, $m_{AC}$ and $m_{BC}$ constrained by the requirement that the triple composite must be a consistent rule diagram. Evidently, associativity of the composition on rule diagrams $\fd_A,\fd_B,\fd_C\in \fD$ carries over to arbitrary linear combinations of rule diagrams via multilinearity. Finally, the statement that $d_{\emptyset}$ is the unit under composition follows directly from our definition of the composition operation $*_{\cD}$.
	\end{proof}
\end{thm}

\subsection{Definition of the reduction morphisms}

As mentioned in the introduction, the main idea behind our paradigm of a diagram algebraic construction is to encode a given associative algebra composition operation into a diagrammatic composition operation followed by a form of reduction operation. We have already defined the first part of this setup, in that for any of the four conceivable types of rule algebra composition $*_{\cR_T}$ we define the insertion morphism $\psi:Irr(\cD)\rightarrow \cD$ as a vector space homomorphism, which is consistent because the underlying vector space for the $\bK$-algebras $\cR_T$ will be $Irr(\cD)$ for all four types $T$. We will thus have to define four variants of a reduction operation that map a generic rule diagram to an irreducible one. We will proceed in two steps: firstly, there is a natural operation available (which we call the \emph{boundary map}) which pre-reduces a generic rule diagram to an irreducible pre-diagram. The latter will then be modified via four different types of \emph{fixing morphisms} into consistent irreducible rule diagrams, thereby completing the definition of the reduction maps.\\

We begin our construction by recalling that a generic basis diagram $\fd\in\fD$ encodes a succession of individual graph transformations that potentially overlap at some vertices and/or edges. As a first step towards a notion of reduction, we define the so-called boundary map:
\begin{defn}[Boundary map $\partial$]
Let $\fd\equiv(I,O,r,m,s,t)$ be a basis diagram. Then we define the \emph{boundary map} $\partial$ via 
\begin{equation}
	\begin{aligned}
		\partial(\fd)&:=(\cI(\fd),\cO(\fd),\partial(r),\emptyset,\partial(s),,\partial(t))\\
	\partial(r)&:=id_{\cO(\fd)}\circ \omega_r(\fd)\circ id_{\cI(\fd)}\,,\quad \omega_r(\fd):=r\circ(m\circ r)^{*}\\
		\partial(s)&:=id_{\cI(\fd)}\circ \omega_I\circ s\circ id_{\cI(\fd)}\uplus id_{\cO(\fd)}\circ \omega_O\circ s\circ id_{\cO(\fd)}\\
		\partial(t)&:=id_{\cI(\fd)}\circ \omega_I\circ t\circ id_{\cI(\fd)}\uplus id_{\cO(\fd)}\circ \omega_O\circ t\circ id_{\cO(\fd)}\\
		\omega_I &:=(r^{\rConv}\circ m^{\rConv})^{*}\,,\quad \omega_O:=(r\circ m)^{*}\,.
	\end{aligned}
\end{equation}	
We extend this map to generic linear combinations of pure rule diagrams by linearity, and denote the vector space of isomorphism classes of possible outcomes of this operation $Irr_{pre}(\cD)$, aka the space of \emph{rule pre-diagrams}. 
\end{defn}

According to our definition, we have that
\begin{equation}
Irr(\cD)\subsetneq Irr_{pre}(\cD)\,,	
\end{equation}
whence in general applying $\partial$ to a basis diagram $\fd\in \cD$ may not lead to a consistent irreducible rule diagram. More precisely, it may happen that $\partial(\fd)$ contains what is known in the graph rewriting literature as \emph{dangling edges}, manifesting themselves as edges $e\in E_{\partial(\fd)}$ for which either $\partial(s)(e)=\emptyset$ and/or $\partial(t)(e)=\emptyset$. As a simple example for such a case, consider the application of the boundary map to a composite of a diagram of an edge preserving rule with a vertex deletion and a vertex creation rule:
\begin{equation}
\partial\left(\includetikz{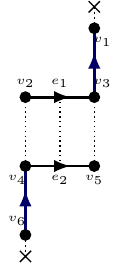}\right)=
\includetikz{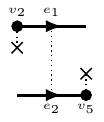}\,,
\end{equation}
whence the edge $e_1$ becomes a dangling edge via $\partial$ because $\partial(t)(e_1)=\emptyset$, and $e_2$ because of $\partial(s)(e_2)=\emptyset$.\\

It is evident that elements of $Irr_{pre}(\cD)$ containing dangling edges do not possess a consistent interpretation as linear rules. The final part of the definition of rule algebras thus consists in the construction of so-called \emph{fixing morphisms} that will allow to ``repair'' some or all of the diagrams $\partial(\fd)\in Irr_{pre}(\cD)\setminus Irr(\cD)$ according to four possible strategies, and discard those that are not ``fixable'' according to the given strategy:
\begin{defn}
The \emph{fixing morphisms} $\cF_T: Irr_{pre}(\cD)\rightarrow Irr(\cD)$ of types \[
T\in\{DPO,SPO_A,SPO_B,SPO_{AB}\}
\]
are defined as follows: we first define the \emph{partial fixing morphisms} $f_{SPO_A}$ and $f_{SPO_B}$ via their action on basis pre-rule diagrams $\fd\in Irr_{pre}(\cD)$,
\begin{equation}
	\begin{aligned}
		f_{SPO_A}&:Irr_{pre}(\cD)\rightarrow Irr_{pre}(\cD):\\
		&\fd\equiv (I,O,r,\emptyset,s,t)\mapsto f_{SPO_A}(\fd):=(I,O',r',\emptyset,s',t')\\
		O'&:=O\setminus \left\{
e\in E_O\mid s(e)=\emptyset \lor t(e)=\emptyset\right\}\\
r'&:=id_O'\circ r\\
s'&:=id_{(I\uplus O')}\circ s\circ id_{(I\uplus O')}\,,\;
t':=id_{(I\uplus O')}\circ t\circ id_{(I\uplus O')}\\
\\
    f_{SPO_B}&:Irr_{pre}(\cD)\rightarrow Irr_{pre}(\cD):\\
    &\fd\equiv (I,O,r,\emptyset,s,t)\mapsto f_{SPO_B}(\fd):=(I'',O,r'',\emptyset,s'',t'')\\
    I''&:I\setminus \left\{
e\in E_I\mid s(e)=\emptyset\lor t(e)=\emptyset\right\}\\
r''&:=r\circ id_{I''}\\
s''&:=id_{(I''\uplus O)}\circ s\circ id_{(I''\uplus O)}\,,\;
t'':=id_{(I''\uplus O)}\circ t\circ id_{(I''\uplus O)}
\,.
	\end{aligned}
\end{equation}
Their action on generic irreducible pre-rule diagrams is then defined by multilinear extension.\\

Next, we define the \emph{projection morphism}, which discards all irreducible pre-rule diagrams that are not consistent irreducible rule diagrams. It it is defined to act on basis diagrams $\fd\in \cD$ as
\begin{equation}
	\cP: Irr_{pre}(\cD)\rightarrow Irr(\cD): \fd\mapsto \begin{cases}
	\fd\,,\quad &\text{if } \fd\in Irr(\cD)\\
	0\cdot \fd_{\emptyset}\,, \quad &\text{else.}	
 \end{cases}\,,
\end{equation}
and is extended to all of $Irr_{pre}(\cD)$ by multilinearity. With these preparations, the four \emph{fixing morphisms} $\cF_T:Irr_{pre}(\cD)\rightarrow Irr(\cD)$ are then defined as follows:
\begin{equation}
	\begin{aligned}
		\cF_{DPO}&:= \cP\\
		\cF_{SPO_A}&:= \cP\circ f_{SPO_A}\\
		\cF_{SPO_B}&:= \cP\circ f_{SPO_B}\\
		\cF_{SPO_{AB}}&:= \cP\circ f_{SPO_A}\circ f_{sPO_B}\,.
	\end{aligned}
\end{equation}
\end{defn}
In other words, $f_{SPO_A}$ ``fixes'' edges with missing endpoint vertices in the right hand side graph $R$ of an irreducible pre-diagram by removing them (and also by restricting the partial morphism $t$ to the new right hand side graph $R'$),  while dually $f_{SPO_B}$ ``fixes'' edges with missing endpoint vertices in the left hand side graph analogously. Both maps are extended to all of $\cD$ via multilinearity as usual.\\

The four different variants of the ``fixings'' have been defined in order to precisely mirror the analogous definitions in ordinary graph rewriting where possible -- the experts in graph rewriting may recognize the feature that in $DPO$ graph rewriting it is not permitted to delete a vertex from a graph to which edges are attached unless these edges are also simultaneously deleted (because otherwise a ``dangling edge'' would result), while e.g.\ in $SPO$-rewriting (which we call $SPO_A$-rewriting) deletion of a vertex with incident edges ``automatically'' also deletes the incident edges. Interestingly, the other two feasible variants of $SPO_B$- and $SPO_{AB}$-rewriting appear to not have been introduced in the literature before. The precise relationship between our diagram algebraic approach and the more conventional category-theoretic approach to graph rewriting will be presented elsewhere.

\subsection{The four types of rule algebras $\cR_T$}

The second central result of this paper is the definition and characterization of the four different possible types of rule algebras $\cR_T$:
\begin{defn}
We denote by $\cR_T\equiv (Irr(\cD),+,\cdot,*_{\cR_T})$ the $\bK$-algebras that are called the \textbf{\emph{rule algebras}} of types $T\in\{DPO,SPO_A,SPO_B,SPO_{AB}\}$, with composition operations defined as
\begin{equation}
	*_{\cR_T}:\cR_T\times \cR_T\rightarrow \cR_T: (r_1,r_2)\mapsto r_1*_{\cR_T}r_2:= \bar{\varphi}_T\circ(\psi(r_1)*_{\cD}\psi(r_2))\,,
\end{equation}	
with the \emph{reduction morphisms} $\bar{\varphi}_T:\cD\rightarrow \cR_T$ defined as
\begin{equation}
	\bar{\varphi}_T:=\cF_T\circ \partial\,.
\end{equation}
\end{defn}

As an immediate corollary of our definitions, the moniker ``irreducible'' for the diagrams in $Irr(\cD)$ is now precisely justified:
\begin{cor}\label{corr:IRD}
Let $d\in Irr(\cD)$ be an irreducible rule diagram. Then $\bar{\varphi}_T(d)=d$.
\begin{proof}
	Trivially, since any basis diagram $\fd\in Irr(\cD)$ is of the form
	\[
	\fd\equiv(I,O,r,\emptyset,s,t)\,,
	\]
	a direct application of our definitions reveals that $\partial(\fd)=\fd$. By the same token, $\fd$ does not contain any ``dangling edges'', whence $\cF_T(fd)=\fd$, from which (together with the usual  extension to arbitrary diagrams $d\in Irr(\cD)$ via extension to multilinearity) the claim follows.
\end{proof}	
\end{cor}

We are now in a position to state the second main result of this work:
\begin{thm}\label{thm:mainAssoc}
The reduction morphisms $\bar{\varphi}_T$ are \emph{algebra homomorphisms}, whence
the rule algebras $\cR_T\equiv(Irr(\cD),+,\cdot,*_{\cR_T})$ are \emph{associative, unital $\bK$-algebras} with unit $d_{\emptyset}$.
\begin{proof}
The claimed algebra homomorphism property of $\bar{\varphi}_T:\cD\rightarrow \cR_T$ amounts to
\begin{equation}
\begin{aligned}\label{eq:RAassocToProve}
	\bar{\varphi}_T(d_1*_{\cD}d_2)&=\bar{\varphi}_T(d_1)*_{\cR_T}\bar{\varphi}_T(d_2)\\
	&=\bar{\varphi}_T(\psi(\bar{\varphi}_T(d_1))*_{\cD}
	\psi(\bar{\varphi}_T(d_2)))	
	\quad \forall d_1,d_2\in \cD\,.
\end{aligned}
\end{equation}
If it were not for the nontrivial actions of the fixing morphisms $\cF_T$ followed by the projection morphisms, the claim would follow immediately from the associativity of the rule diagram algebra $\cD$. Due to multilinearity, it is sufficient to check the claim on basis diagrams $\fd_1,\fd_2\in \cD$. The proof proceeds in two steps: we first recall that any admissible match $m_{AB}\in \cM_{\fd_1}(\fd_2)$ must be an injective partial graph homomorphism from $\cO(\fd_2)$ to $\cI(\fd_1)$, and whence more precisely from $\cO(\fd_2)'$ to $\cI(\fd_1)'$, where $\cO(\fd_2)'$ and $\cI(\fd_1)'$ are the graphs that result upon deleting from the interfaces $\cO(\fd_2)$ and $\cI(\fd_1)$ any ``dangling edges''. Consequently, if $\bar{\varphi}_T(\fd_1)\neq 0\cdot d_{\emptyset}$ and $\bar{\varphi}_T(\fd_1)\neq 0\cdot d_{\emptyset}$, we have that
\[
\cM_{\fd_1}(\fd_2)=\cM_{\psi(\bar{\varphi}_T(\fd_1))}(\psi(\bar{\varphi}_T(\fd_2)))\,,
\]
because the fixing morphisms $\cF_T$ act only by possibly (depending on the type $T$) deleting ``dangling edges'' and performing a projection to the space of irreducible rule diagrams. Since by definition the matches $\cM_{\fd_1}(\fd_2)$ do not involve the ``dangling edges'', composing $\fd_1$ with $\fd_2$ along any match $m_{12}\in \cM_{\fd_1}(\fd_2)$ cannot modify the presence of any ``dangling edges'' in $\fd_1$ nor in $\fd_2$ existing prior to the composition. In other words, The set of ``dangling edges'' of the reduction of the composite $\partial(\comp{\fd_1}{m_{12}}{\fd_2})$ can be uniquely decomposed into three disjoint sets $\varepsilon_X\subset E_{\partial(\fd_X)}$ ($X\in\{1,2\}$) and $\varepsilon_{12}\subset \partial(\comp{\fd_1}{m_{12}}{\fd_2})$, with $\varepsilon_X\cap \varepsilon_Y=\emptyset$ for $X\neq Y$ and $X,Y\in \{1,2,12\}$. This proves the claim for the case of non-empty reductions of $\fd_1$ and $\fd_2$. By the an analogous argument, in case that $\bar{\varphi}_T(\fd_1)= 0\cdot d_{\emptyset}$ and/or $\bar{\varphi}_T(\fd_2)= 0\cdot d_{\emptyset}$, we are guaranteed to also have that $\bar{\varphi}_T(\fd_1*_{\cD}\fd_2)=0\cdot d_{\emptyset}$, whence the claim of the homomorphism property follows.\\

The associativity of the rule algebras $\cR_T$ follows directly the homomorphism property of the reduction morphisms $\bar{\varphi}_T$:
\begin{align*}
	(r_1*_{\cR_T}r_2)*_{\cR_T}r_3&=
	\bar{\varphi}_T(\psi(\bar{\varphi}_T(\psi(r_1)*_{\cD}\psi(r_2))*_{\cD}\psi(r_3) &\\
&=
  \bar{\varphi}_T(\psi(\bar{\varphi}_T(\psi(r_1)*_{\cD}\psi(r_2))*_{\cD}\psi(\bar{\varphi}_T(\psi(r_3)))& (*)\\
&=
  \bar{\varphi}_T(\psi(r_1)*_{\cD}\psi(r_2)*_{\cD}\psi(r_3))&\text{(via~\eqref{eq:RAassocToProve})}\\
&=
  \bar{\varphi}_T(\psi(\bar{\varphi}_T(\psi(r_1))*_{\cD}\psi(\bar{\varphi}_T(\psi(r_2)*_{\cD}\psi(r_3)))&\\
&=
  \bar{\varphi}_T(\psi(r_1)*_{\cD}\psi(\bar{\varphi}_T(\psi(r_2)*_{\cD}\psi(r_3)))& (*)\\
&=r_1*_{\cR_T}(r_2*_{\cR_T}r_3)\,,
\end{align*}%
where in the steps marked $(*)$ we have used that according to Definition~\ref{def:insertionMorph} $\psi(r)\in Irr(\cD)$ for all $r\in \cR_T$, whence by Corollary~\ref{corr:IRD} $\psi(r)=\psi(\bar{\varphi}_T(\psi(r))$.\\

Finally, the rule algebras $\cR_T$ are unital since by definition
\[
r*_{\cR_T}r_{\emptyset}=
r_{\emptyset}*_{\cR_T}r=r\quad\forall r\in \cR_T\,,
\] 
whence for each of the algebras $\cR_T$, $r_{\emptyset}$ is the unit element.
\end{proof}
\end{thm}

It will depend on the particular application at hand whether it is more convenient to compute the triple compositions via performing an actual triple composition in $\cD$ followed by application of the appropriate reduction morphism, or whether it is more convenient to perform two consecutive binary compositions, since evidently the former variant produces all the matches that are effectively discarded in either of the binary compositions. On the other hand, in particular if we have a small number of concrete rule algebra elements given of which we want to compute higher composites, it might be advantageous to perform the computations via higher compositions in $\cD$, since this route illustrates more clearly the combinatorics of this particular problem.

\subsection{Visualizations: diagrams and worldlines}\label{sec:viz}

We conclude this section with a visualization of the previously introduced concepts. First of all, let us illustrate why we like to think of a rule diagram $\fd\in \fD$ associated to which is the rule algebra element $\fd\in\cD$ (i.e.\ the associated basis element $\fd\equiv\delta(\fd)$ of $\cD$ with prefactor $1_{\bK}$) as a \emph{diagram}. Consider thus a generic rule diagram $\fd\in \fD$, with
\[
\fd\equiv\left(\biguplus_{i=1}^n I_i,\biguplus_{i=1}^n O_i,\biguplus_{i=1}^n r_i,\biguplus_{\stackrel{i,j=1}{i<j}}^n\{m_{i,j}\},s,t\right)\,.
\]
We may visualize this data (or, more precisely, a particular representative of the isomorphism class $\fd\in\fD$) as follows:
\begin{itemize}
    \item For each subpart of the data $r_i:\GRule{I_i}{O_i}$ (which by definition is a connected linear rule), draw $O_i$ above $I_i$, and link vertices and edges that are nontrivially mapped by $r_i$ with dotted lines; from all remaining vertices and edges of $I_i$, draw a dotted line upwards ending in a $\times$ symbol in the space between $I_i$ and $O_i$, and moreover draw a dotted line from all remaining vertices and edges of $O_i$ downwards, also ending in a $\times$ symbol in the space between $I_i$ and $O_i$. Each such configuration presents an instance of an \emph{atomic rule diagram}.
    \item Each match entry $m_{i,j}\equiv(m^V_{i,j},m^E_{i,j})$ consists of an injective partial map from $V_{O_j}$ to $V_{I_i}$ and of an injective partial map from $E_{O_j}$ to $E_{I_i}$; draw a solid line decorated with an arrow from vertices and edges of $O_j$ to vertices and edges of $I_i$ that are nontrivially mapped to one another via $m_{i,j}$.
\end{itemize}
Presented below is an example of the above procedure:
\begin{equation}\label{eq:rdIllustration2}
\includetikz{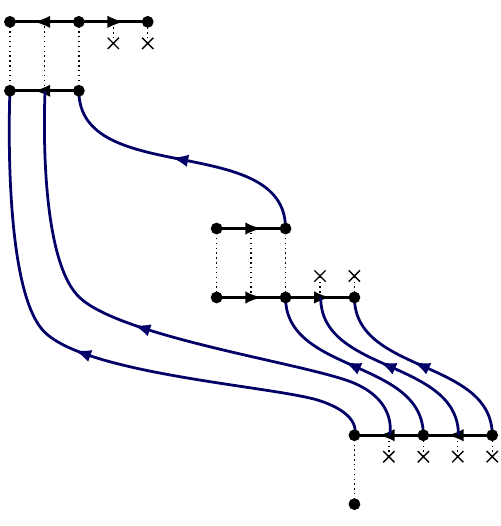}
\end{equation}
In this particular example, the rule diagram is a composite of three atomic rule diagrams $\fd_i\in \frak{Atom}(\fD)$ ($i\in\{1,2,3\}$, numbering from top to bottom), composed along the matches $m_{1,2}$, $m_{1,3}$ and $m_{2,3}$. We have moreover chosen this example to illustrate that the ``delayed edge morphism condition'' entails that ``delayed matches'' of edges (such as the one implemented via $m_{1,3}$) are admissible as long as the respective endpoint vertices are matched consistently as well (here via ``being threaded through'' diagram $\fd_2$ via the matches $m_{2,3}$ and $m_{1,2}$). Note that we have chosen to omit any vertex or edge labels in this drawing in order to enhance visual clarity -- nevertheless, we may think of any vertex or edge of the representative of the composite rule diagram as being implicitly uniquely labelled by its position in the plane.\\

Another rather illuminating way to think about the data contained in a given rule diagram $\fd\in \fD$ is to think of it as a \emph{worldline history}: each vertex and edge due to injectivity has a precise ``history'' within a given rule diagram, i.e.\ it starts its ``existence'' in one of the graphs $I_i$ or $O_i$ and is then evolved via partial maps $r_i$ and matches $m_{i,j}$. In fact, one may think of each vertex and edge as having their unique \emph{worldlines}, marked with ``events'' that are the concrete instances that a vertex or edge occurs in a given rule diagram (in the sense of which explicitly labelled vertices or edges in a given representative of $\fd$ are linked via $r_i$'s and $m_{i,j}$'s). The most efficient way of drawing what we call the \emph{worldline history} of a given $\fd\in \fD$ is to extract from the transitive reflexive closure of
\begin{equation}
	W:=m\cup r\cup id_{\cI(\fd)}\cup id_{\cO(\fd)}\subset \left(V_I\cup V_O\cup E_I\cup E_O\right)^{\times\:2}
\end{equation}
all the connected chains of related vertices and edges, excluding self-loops -- note that due to the inherent directedness of the partial maps $r_i:\GRule{I_i}{O_i}$ and $m_{i,j}:\GRule{O_j}{I_i}$, the individual elements of a given chain are \emph{totally ordered}. For each chain of vertices, we draw a solid vertical line, with the first vertex of a chain marked at the bottom and the last one at the top (following the direction of the ``optimistic arrow of time''). Furthermore, in case the first vertex of a chain is an element of $V^O$, we decorate the start of the vertex worldline by a $\times$ symbol, and dually if the last vertex of a chain is an element of $V^I$, we decorate the end of the worldline with a $\times$ symbol. We proceed analogously for the edge worldlines (which will be drawn as wavy lines for better graphical distinction), only that we indicate in addition graphically the endpoint vertices' worldlines for each edge worldline (by a dotted horizontal line somewhere in the worldline picture).\footnote{For every edge worldline, due to the consistency conditions for rule diagrams, the vertex worldlines passing through the endpoint vertices of the first edge of a given edge worldline are guaranteed to pass through the endpoint vertices of all other edges along the edge worldline.} For each pair of vertices or edges included in $id_{\cI(\fd)}$ or $id_{\cO(\fd)}$, we only mark the ``position'' of the vertex or edge once on the respective worldline. We call the resulting graphical description in terms of vertex and edge worldlines a \emph{worldline history}. For example, the worldline history for the rule diagram depicted in~\eqref{eq:rdIllustration2} reads
\begin{equation}\label{eq:wlIllustration1}
\includetikz{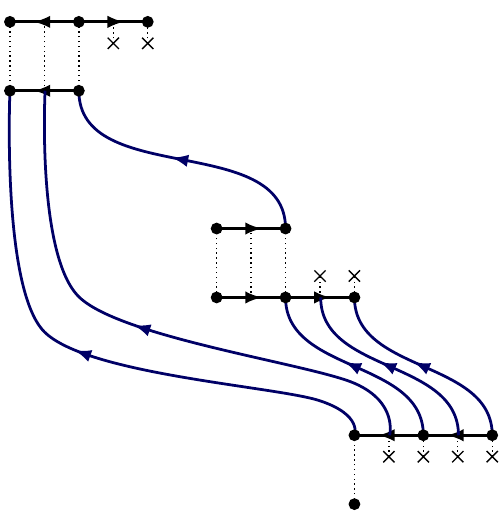}
\widehat{=}\quad\vcenter{\hbox{\includegraphics[width=0.45\textwidth]{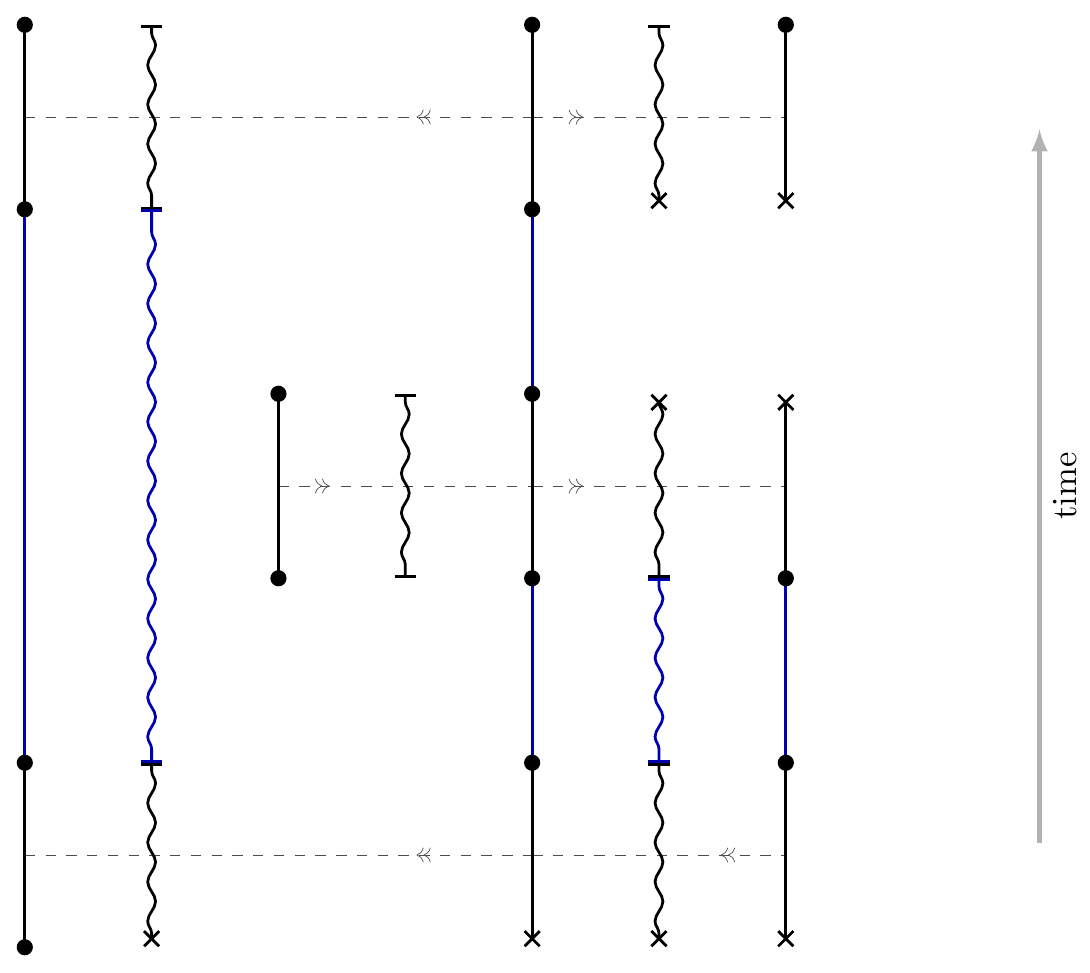}}}
\end{equation}
Only the absolute orders along each individual worldline carries any immutable meaning for a given rule diagram, but we have grouped the incidences of vertices and edges relative to one another across worldlines to improve the intuitive understanding of the ``time evolution'' of vertices and edges encoded in the rule diagram. This example already hints at the practical complications in drawing worldline histories -- in particular the edge worldlines may be difficult to mark clearly in the diagrams, since one always has to keep track of their endpoint vertices.\\

We may thus picture a given rule diagram quite intuitively as a sort of ``transformation movie script'' of vertices and edges, and compositions of rule diagrams amount to concatenation of these ``movie scripts''. Moreover, the somewhat cryptic conditions on possible matches of two rule diagrams are now clearly visualizable: if one matches two edges anywhere in a given composition of two or more rule diagrams, their respective endpoint vertex worldlines must be matched as well, up to a possible ``delay'' in the form of intermediate matches of the endpoint vertex worldlines into other endpoint vertex worldlines. An example of such a delayed morphism of an edge may be found in the leftmost edge worldline in~\eqref{eq:wlIllustration1}.\\

Another interesting observation is the fact that the application of the boundary map $\partial:\cD\rightarrow Irr_{pre}(\cD)$ amounts in the worldline history picture to ``shortening'' the worldlines for a given rule diagrams to their absolute start and end points, thereby producing a worldline history akin to the one of an irreducible rule diagram, and discarding any worldlines that both start and end in a symbol $\times$. We may thus generically end up with edge worldlines that are quite literally missing one or both of their entire endpoint vertex worldlines (in case these had been starting and ending in symbols $\times$). Equivalently, these situations occur if in the worldline history we have one of the following four possible configurations that are inconsistent with the shapes of worldline histories of irreducible rule diagrams:
\begin{equation}\label{eq:inconsistencies}
\includetikz{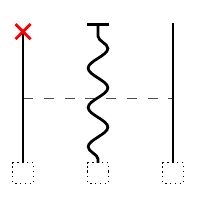}\,,\qquad \includetikz{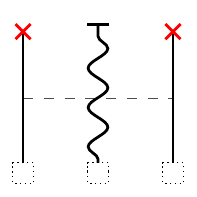}\,,\quad \includetikz{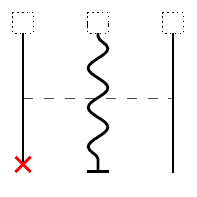}\,,\qquad \includetikz{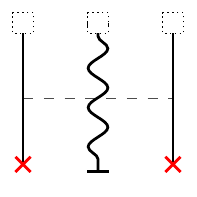}\,.
\end{equation}
Here, the dashed boxes indicate that the ends they mark are either of type $|$ or $\times$, whence the diagrams represent the situations of an edge worldline with one or two ``missing endpoint vertices'' at their ``future end'', or one or two ``missing endpoint vertices'' at their ``past end'', respectively.\\

It may then be easily seen that the two nontrivial partial fixing morphisms $f_{SPO_A}$ and $f_{SPO_B}$ amount graphically to the following operations on worldlines:
\begin{equation}\label{eq:fixing}
\includetikz{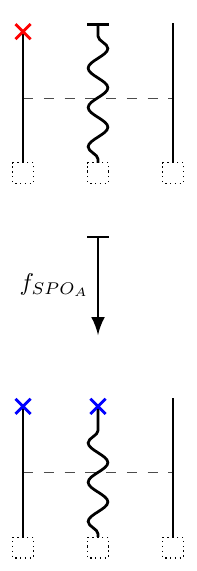}\,,\qquad \includetikz{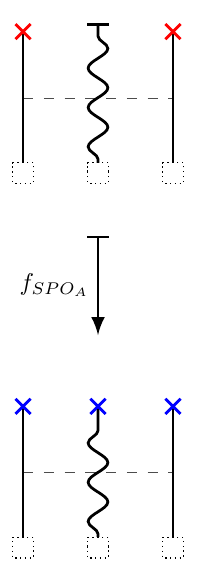}\,,\quad \includetikz{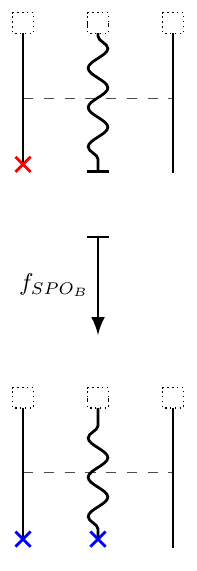}\,,\qquad \includetikz{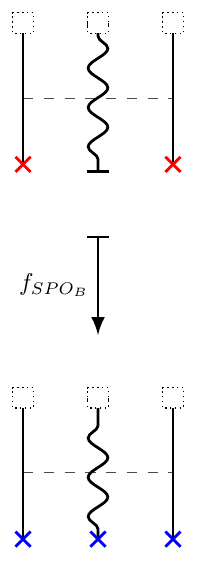}\,.
\end{equation}
The action of the fixing morphisms thus modifies the edge worldlines as marked with the blue crosses, thereby rendering the previously inadmissible vertex worldline configurations (marked with red crosses) admissible (marked with blue crosses).\\

In practice, the visualization in terms of worldline histories has the advantage that it may be directly implemented as an algorithm, in a form that will be presented elsewhere. However, we base our mathematical framework on the notion of rule diagrams instead, since this concept is more directly amenable to a combinatorial analysis, as will be demonstrated in the next section.

\subsection{Dualization of rule diagrams}

One operation that suggests itself in view of our current description of rule diagrams is the operation of reversing all partial maps\footnote{For this subsection only, we will be precise again about the instances of elements $\fd\in\fD$ of the \emph{set} of rule diagrams and the associated rule diagram \emph{algebra elements} $\delta(\fd)\in\cD$.}:
\begin{defn}[Dualization of rule diagrams]
	The \emph{dualization map}
	\begin{equation}
		\dag:\cD\rightarrow \cD
	\end{equation}
	assigns to each basis diagram 
	\[
	d\equiv\delta(\fd)\equiv\delta\big((I,O,r,m,s,t)\big)\in \cD
	\]
	the dual diagram
	\[
	d^{\dag}:=\delta(\fd^{\dag}):=\delta\big((O,I,r^{\dag},m^{\dag},s,t)\big)\,,
	\]
	with 
	\begin{equation}
		r^{\dag}:=r^{\rConv}\,,\quad m^{\dag}:=m^{\rConv}\,.
	\end{equation}
	The dualization operation is extended to all of $\cD$  via the definition
	\begin{equation}
		\left(\sum_{i\in \cI}\kappa_i d_i\right)^{\dag}:=\sum_{i\in \cI}\kappa_i^{*} d_i^{\dag}\qquad (\kappa_i\in \bK)\,,
	\end{equation}
	where $d_i\equiv\delta(\fd_i)$ are basis diagrams, and
	where ${}^{*}:\bK\rightarrow \bK$ is an (arbitrarily, but globally chosen) \emph{involutory automorphism}\footnote{For example, ${}^{*}$ for the reals is just the identity mapping, while the field $\bC$ admits the typical choices of the identity mapping or of complex conjugation. We are tacitly assuming that $\bK$ is a field of $char(\bK)\neq2$ int he generic case.}.
\end{defn}

This definition entails the following crucial fact about the rule diagram algebra and also about the four types of rule algebras:
\begin{thm}
$(i)$ The dualization map $\dag:\cD\rightarrow\cD$ is a well-defined endomorphism of $\cD$, and moreover an \emph{involution}\footnote{Strictly speaking, $\dag$ is an involution on the $\cD$-module, but an anti-involution on $\cD$ itself; it is however quite common to refer to such a map as involution nonetheless.}, i.e.\
\[
\dag\circ \dag =Id_{\cD}\,.
\]	
The rule diagram algebra together with the dualization map, $\cD\equiv(\fD,+,\cdot,*_{\cD},d_{\emptyset},\dag)$, is a unital associative algebra with involution, and hence a (non commutative) \emph{$*$-algebra}, with the usual properties
\begin{equation}
\begin{aligned}
	(\lambda_1\cdot d_1+\lambda_2\cdot d_2)^{\dag}&=
	\lambda_1^{*}\cdot d_1^{\dag}
	+\lambda_2^{*}\cdot d_2^{\dag}\\
	(d_1*_{\cD}d_2)^{\dag}&= 
	d_2^{\dag}*_{\cD}d_1^{\dag}\\
	d_{\emptyset}^{\dag}&=d_{\emptyset}\,,
\end{aligned}	
\end{equation}
for all $d_1,d_2\in \cD$ and $\lambda_1,\lambda_2\in \bK$. Here, we have denoted again by ${}^{*}:\bK\rightarrow\bK$ the globally chosen involutory automorphism of the field $\bK$.\\

$(ii)$ The dualization map may be properly restricted to $Irr(\cD)$, 
\[
\dag:Irr(\cD)\rightarrow Irr(\cD)\,,
\]
whence the associative unital rule algebras $\cR_T$ endowed with $\dag$ become $*$-algebras.
\begin{proof}
	Evidently, the converse of an injective partial map is again an injective partial map, whence the outcome of applying $\dag$ to any basis diagram $d\equiv\delta(\fd)\in \cD$ is trivially again a rule diagram. Moreover, the structure of the diagram is not changed beyond the reversal of the roles of input and output graphs and of the partial maps $t$ and $m$, whence in particular an irreducible rule diagram remains irreducible. As for the composition of two basis diagrams, each given contribution $\comp{\fd_1}{m_{12}}{\fd_2}$ to a composition $\delta(\fd_1)*_{\cD}\delta(\fd_2)$ is again a rule diagram; applying $\dag$ to this rule diagram, we obtain a rule diagram which may be written as $\comp{\fd_2^{\dag}}{m_{12}^{\dag}}{\fd_1^{\dag}}$ (simply by reversing $t$'s and $m$'s and keeping track of the subdiagrams), whence we obtain the (anti-)involution property. In particular, if the match $m_{12}$ fulfills the ``delayed edge match'' condition, so dies $m_{12}^{\smallsmile}$ for the dualized composite diagram. The empty rule diagram is invariant under $\dag$, and $\dag$ also does not affect any $\bK$-prefactors of a given rule diagram.\\
	
	Finally, the proof for the (anti-)involution property of $\dag$ on each of the rule algebras $\cR_T$ is entirely analogous to the proof of the property for $\cD$. 
\end{proof}
\end{thm}

\subsection{Extensions to other types of graphs}

It is a straightforward exercise to extend the rule algebraic formalism to colored and/or undirected multi-graphs, or even general graphs with both directed and undirected edges. To this end, one only needs to employ the standard definitions of injective partial graph morphisms and graph isomorphisms for the respective types, and in addition require matches to respect colors and types of edges. We conclude this chapter by commenting that it is also possible to lift the definitions from graphs to hypergraphs, with the evident induced generalizations of morphisms and matches. The details are left to the interested reader.

\section{Combinatorial Hopf algebra structure of the rule diagram algebra}\label{sec:RDhopf}

The derivation of the compatibility of the combinatorial Hopf algebra structure with the definition of the rule diagram algebra as presented in Definition~\ref{def:ruleDiagramAlgebra} will follow the standard hierarchical development of algebraic concepts (cf.\ e.g.~\cite{manchon2008hopf,blasiak2010combinatorial}), starting from a vector space (in our case the $\bK$-vector space $\cD\equiv(\cD,+,\cdot)$ of isomorphism classes of rule diagrams), and endowing it step by step with further structures: 
\begin{enumerate}
\item Check that the vector space endowed with a suitable \emph{composition operation} satisfies the axioms of a \emph{unital associative algebra}.
\item Check that with the definition of a suitable \emph{decomposition operation} it becomes a \emph{counital coassociative coalgebra}.
\item Confirm that both structures are compatible, such that we have in fact the structure of a \emph{bialgebra}.
\item Introduce a concept of \emph{filtration} of elements of the bialgebra.
\item Via the filtration, construct the (unique) \emph{antipode} of the bialgebra, and thereby demonstrate that one thus obtains the structure of a \emph{Hopf algebra}.
\end{enumerate}
While we will follow in spirit quite closely the derivation of the combinatorial Hopf algebra structure of the Heisenberg-Weyl diagram algebra as presented in~\cite{blasiak2010combinatorial} for the first three steps, we will employ the theorem on the existence of a unique antipode for any filtered bialgebra as presented in~\cite{manchon2008hopf} in order to prove the existence of the Hopf algebra structures.

\subsection{Algebra structure}

We have already introduced in the previous section the concept of the rule diagram algebra $\cD\equiv(\cD,+,\cdot,*_{\cD})$ and have demonstrated in~\ref{thm:RDassoc} that it is an \emph{associative unital $\bK$-algebra}. Therefore, the only work left to do for this aspect of the Hopf algebra construction is to formulate these results in the form conventional in the Hopf algebra literature (cf.\ Appendix~\ref{app:Algebra}):
\begin{defn}[Rule diagram algebra $\cD$]\label{def:ruleDiagramAlgebra}
Let $\cD\equiv(\cD,+,\cdot,*_{\cD})$ denote the associative unital $\bK$-algebra of rule diagrams. In the following, we will refer to the rule diagram algebra alternatively (but equivalently) as $\cD\equiv(\cD,\mu,\eta)$, with \emph{multiplication map}
\begin{equation}
\mu:\cD\otimes \cD\rightarrow \cD: d_1\otimes d_2\mapsto d_1*_{\cD}d_2
\end{equation}
and \emph{unit map}
\begin{equation}
\eta: \bK\rightarrow \cD: k\mapsto k \cdot d_{\emptyset}\,.
\end{equation}
Here, the special element $d_{\emptyset}$, i.e.\ the (unique) isomorphism class of the empty rule diagram, is the (unique) \emph{unit under composition} as before,
\begin{equation}
d_{\emptyset}*_{\cD}= d*_{\cD}d_{\emptyset}=d\qquad \forall d\in \cD\,.
\end{equation}
\end{defn}

\subsection{Coalgebra structure}

In this section, we will have to make frequent use of the relationship between a rule diagram $\fd\in \fD$ and its associated basis vector $\delta(\fd)\in \cD$, whence for this subsection we will make the distinction explicit.\\

We will demonstrate now that the rule diagram algebra $\cD$ can be equipped with the structure of a \emph{coassociative counital coalgebra} (cf.\ Appendix~\ref{app:Coalgebra}). Following in spirit the ideas presented in~\cite{blasiak2010combinatorial}, while the algebra structure was related to \emph{composition} of rule diagrams, the coalgebra structure will be related to \emph{decompositions} of rule diagrams.\\

More precisely, consider the following definition of decompositions of rule diagrams:
\begin{defn}
Let
\[
\fD\ni \fd\equiv(I,O,r,m,s,t)
\]	
be an arbitrary rule diagram. Then a \emph{decomposition} of $\fd$ into two rule diagrams $\fd_A$ and $\fd_B$ is defined such that
\begin{equation}
	(\fd_A,\fd_B\in \fD)\quad\land\quad  \fd_A\cap \fd_B=\emptyset\quad \land \quad\fd=\fd_A\uplus \fd_B\,.
\end{equation}
Here, we have employed the shorthand notations
\begin{align*}
	&\fd_A\cap \fd_B=\emptyset\\
	&\quad:\Leftrightarrow\quad (I_A\cap I_B,O_A\cap O_B,r_A\cap r_B,m_A\cap m_B,s_A\cap s_B,t_A\cap t_B)\\
	&\quad \quad = (\emptyset,\emptyset,\emptyset,\emptyset,\emptyset,\emptyset)
\end{align*}%
as well as the convenient notation for the \emph{disjoint union} of two pure rule diagrams, 
\begin{equation}
	\fd_A\uplus \fd_B:=(I_A\uplus I_B,O_A\uplus O_B,r_A\uplus r_B,m_A\uplus m_B,s_A\uplus s_B,t_A\uplus t_B)\,.
\end{equation}
\end{defn}
In other words, a decomposition of a pure rule diagram is a separation of the tuple of data of the pure rule diagram into two disjoint tuples of data which are (a) themselves proper pure rule diagrams, (b) disjoint, and (c) whose disjoint union reproduces the original rule diagram. \\

An important property for the coalgebra construction will be that of indecomposability: 
\begin{defn}
	A rule diagram $\fd\in \fD$ for which no decomposition exists, is called \emph{indecomposable}. We refer to \emph{non-empty indecomposable rule diagrams} as \emph{primitive rule diagrams}, and introduce the notation $\frak{Prim}(\fD)\subsetneq \fD$ for the \emph{set of primitive rule diagrams}, as well as
	\begin{equation}
		Prim(\cD):=span_{\bK}(\frak{Prim}(\fD))\subsetneq \cD
	\end{equation}
	for the subvector space of the vector space of rule diagrams spanned by the basis vectors associated to primitive rule diagrams. Our definitions imply that an arbitrary nonempty rule diagram $\fd\in\fD$ may be decomposed uniquely into a superposition of primitive rule diagrams,
	\begin{equation}
		\fd=\fd(1)\uplus\dotsc\uplus \fd(n_c)\,,
	\end{equation}
	where $n_c\in\bZ_{>0}$ is called the \emph{number of connected components} of $\fd$, and where each diagram $\fd(i)$ is a primitive rule diagram.
\end{defn}
According to this definition, the empty rule diagram $\fd_{\emptyset}$ is evidently indecomposable. For a generic rule diagram $\fd\in \fD$, we may sharpen the analysis of decomposability via the following Lemma:
\begin{lem}[Decomposition of rule diagrams]\label{lem:decompRD}
Let $\fd\in \fD$ be a generic rule diagram,
\begin{equation}
\fd\equiv\left(\biguplus_{i=1}^n I_i,\biguplus_{i=1}^n O_i,\biguplus_{i=1}^n r_i,\biguplus_{\stackrel{i,j=1}{i<j}}^n\{m_{i,j}\},\biguplus_{i=1}^n s_i,\biguplus_{i=1}^n t_i\right)\,,
\end{equation}
where we have by virtue of our general definition of rule diagrams that the $r_i:\GRule{I_i}{O_i}$ are \emph{connected} injective partial maps. In other words, the rule diagrams one may construct from each individual single such connected linear rule are \emph{atomic} rule diagrams. As the moniker ``atomic'' already suggests, in the special case $n=1$, where 
\[
\hat{\fd}=(I_1,O_1,r_1,\emptyset,s_1,t_1),
\] 
(i.e.\ when $\fd$ is an atomic rule diagram), this diagram is according to our definitions indecomposable.\\

In the generic case, $\fd$ is \emph{indecomposable} if and only if 
\begin{equation}
\begin{aligned}
&\left(\forall 1\leq i< n:\exists 1<j\leq n:m_{i,j}\neq \emptyset\right)\\
\land
\quad&\left(\forall 1< j\leq n:\exists 1\leq i<n:m_{i,j}\neq \emptyset\right)\,,
\end{aligned}
\end{equation}
which amounts in the graphical presentation of the rule diagram to the statement that all constituent atomic subdiagrams $\fd_i$ (which are the atomic rue diagrams presenting the connected linear rules $r_i:\GRule{I_i}{O_i}$) must be connected by the matches $m_{i,j}$ into an overall connected diagram. Otherwise, the rule diagram is \emph{decomposable}, namely into its connected components.
\end{lem}

With these ingredients, we may now proceed to define the counit the counit map and the coproduct of the rule algebra:
\begin{defn}[Counit and counit map]
The empty rule diagram $d_{\emptyset}\equiv\delta(\fd_{\emptyset})$ is defined as the \emph{counit} of the rule diagram algebra, while the \emph{counit map} of $\cD$ is defined as
\begin{equation}\label{eq:coUnit}
\varepsilon: \cD\rightarrow \bK: d\mapsto \begin{cases}
k\quad &\text{, if } d=k\cdot d_{\emptyset}\\
0\quad &\text{, else.}
\end{cases}
\end{equation}
\end{defn}

\begin{defn}[Coproduct]\label{def:coproduct}
Let $d\equiv\delta(\fd)\in \cD$ be a basis diagram, whence either the empty rule diagram $d=d_{\emptyset}$, or a non-empty diagram
\begin{equation}
d=\biguplus_{i\in \cI}d(i)\,,
\end{equation}
with connected subdiagrams $d(i)\equiv \delta(\fd(i))\in Atom(\cD)$, and where $\cI$ is some set of indices\footnote{It is evident from our definitions that the concrete indexing of the subdiagrams of a given basis diagram is immaterial due to the quotienting by isomorphisms; the concrete indexing is only performed as a convenient means of explicitly defining the coproduct operation.}, $\cI=\{1,2,\dotsc,n_c\}$, with $n_c>0$ the number of connected components of $d$.\\

The \emph{coproduct} of $\cD$ is then defined as the $\bK$-linear map defined on basis diagrams $d\equiv\delta(\fd)\in \cD$ as
\begin{equation}
\Delta: \cD\rightarrow \cD\otimes \cD:d\mapsto \Delta(d):=\sum_{X\subseteq \cI} d_X\otimes d_{X^c}\,,
\end{equation}
with $X^{c}:=\cI\setminus X$, and
\begin{equation}
d_X:=\begin{cases}
d_{\emptyset} \quad&\text{, if } X=\emptyset\\
\delta\left(\biguplus_{i\in X}\fd(i)\right)\quad &\text{, else.}
\end{cases}
\end{equation}
Finally, the definition of $\Delta$ is extended to all of $\cD$ via multilinearity,
\begin{equation}
	\Delta\left(\sum_{i\in\cI}\kappa_i \delta(\fd_i)\right):=\sum_{i\in\cI}\kappa_i \Delta(\delta(\fd_i))\,.
\end{equation}
\end{defn}
For concreteness, the definition of the coproduct entails that
\begin{equation}
\Delta(d_{\emptyset})=d_{\emptyset}\otimes d_{\emptyset}
\end{equation}
while for any other nonempty basis diagram $\fd \in \fD\setminus\{\fd_{\emptyset}\}$ with decomposition
\[
\fd=\fd(1)\uplus\dotsc \uplus \fd(n_c)\,,
\]
where $n_c\in \bZ_{>0}$ denotes the number of connected components of $\fd$:
\begin{equation}
\begin{aligned}
\Delta(\delta(\fd))&=d_{\emptyset}\otimes \delta(\fd)+\delta(\fd)\otimes d_{\emptyset}\\
&+\sum_{k=1}^{n_c-1}\sum_{\sigma\in Sh(k,n_c-k)}\delta\left(\biguplus_{p=1}^k \fd(\sigma(p)))\right)\otimes \delta\left(\biguplus_{q=k+1}^n \fd_(\sigma(q))\right)\,.
\end{aligned}
\end{equation}

The main result of this subsection is the following proposition, demonstrating that our tentative definitions lead to a consistent coalgebra structure:
\begin{prop}\label{prop:RDAcoalgebra}
The diagram algebra $\cD\equiv(\cD,\Delta,\varepsilon)$ is a \emph{coassociative counital and cocommutative $\bK$-coalgebra}.
\begin{proof}
See Appendix~\ref{app:Coalgebra}.
\end{proof}
\end{prop}

\subsection{Bialgebra structure}

Referring the interested readers to Appendix~\ref{app:Bialgebra} for the details, one may prove that the rule diagram algebra equipped with the previously introduced algebra and coalgebra structures in fact forms a bialgebra:
\begin{thm}[Bialgebra structure of $\cD$]\label{thm:bialgRD}
The rule diagram algebra $\cD\equiv(\cD,\mu,\eta,\Delta,\varepsilon)$ is a bialgebra.
\begin{proof}
See Appendix~\ref{app:Bialgebra}.
\end{proof}
\end{thm}

\subsection{Filtration and Hopf algebra structure}

Throughout the proof of Theorem~\ref{thm:bialgRD} as presented in Appendix~\ref{app:Bialgebra}, the indecomposable subdiagrams of a given rule diagram seem to play a very special role in the ability of $\cD$ to carry the structure of a bialgebra. We will now present a concept that emphasizes this point even more (cf.\ Appendix~\ref{app:FilteredHopf}):
\begin{prop}\label{prop:connFilt}
The rule diagram algebra $\cD\equiv(\cD,\mu,\eta,\Delta,\varepsilon)$ is a \emph{connected, filtered bialgebra}, with 
\begin{equation}
\cD^0=span_{\bK}(\{\fd_{\emptyset}\})\,,
\end{equation}
and where the subspace $\cD^n$ is in general given by the linear span of all \emph{basis diagrams with at most $n$ indecomposable subdiagrams},
\begin{equation}
	\cD^n:=span_{\bK}(\{ \fd\in \fD\mid \fd=\fd(1)\uplus\dotsc \fd(n_c)\,,\;\fd(i)\in \frak{Prim}(\fD)\,,\; n_c\leq n\})\,.
\end{equation}
\begin{proof}
See Appendix~\ref{app:FilteredHopf}.
\end{proof}
\end{prop}
In particular, this identifies the subvector space $Prim(\cD)$ spanned by the \emph{primitive basis diagrams} as
\begin{equation}
Prim(\cD)= \cD^1\setminus \cD^0\,.
\end{equation}

Note that while any atomic (aka indecomposable and irreducible) non-empty rule diagram $\fd\in \frak{Atom}(\fD)\setminus\{\fd_{\emptyset}\}$ is evidently by definition also a primitive rule diagram, the converse is not true, since also nontrivial composites of atomic rule diagrams may be indecomposable and hence primitive. In other words,
\begin{equation}
(Atom(\cD)\setminus\{d_{\emptyset}\})\subsetneq Prim(\cD)\,.
\end{equation}

Next, referring to Appendix~\ref{app:FilteredHopf} for motivation and further details, we may define the \emph{$k$-fold convolution product} as
\begin{equation}
(e-Id)^{\ast\:k}:=\mu_k\circ (e-Id)^{\otimes\: k}\circ \Delta_k\,,
\end{equation}
where $\mu_k:\cD^{\otimes\:k}\rightarrow \cD$ is the $k-fold$ multiplication and $\Delta:\cD\rightarrow \cD^{\otimes \: k}$ the $k$-fold comultiplication operation (cf.\ Appendix~\ref{app:MultipleProductsAndCoproducts}), and where the map $e:=\eta\circ\varepsilon$ acts by definition as
\begin{equation}
e(d)=\begin{cases} 
k\cdot d_{\emptyset}\quad &\text{, if $d=k\cdot d_{\emptyset}$}\\
0\cdot d_{\emptyset} &\text{, else.}
\end{cases}
\end{equation}

With these preparations, employing the well-known mathematical fact that any connected filtered bialgebra is a Hopf algebra (cf.\ \cite{manchon2008hopf}, Ch.~4.3, Corr.~5; presented for convenience as Corollary~\ref{corr:bialdCFhopf} in Appendix~\ref{app:FilteredHopf}), and since Theorem~\ref{thm:bialgRD} states that the rule diagram algebra $\cD$ is indeed a connected, filtered $\bK$-bialgebra, we immediately conclude from Corollary~\ref{corr:bialdCFhopf} that it is also a Hopf algebra:
\begin{thm}\label{thm:RDhopf}
The rule diagram algebra $\cD\equiv(\cD,\mu,\eta,\Delta,\varepsilon,S)$ is a \emph{Hopf algebra}, with \emph{antipode $S$} given by
\begin{equation}\label{eq:antipodeRD}
\begin{aligned}
S(k\cdot d_{\emptyset})&:=e(k\cdot d_{\emptyset})=k\cdot d_{\emptyset}\\
S(d)&:=Id^{\ast\:-1}(d)=e(d)+\sum_{k=1}^{n}(e-Id)^{\ast\: k}(d)\qquad \forall d\in\cD^n\setminus\{d_{\emptyset}\}\quad (n>0)\,.
\end{aligned}
\end{equation}
\end{thm}
To give a more explicit formula for the antipode and to define the notion of the convolution product $(\dotsc)^{*}$ explicitly, consider the cases for the first few filtration components $\cD^n$ for $n=1,2,3$. A useful auxiliary formula is
\begin{equation}\label{eq:auxAntipode}
(e-Id)(d)=\begin{cases}
0\cdot d_{\emptyset}\quad &\text{, if $d=k\cdot d_{\emptyset}$}\\
-d &\text{, else}
\end{cases}
\end{equation}
from which we immediately conclude that
\begin{equation}
d\in \cD^1\quad \Rightarrow \quad S(\theta)=e(d)+(e-Id)(d)=
\begin{cases}
k\cdot d_{\emptyset}\quad &\text{, if $d=k\cdot d_{\emptyset}$}\\
-d &\text{, else.}
\end{cases}
\end{equation}
Due to~\eqref{eq:auxAntipode}, it will prove useful to  introduce the notations
\begin{equation}
\Delta_k(d)\equiv \Delta_k'(d)+\tilde{\Delta}_k(d)\,,
\end{equation}
where $\Delta_k'(d)$ contains all contributions to the $k$-fold coproduct of $d$ that contain at least one factor of $d_{\emptyset}$ in each term, such that
\begin{equation}
\mu_k\circ (e-Id)^{\otimes\: k}\circ \Delta'_k(d)=0\,,
\end{equation}
and where consequently $\tilde{\Delta}_k(d)$ contains all terms with no occurrences of $d_{\emptyset}$ in the tensor product factors. Introducing the notation
\begin{equation}
\begin{aligned}
Sh(p_1,\dotsc,p_k)&:=\bigg\{
\sigma\in S_{p_1+p_2+\dotsc +p_k}\mid \\
&\qquad\qquad \sigma(1)<\dotsc<\sigma(p_1) \land
\sigma(p_1+1)<\dotsc <\sigma(p_1+p_2)\\
&\qquad \qquad \land \dotsc\\
&\qquad \qquad \land \sigma(p_1+\dotsc +p_{k-1}+1)<\dotsc <\sigma(p_1+\dotsc+p_k)
\bigg\}\,.
\end{aligned}
\end{equation}
Then we may write explicitly for a generic rule diagram $\fd\in\fD$
\begin{equation}
\tilde{\Delta}_k(\delta(\fd))=\sum_{\stackrel{p_1,\dotsc,p_k}{\sum_i p_i=n\land \prod_i p_i\neq 0}}
\sum_{\sigma\in Sh(p_1,\dotsc,p_k)} \bigotimes_{q=1}^k \delta\left(\biguplus_{i_q=1}^{p_q}   \fd(i_q+\sum_{r=1}^{k-1}p_r)\right)\,,
\end{equation}
and consequently
\begin{equation}
\begin{aligned}
&\mu_k\circ (e-Id)^{\otimes\: k}\circ \tilde{\Delta}_k(\delta(\fd))\\
&\quad=
(-1)^k\sum_{\stackrel{p_1,\dotsc,p_k}{\sum_i p_i=n\land \prod_i p_i\neq 0}}
\sum_{\sigma\in Sh(p_1,\dotsc,p_k)} \prod_{q=1}^k \delta\left(\biguplus_{i_q=1}^{p_q} \fd(i_q+\sum_{r=1}^{k-1}p_r)\right)\,.
\end{aligned}
\end{equation}
We thus obtain the following explicit formulae for the antipode of a basis diagram $d=\delta(\fd)\in \cD^n\setminus \cD^{n-1}$, i.e.\ for basis diagrams with exactly $n$ superposition factors, $\fd=\uplus_{i=1}^n \fd(i)$, for the first few values $n=0,1,2,3$ (with $d(i)\equiv \delta(\fd(i))$):
\begin{gather}
\begin{aligned}\label{eq:AntipodeExpl}
d&\in\cD^0&\qquad S(d)&=d\\
d&\in\cD^1\setminus \cD^0&\qquad S(d)&=-d\\
d&\in\cD^2\setminus\cD^1&\qquad S(d)&=d(1)*d(2)+d(2)*d(1)-d\\
d&\in\cD^3\setminus\cD^2&\qquad S(d)&=
-\sum_{\sigma\in S_3}d(\sigma(1))*d(\sigma(2))*d(\sigma(3))\\
&&&\quad
+\sum_{\sigma\in Sh(1,2)} d(\sigma(1))*(d(\sigma(2))\uplus d(\sigma(3)))\\
&&&\quad+\sum_{\sigma\in Sh(2,3)}
(d(\sigma(1))\uplus d(\sigma(2)))*d(\sigma(3))\\
&&&\quad-d
\end{aligned}
\end{gather}%

\section{Structural insights via construction of the Lie- and universal enveloping algebras of the rule diagram algebra}
\label{sec:LieUnAlg}

It is quite amusing that we can carry out the analogue of Blasiak et al.'s analysis of relationships and mathematical structures~\cite{blasiak2010combinatorial} for our very general combinatorial Hopf algebra of rule diagrams by simply following the fundamental mathematical constructions of the \emph{Lie algebra of an associative algebra} and of its \emph{universal enveloping algebra}. More precisely, we wish to complete the diagram of relations (in partial analogy to~\cite{blasiak2010combinatorial}):
\begin{equation}\label{eq:algConnectionsB}
\includetikz{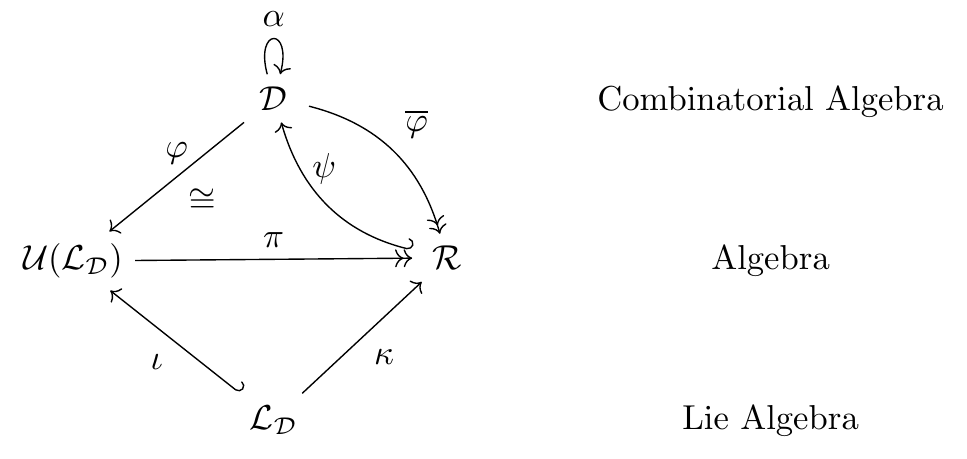}
\end{equation}
which was already mentioned in the introduction. As a small, but beautiful side result, we will demonstrate in Section~\ref{sec:HW} that our very general construction nicely specializes to the case of the Heisenberg-Weyl diagram algebra, thereby clarifying some of the more subtle points of Blasiak et al.'s constructions. Notably, we will argue there that it would be more appropriate also in that special case to speak of the ``Lie algebra of the Heisenberg-Weyl \emph{diagram} algebra'', as opposed to ``the Lie algebra of the Heisenberg-Weyl algebra''. It would be interesting to investigate whether this slight misunderstanding might have prevented analogous constructions for other algebras.

\subsection{The Lie algebra of the rule diagram algebra}

Recall from Proposition~\ref{prop:connFilt} that the \emph{primitive} rule diagrams are defined as those rule diagrams $\fd\in\fD$ which are neither empty nor decomposable, which evidently imposes a particular form onto the coproducts of their associated basis diagrams:
\begin{equation}
\forall \fd\in \frak{Prim}(\fD)\subsetneq\fD: \Delta(\delta(\fd))=d_{\emptyset}\otimes \delta(\fd)+\delta(\fd)\otimes d_{\emptyset}\,.
\end{equation}
We will now employ a standard result from the mathematics literature, namely that for any associative algebra one may define a so-called \emph{Lie-bracket}:
\begin{defn}[Lie-bracket]
Let $\cA\equiv(\cA,+,\cdot,\mu)$ be an associative $\bK$-algebra. Then the \emph{Lie-bracket} of $\cA$ is defined as
\begin{equation}
[\cdot,\cdot]:\cA\times \cA\rightarrow \cA: (a_1,a_2)\mapsto [a_1,a_2]:=a_1*a_2-a_2*a_1\,.
\end{equation}
\end{defn}
While we could of course construct the Lie-bracket for the whole of the rule diagram algebra, it will prove to be much more insightful to construct the Lie algebra over the primitive rule diagrams only, which happen to be closed under the operation of taking the Lie bracket:
\begin{prop}[The Lie algebra of the rule diagram algebra]\label{prop:LieRD}
The \emph{Lie algebra $\cL_{\cD}\equiv(Prim(\cD),+,\cdot,[\cdot,\cdot])$ of the rule diagram algebra} is the Lie algebra defined over the $\bK$-vector space $Prim(\cD)$ of primitive rule diagrams via the Lie-bracket
\begin{equation}
[\cdot,\cdot]: Prim(\cD)\times Prim(\cD)\rightarrow Prim(\cD): (d,d')\mapsto 
[d,d']:= d*_{\cD}d'-d'*_{\cD}d\,,
\end{equation} 
where the nontrivial claim consists in the property that $[\cdot,\cdot]$ closes on $Prim(\cD)$.
\begin{proof}
For two primitive rule diagrams $\fd,\fd'\in \frak{Prim}(\fD)$, we have that
\begin{equation}
\begin{aligned}
\delta(\fd)*_{\cD}\delta(\fd')&= \delta(\fd\uplus \fd') +\sum_{\stackrel{m\in\cM_{\fd}(\fd')}{m\neq \emptyset}} \delta(\comp{\fd}{m}{\fd'})\\
\delta(\fd')*_{\cD}\delta(\fd)&= \delta(\fd'\uplus \fd) +\sum_{\stackrel{m'\in\cM_{\fd'}(\fd)}{m'\neq \emptyset}} \delta(\comp{\fd'}{m'}{\fd})\,.
\end{aligned}
\end{equation}
Since we have in $\cD$ due to the definitions of isomorphisms of rule diagrams that
\[
 \delta(\fd\uplus \fd') =  \delta(\fd'\uplus \fd)\,,
\]
and since the composite of two primitive rule diagrams along a nontrivial match is again a primitive rule diagram (which is moreover guaranteed not to be the empty rule diagram), we immediately conclude that
\begin{equation}
[\delta(\fd),\delta(\fd')]\in Prim(\cD)\quad \forall \fd,\fd'\in \frak{Prim}(\fD)\,.
\end{equation}
The claim for generic primitive rule diagram algebra elements $d,d'\in Prim(\cD)$ then follows from multilinearity.
\end{proof}
\end{prop}

At this point, it is not yet obvious why it is meaningful to define $\cL_{\cD}$ over the $\bK$-vector space $Prim(\cD)$ instead of over the whole of $\cD$, but the reason for this choice will be eminent from the relationship between the universal enveloping algebra $\cU(\cL_{\cD})$ of $\cD$, the Lie algebra $\cL_{\cD}$ and $\cD$ itself, to be presented in the following subsection.\\

We conclude this subsection by defining the auxiliary inclusion morphism
\begin{equation}
\bar{\kappa}: \cL_{\cD}\hookrightarrow \cD: d\mapsto d\,,
\end{equation}
which simply states that the $\bK$-vector space $Prim(\cD)$ that underlies $\cL_{\cD}$ according to Proposition~\ref{prop:LieRD} is evidently a subspace of the $\bK$-vector space $\cD$ that underlies the rule diagram algebra $\cD$. Finally, the morphism $\kappa$ that appears in~\eqref{eq:algConnectionsB} is defined as
\begin{equation}
\kappa:=\bar{\varphi}\circ\bar{\kappa}: \cL_{\cD}\rightarrow \cR_T\,.
\end{equation}
It simply expresses the fact that any primitive rule diagram considered as an element of the Lie algebra $\cL_{\cD}$ may be mapped to its reduction, i.e.\ to an irreducible rule diagram via first embedding it into $\cD$, followed by applying (the appropriate variant of) the reduction morphism $\bar{\varphi}_T$. It is important to emphasize that the morphism $\kappa:\cL_{\cD}\rightarrow \cR_T$ is \emph{not} an inclusion of vector spaces, simply because generic primitive rule diagrams $\fd\in \frak{Prim}(\fD)$ are not necessarily irreducible (but instead generically nontrivial composites of atomic rule diagrams), whence in particular
\begin{equation}
\frak{Prim}(\fD)\not\subset \frak{Irr}(\fD\,.
\end{equation}

\subsection{The universal enveloping algebra of the rule diagram algebra} We have already seen in Section~\ref{sec:RDpart} that a generic rule diagram $\fd\in\cD$ is a composite of $n$ atomic rule diagrams $\fd_i$, which in general is in the isomorphism class of a superposition of a number of primitive rule diagrams $\fd(j)$:
\begin{equation}
\begin{aligned}
\fD\ni \fd&\equiv \left(\biguplus_{i=1}^n I_i,\biguplus_{i=1}^n O_i,\biguplus_{i=1}^n r_i,\biguplus_{\stackrel{i,j=1}{i<j}}^n \{m_{i,j}\},\biguplus_{i=1}^n s_i,\biguplus_{i=1}^n t_i\right)\equiv\biguplus_{j=1}^{n_c} \fd(j)\,.
\end{aligned}
\end{equation}
Here, $n_c$ is the \emph{number of connected components} aka \emph{primitive rule diagrams} of the composite rule diagram. Note that we have generically that $n_c\leq n$, with $n_c=n$ in case the rule diagram is the superposition of $n_c$ atomic rule diagrams. In other words, each generic primitive rule diagram itself is either an atomic rule diagram (i.e.\ irreducible and indecomposable), or it is an indecomposable composite of atomic rule diagrams:
\begin{equation}
\fd(j)\equiv\left(\biguplus_{k=1}^{n_j} I_{j_k},\biguplus_{k=1}^{n_j} O_{j_k},\biguplus_{p=1}^{n_j}r_{j_p},\biguplus_{\stackrel{p,q=1}{p<q}}^{n_j}\{ m_{j_p,j_q}\},\biguplus_{k=1}^{n_j} s_{j_k},\biguplus_{k=1}^{n_j} t_{j_k}\right)\,,
\end{equation}
with $\fd(j)\in \frak{Prim}(\fD)$, where the $r_{j_i}:\GRule{I_{j_i}}{O_{j_i}}$ are connected linear rules, and where $\fd(j)\in \frak{Prim}(\fD)$ necessitates that the matches of $\fd(j)$ lead to a fully connected diagram. This raises the natural question: can we express a generic basis diagram as a linear combination of compositions of primitive basis diagrams? More concretely, instead of specifying the precise set of matches $m_{i,j}$, can we alternatively just specify the full compositions (which are generically linear combinations of the composites along all possible matches)?	\\

Let us introduce the following convenient notation:
\begin{defn}
Let $\fd_1,\fd_2\in\fD$ be two  rule diagrams, and let $d_i:=\delta(\fd_i)\in\cD$ ($i=1,2$) be the associated basis vectors (aka basis diagrams). Then the \emph{nontrivial part of their composition}, denoted $d_1\ntC d_2$, is defined as
\begin{equation}
\begin{aligned}
	d_1\ntC d_2&:=	d_1*_{\cD} d_2-d_1\uplus d_2\\
&=\sum_{\stackrel{m_{12}\in\cM_{\fd_1}(\fd_2)}{m_{12}\neq \emptyset}}\delta(\comp{\fd_1}{m_{12}}{\fd_2})\,.\end{aligned}
\end{equation}	
\end{defn}
Moreover, for the remainder of this section we will use the notational abbreviation
\[
	*\equiv *_{\cD}\,.
\]

We begin our derivation by the observation that the composition of two primitive pure rule diagrams $d(i)=\delta(\fd(i))\in Prim(\cD)$ ($i=1,2$) may be expressed as
\begin{equation}
d(1)*d(2)=d(1)\uplus d(2) + d(1)\ntC d(2)\,.
\end{equation} 
But as already noted in the construction of the Lie algebra $\cL_{\cD}$ of $\cD$, the nontrivial part of the composition of two primitive basis diagrams consists only of linear combinations of primitive basis diagrams! Therefore, we may express the superposition of two primitive basis diagrams $d(i):=\delta(\fd(i))\in Prim(\cD)$ ($i=1,2$) as
\begin{equation}\label{eq:supA}
d(1)\uplus d(2)= d(1)* d(2)- d(1)\ntC d(2)
= d(2)* d(1)+[d(1),d(2)]-d(1)\ntC d(2)\,.
\end{equation}
Here, in the second equality, we have simply inserted the definition of the Lie bracket. On the other hand, we could have used the isomorphic representative $d(2)\uplus d(1)$, whence
\begin{equation}\label{eq:supB}
d(2)\uplus d(1)= d(2)*d(1)- d(2)\ntC d(1)
= d(1)*d(2)+[d(2),d(1)]-d(2)\ntC d(1)\,.
\end{equation}
While the two sets of expressions~\eqref{eq:supA} and~\eqref{eq:supB} do not look quite compatible at first sight, their equality follows from
\begin{equation}\label{eq:commAux}
[d_1,d_2]=d_1\ntC d_2-d_2\ntC d_1\,.
\end{equation}
Entirely analogously, we may derive formulae for $n$-fold superpositions of primitive basis diagrams from linear combinations of $p$-fold composites of primitives, with $1\leq p\leq n$. For example, for $n=3$ and $d(i):=\delta(\fd(i))\in Prim(\cD)$ ($i=1,2,3$) we have generically:
\begin{equation}
\begin{aligned}
d(1)*d(2)*d(3)&= d(1)\uplus(d(2)*d(3))+ d(1)\ntC(d(2)*d(3))\\
&= d(1)\uplus d(2)\uplus d(3)+d(1)\uplus(d(2)\ntC d(3))\\
&\quad 
+ d(1)\ntC(d(2)\uplus d(3))+ d(1)\ntC(d(2)\ntC d(3))\\
&= d(1)\uplus d(2)\uplus d(3)+d(1)*(d(2)\ntC d(3))\\
&\quad 
+ d(1)\ntC(d(2)\uplus d(3))\,.
\end{aligned}
\end{equation}
The only term that needs more attention is the following one:
\begin{equation}
\begin{aligned}
d(1)\ntC(d(2)\uplus d(3))&=
(d(1)\ntC d(2))\uplus d(3)+(d(1)\ntC d(3))\uplus  d(2)\\
&\quad+d(1)^{\ntC}(d(2)\uplus d(3))\\
&\overset{(*1)}{=}
(d(1)\ntC d(2))* d(3)+(d(1)\ntC d(3))\uplus d(2)\\
&=(d(1)\ntC d(2))* d(3)+(d(1)\ntC d(3))*d(2)\\
&\quad
-(d(1)\ntC d(3))\ntC d(2)\\
&\overset{(*2)}{=}
(d(1)\ntC d(2))\uplus d(3)+(d(1)\ntC d(3))*d(2)\\
&=
(d(1)\ntC d(2))* d(3)+(d(1)\ntC d(3))*d(2)\\
&\quad-
(d(1)\ntC d(2))\ntC d(3)\,.
\end{aligned}
\end{equation}
Here, the somewhat awkward notation $d(1)^{\ntC}(d(2)\uplus d(3))$ used in the first line stands for all contributions that have non-trivial matches between $\fd_1$ and $\fd_2$ as well as between $\fd_1$ and $\fd_3$. Moreover, the two variants $(*1)$ and $(*2)$ amount to the two different variants to interpret the first line (and we could additionally have permuted the superposition factors in the two variants). In summary, one of the possible presentations of the triple composition reads
\begin{equation}
\begin{aligned}
 d(1)*d(2)*d(3)&= d(1)\uplus d(2)\uplus d(3)
+ d(1)*( d(2)\ntC d(3))\\
&\quad
+ ( d(1)\ntC d(2))* d(3)+( d(1)\ntC d(3))* d(2)\\
&\quad
-\tfrac{1}{2}\left[( d(1)\ntC d(3))\ntC d(2)
+
( d(1)\ntC d(2))\ntC d(3)\right]\,.
\end{aligned}
\end{equation}

In other words, even though it is somewhat unwieldy to compute a presentation already for the triple superposition of primitive rule diagrams in this fashion, it is immediately obvious that it is always possible to inductively derive such a presentation for superpositions of arbitrary numbers of primitive rule diagrams! This motivates the following definition:
\begin{defn}[Autoequivalence morphism in $\cD$]\label{eq:RDautoEq}
Let $d\equiv\delta(\fd)\in\cD$ be a generic basis diagram, whence
\begin{equation}
d=\biguplus_{i=1}^{n_c} d(i)\qquad (d(i)\equiv\delta(\fd(i))\in Prim(\cD))\,.
\end{equation}
Then the \emph{autoequivalence morphism}
\begin{equation}
\alpha: \cD\rightarrow \cD: d=\biguplus_{i=1}^{n_c} d(i) \mapsto \alpha_{n_c}(d)
\end{equation}
is defined inductively as follows:
\begin{itemize}
\item $n_c=0$ or $n_c=1$: 
\begin{equation}
\alpha_{0}(d)\equiv\alpha_0(d_{\emptyset}):=d_{\emptyset}\,,\quad
\alpha_1(d(1)):=d(1)\,.
\end{equation}
\item $n_c=2$: 
\begin{equation}
\alpha_2(d(1)\uplus d(2)):=d(1)*d(2)- d(1)\ntC d(2)=\alpha_2(d(2)\uplus d(1))\,.
\end{equation}
The last equality follows directly from the definitions of $[\cdot,\cdot]$ and of $\ntC$.
\item $n_c\to n_c':=n_c+1$: We employ the formulae for $\alpha_p$ with $1\leq p\leq n_c$ to decompose the sequential composition of $(n_c+1)$ primitive rule diagrams, which yields a formula for the superposition of $(n_c+1)$ primitive rule diagrams.
\end{itemize}
Moreover, $\alpha$ is extended to all of $\cD$ by multilinearity.
\end{defn}

It is then immediately obvious that we may formulate the following generalization of the Poincar\'{e}-Birkhoff-Witt Theorem:
\begin{thm}[Generalized PBW-theorem for $\cD$]\label{thm:gPBWrD}
Introduce an arbitrary \emph{total order on the set of isomorphism classes of primitive rule diagrams} $\fd_{p_i}\in \frak{Prim}(\fD)$, 
\begin{equation}
\fd_{1} \prec \fd_{2}\prec\dotsc\,,
\end{equation}
and let the same symbol $\prec$ also stand for the induced total order of the associated basis vectors $d_{i}\equiv \delta(\fd_{i})\in Prim(\cD)$ aka primitive pure rule diagrams. Then an arbitrary pure rule diagram $d=\uplus_{i=1}^{n_c}d_{p_i}$ (where repetitions in the indices $p_i$ are explicitly allowed) may be \emph{uniquely represented} as
\begin{equation}
\alpha(d)=t_0\cdot d_{\emptyset}+\sum_{i=1}^{n_c}\sum_{\stackrel{p_{1},\dotsc,p_{i}\in \cI_{\cP}}{p_{1}\leq p_{2}\leq\dotsc\leq p_i}} t_{p_1\dotsc p_i} d_{p_1}*\dotsc*d_{p_i}\,,
\end{equation}
where $\cI_{\cP}$ denotes the set of indices for the isomorphism classes $\fd_{i}\in \frak{Prim}(\fD)$, and where the coefficients are numbers (i.e.\ with values in $\bK$).
\end{thm}

In a sense, this generalized form of the PBW theorem entails that there exists an \emph{alternative basis} for the vector space of isomorphism classes of rule diagrams that is expressed in terms of \emph{``normal-ordered'' words in primitive pure rule diagrams modulo the relations induced by the Lie bracket}. We will encounter some explicit realizations of this isomorphism in the study of subalgebras of $\cD$ in Section~\ref{sec:structureTheory}.\

It seems intuitive to employ a choice of basis in terms of superpositions of primitive rule diagrams, 
\begin{equation}
\delta(\fd)=\delta\left(\biguplus_{\stackrel{j=1}{i_1\leq i_2\leq \dotsc \leq i_{n_c}}}^{n_c} \fd_{i_j}\right)\,,
\end{equation}
where we again fixed some arbitrary, but global choice of total order on the isomorphism classes of primitive rule diagrams. There even existed an alternative choice of algebra operation in the form of the superposition operation that would render a \emph{commutative algebra},
\begin{equation}
Poly(\cD):=(span_{\bK}(\fD),\uplus)\cong \bK[\fD]\,,
\end{equation}
with
\begin{equation}
\delta(\fd_1)\uplus \delta(\fd_2):=\delta(\fd_1\uplus \fd_2)\,,
\end{equation}
and where $\bK[\fD]$ denotes the polynomial ring over $\bK$. However, clearly the rule diagram algebra is \emph{not} commutative, and one would have to encode horrifically complex relations on the rule diagrammatic compositions of superpositions of rule diagrams in order to encode the non-commutativity, such as for example (with $d_i\equiv \delta(\fd_i)\in \cD$)
\begin{align*}
(d_1\uplus d_2)*_{\cD}d_3&= d_1\uplus d_2\uplus d_3+(d_1\ntC_{\cD}d_2)\uplus d_3+\dotsc\,.
\end{align*}%
In other words, only the PBW-like choice of basis promises for applications of ideas from combinatorics, where in particular one tries to derive combinatorial rules from the commutation relations of the primitive elements.\\

Finally, we may define the \emph{universal enveloping algebra} of the Lie algebra of the rule diagram algebra by specializing the standard construction (where $\cT(V)$ is the \emph{tensor algebra} over a vector space $V$)
\begin{equation}
\cU(\cL):=\frac{\cT(\cL)}{\langle x\otimes y-y\otimes x -[x,y]\rangle}
\end{equation}
to the case at hand:
\begin{defn}\label{def:UA}
The \emph{universal enveloping algebra} $\cU(\cL_{\cD})$ of the Lie algebra $\cL_{\cD}$ of the rule diagram algebra $\cD$ is defined as 
\begin{equation}
\cU(\cL_{\cD}):=\frac{\cT(\cL_{\cD})}{\langle d_a\otimes d_b-d_b\otimes d_a -[d_a,d_b]\rangle}\,.
\end{equation}
Here, the indices in the denominator are understood as running over any possible pair of indices. As conventional in the mathematics literature, we will not always explicitly write out the tensor product symbols $\otimes$ when specifying elements of $\cU(\cL_{\cD})$.
\end{defn}

It is then immediately clear from our definitions that we may specify an \emph{isomorphism} 
\begin{equation}
	\varphi:\cU(\cL_{\cD})\xrightarrow{\cong}\cD: d_1\otimes d_2\dotsc \otimes d_n\mapsto  d_1*_{\cD} d_2*_{\cD}\dotsc*_{\cD} d_n\,,
\end{equation}
which is extended to all of $\cD$ by multilinearity. Quite importantly, this constitutes a very powerful a posteriori consistency check in favor of our construction of the rule diagram algebra $\cD$, since it suggests that our construction is (up to choice of basis) essentially canonical.\\

Another instructive exercise consists in computing the effect of the autoequivalence $\alpha:\cD\rightarrow \cD$ onto the formula for the antipode, as explicitly presented in~\eqref{eq:AntipodeExpl}. Noticing that since the antipode is a multilinear mapping, we may e.g.\ compute for $d=d_1\uplus d_2$ (with $d_i\equiv \delta(\fd_i)\in Prim(\cD)$ primitive basis diagrams):
\begin{equation}
\begin{aligned}
S(d_1\uplus d_2)&\overset{\eqref{eq:AntipodeExpl}}{=}d_1*d_2+d_2*d_1-(d_1\uplus d_2)=d_1*d_2+d_2\ntC d_1\\
&= S(d_1*d_2-d_1\ntC d_2)=S(d_1*d_2)-S(d_1\ntC d_2)\\
&=S(d_1*d_2)+d_1\ntC d_2\\
\Leftrightarrow\quad S(d_1*d_2)&=d_1*d_2-[d_1,d_2]
=d_2\uplus d_1+d_2\ntC d_1=d_2*d_1\,.
\end{aligned}
\end{equation}
A very tedious computation reveals moreover that (for $d_i\equiv\delta(\fd_i)\in Prim(\cD)$)
\begin{equation}
S(d_1*d_2*d_3)=-d_3*d_2*d_1\,,
\end{equation}
illustrating the \emph{antihomomorphism property} of the antipode, i.e.\
\begin{equation}
S(d_1*(d_2*d_3))=S(d_2*d_3)*S(d_1)=-d_3*d_2*d_1\,.
\end{equation}
By induction, we thus conclude that
\begin{equation}\label{eq:antipodeRDA}
S(d_1*\dotsc*d_n)=(-1)^nd_n*\dotsc*d_1\qquad \forall n>0, d_i\equiv\delta(\fd_i)\in Prim(\cD)\,.
\end{equation}
But by virtue of Theorem~\ref{thm:gPBWrD}, any arbitrary rule diagram $d\in \cD$ may be represented as a linear combination of compositions of primitive basis diagrams, which demonstrates that 
\begin{equation}
S\circ S(d)=d\qquad \forall d\in \cD\,,
\end{equation}
whence we have obtained the following refinement of Theorem~\ref{thm:RDhopf}:
\begin{thm}\label{thm:RDhopfV2}
The rule diagram algebra $\cD\equiv(\fD,\mu,\eta,\Delta,\varepsilon,S)$ is an \emph{involutive Hopf algebra} and whence a \emph{$*$-algebra}, with \emph{antipode $S$} given by~\eqref{eq:antipodeRD}, which is also an \emph{involution}.
\end{thm}

It is finally easy to read off the antipode $S_{\cU}$ of the universal enveloping algebra $\cU(\cL_{\cD})$:
\begin{equation}
S_{\cU}:\cU(\cL_{\cD})\rightarrow\cU(\cL_{\cD}):\bigotimes_{i=1}^n d_i\mapsto (-1)^n\bigotimes_{i=1}^n d_{n-i+1}\,.
\end{equation} 

\section{The prototypical example: the Heisenberg-Weyl and the HW diagram algebra}\label{sec:HW}

The so-called (single-type) \emph{Heisenberg-Weyl (HW) algebra} is the algebra with generators $a$, $a^{\dag}$ and $I$ and presented by the relation
\begin{equation}\label{eq:commHW}
[a,a^{\dag}]=I\,.
\end{equation}
It is one of the archetypical algebras used in mathematical combinatorics, and also plays a fundamental role in the study of the harmonic oscillator in quantum mechanics. For the purposes of the present paper, as we will present in this section the algebra may also be interpreted as one of the simplest conceivable subalgebras of the rule algebras $\cR_T$! We will present the analysis of this algebra in some detail, since it will serve as a first \emph{a posteriori} consistency check for our rule algebra constructions, in particular by comparing it to the results of Blasiak et al.~\cite{blasiak2010combinatorial}.\\ 

According to our definition of the rule algebra $\cR_T$, a realization of the HW algebra may be specified via the identifications
\begin{equation}
\begin{aligned}\label{eq:HWAlgRA}
a\leftrightarrow r_a&:=r_{\GRule{\bullet}{\emptyset}}\equiv \includetikz{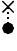}\,,
\quad a^{\dag}\leftrightarrow r_{a^{\dag}}:= r_{\GRule{\emptyset}{\bullet}}\equiv \includetikz{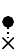}\,,
\quad
I\leftrightarrow r_I:= r_{\emptyset}\,.
\end{aligned}
\end{equation}
Note that if we only study the algebra generated by these specific rule algebra elements (and not its representations), the type $T$ of the rule algebra $\cR_T$ is immaterial, because the generating rule algebra elements only involve vertex transformations, while the four types of rule algebra differ only in their treatment of edge transformations (this general property of the rule algebras will be further elaborated upon in Section~\ref{sec:structureTheory}). We will thus in the following for brevity denote ``the'' rule algebraic realization of the HW algebra by $\cR_{\cH \cW}\subset \cR_T$, dropping the explicit mention of the type $T$ (we will return to the distinction between the four types of HW algebra in the rule algebraic formulation in~\cite{bd2016}, where we will study the interplay between the type of the rule algebra $\cR_T$ and different special cases of representations of the subalgebra $\cR_{\cH \cW}$).\\

We will now reproduce the analysis of Blasiak et al.~\cite{blasiak2010combinatorial} in terms of the rule algebraic realization $\cH\cW$ of the Heisenberg-Weyl algebra. To this end, we also define the \emph{Heisenberg-Weyl rule diagram algebra} $\cD_{\cH \cW}$ as the subalgebra of $\cD$ spanned by the primitive basis diagrams
\begin{equation}\label{eq:defHWrd}
d_a:=\includetikz{HWdA}\,,\quad d_{a^{\dag}}:=  \includetikz{HWdAD}\,,\quad d_I:=d_{\emptyset}\,,\quad 
d_e:=\includetikz{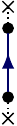}\,.
\end{equation}

\subsection{The HW diagram algebra} 

Let us first study the properties of the HW rule diagram algebra $\cD_{\cH\cW}$. It is evident from the definitions that the only nontrivial composite of the generators of $\cD_{\cH\cW}$ is given by
\begin{equation}
d_a\ntC_{\cD} d_{a^{\dag}}=d_e\,,
\end{equation}
or diagrammatically 
\begin{equation}\label{eq:HWrdNonTriv}
\includetikz{HWdA}\ntC_{\cD}  \includetikz{HWdAD}=\includetikz{HWdE}\,.
\end{equation}
In particular, we have that all other conceivable compositions of the basis elements only have contributions along trivial matches, whence if we denote as usual by $\ntC_{\cD}$ the contributions to a composition $*_{\cD}$ along nontrivial matches, we obtain an extremely simple ``multiplication table'' for $\cD_{\cH\cW}$:
\begin{equation}\label{eq:HWdAdagA}
\begin{array}{c|ccc}
\downarrow d_1\ntC_{\cD}d_2\to & d_a & d_e & d_{a^{\dag}}\\\hline
d_{a^{\dag}} & 0 & 0 & 0\\
d_e & 0 & 0 & 0\\
d_a & 0 & 0 & d_e
\end{array}
\end{equation}
Noticing that coincidentally all generators of $\cD_{\cH\cW}$ are \emph{primitive basis diagrams} and invoking the isomorphism between the universal enveloping algebra $\cU(\cL_{\cD})$ and the rule diagram algebra, we immediately obtain the \emph{normal ordered form} for an arbitrary element of ${\,\cU(\cL_{\cD})}$ via employing the relation of the universal enveloping algebra $d_1\otimes d_2=d_2\otimes d_1+[d_1,d_2]$ repeatedly, and by choosing the partial order
\begin{equation}
d_{a^{\dag}}\prec d_{a} \prec d_e\,.	
\end{equation}
More explicitly:
\begin{equation}
d_{i_1}\otimes\dotsc \otimes d_{i_n}=f_{0}^{i_1\dotsc i_n}d_{\emptyset}+
\sum_{\stackrel{r,s,t=0}{r+s+t>0}}^{n} f_{rst}^{i_1\dotsc i_n} d_{a^{\dag}}^{\otimes\:r}\otimes d_a^{\otimes\: s}\otimes d_e^{\otimes\: t}\,,
\end{equation}
where we define that $d_a^{\otimes0}=d_{a^{\dag}}^{\otimes0}=d_e^{\otimes0}=1_{\bK}$, and where the coefficients $f_{rst}^{i_1\dotsc i_n}$ are called the \emph{normal ordering coefficients}. Therefore, we have in particular also that (applying the isomorphism back to the rule diagram algebra)
\begin{equation}
d_{i_1}*_{\cD}\dotsc *_{\cD}d_{i_n}=f_{0}^{i_1\dotsc i_n}d_{\emptyset}+\sum_{\stackrel{r,s,t=0}{r+s+t>0}}^{n}f_{rst}^{i_1\dotsc i_n} 
\underbrace{d_{a^{\dag}}^{*r}*_{\cD}d_a^{*s}*_{\cD} d_e^{*t}}_{=d_{a^{\dag}}^{\uplus\:r}\uplus d_a^{\uplus\: s}\uplus  d_e^{\uplus\: t}}\,.
\end{equation}
Notice that for the case of the HW diagram algebra, the chosen total order of basis elements is advantageous, since in the form presented above all compositions only have contributions along trivial matches. In other words: 
\begin{center}
\emph{The standard normal ordered form of the HW diagram algebra is one in which all elements are presented as \emph{superpositions} of the basis elements $d_{a^{\dag}}$, $d_a$ and $d_e$!}
\end{center}
This observation will be taken as the fundamental idea for defining a generalized normal ordering in Section~\ref{sec:structureTheory}.\\

A very interesting computation that may be performed in the rule diagram algebra consists now in deriving a formula for the normal ordering coefficients of a composition of two normal ordered expressions in $\cD_{\cH\cW}$. Due to multilinearity, it suffices to derive the explicit formula for the composition of two normal ordered expressions of the form
\begin{equation}
d(r,s,t):= d_{a^{\dag}}^{*\:r}*_{\cD} d_{a}^{*\:s}*_{\cD} d_{e}^{*\:t}=
 d_{a^{\dag}}^{\uplus\:r}\uplus d_{a}^{\uplus\:s}\uplus d_{e}^{\uplus\:t}\quad (r,s,t\in \bZ_{\geq0})\,,
\end{equation}
where the second form of the expression follows because of~\eqref{eq:HWdAdagA}. Composing thus a rule diagram $d(r_1,s_1,t_1)$ with a rule diagram $d(r_2,s_2,t_2)$,
\begin{equation}\label{eq:HWnoCoeffRD}
\begin{aligned}
&d(r_1,s_1,t_1)*_{\cD}d(r_2,s_2,t_2)\\
&\quad=\left(
\overbrace{
\includetikz{HWdAD}\dotsc \includetikz{HWdAD}
}^{\text{$r_1$ times}}\;\overbrace{
\includetikz{HWdA}\dotsc \includetikz{HWdA}
}^{\text{$s_1$ times}}\;
\overbrace{\includetikz{HWdE}\uplus\dotsc\uplus\includetikz{HWdE}}^{\text{$t_1$ times}}
\right)*_{\cD}\left(
\overbrace{
\includetikz{HWdAD}\dotsc \includetikz{HWdAD}
}^{\text{$r_2$ times}}\;\overbrace{
\includetikz{HWdA}\dotsc \includetikz{HWdA}
}^{\text{$s_2$ times}}\;
\overbrace{\includetikz{HWdE}\dotsc\includetikz{HWdE}}^{\text{$t_2$ times}}
\right)\,,
\end{aligned}
\end{equation}
we see that the contribution of the composition along the trivial match is just the total superposition,
\begin{equation}
\begin{aligned}
	\comp{d(r_1,s_1,t_1)}{\emptyset}{d(r_2,s_2,t_2)}&=d(r_1,s_1,t_1)\uplus d(r_2,s_2,t_2)\\
&=d(r_1+r_2,s_1+s_2,t_1+t_2)\,,
\end{aligned}
\end{equation}
where in the last step we have simply used that superpositions are totally commutative, whence we are free to present the isomorphism class as indicated. The nontrivial matches consist in forming vertex-vertex matches, whence for each such match a component $d_{a}$ of $d(r_1,s_1,t_1)$ needs to be matched with a component $d_{a^{\dag}}$ of $d(r_2,s_2,t_2)$ (and there are no other possibilities for nontrivial matches). Each individual matching of a component $d_{a}$  with a component $d_{a^{\dag}}$ simply leads to an instance of $d_e$ under composition along the match, with the remaining diagrams unaffected. Moreover, since we need one-to-one matches of vertices, it is clear that we can at most form $min(s_1,r_2)$ many such pairs, and finally all copies of $d_{a}$ and $d_{a^{\dag}}$ are \emph{indistinguishable} (because we work with \emph{isomorphism classes} of rule diagrams), which may all be summarized into the following succinct formula:
\begin{equation}
\begin{aligned}\label{eq:HWDAnorm}
&d(r_1,s_1,t_1)*_{\cD} d(r_2,s_2,t_2)\\
&\quad=\sum_{n=0}^{min(s_1,r_2)}\frac{s_1! r_2!}{(s_1-n)! n!(r_2-n)!}d(r_1+r_2-n,s_1+s_2-n,t_1+t_2+n)\,.
\end{aligned}
\end{equation}
In words: there are $s_1!/(s_1-n)!$ possibilities to pick $n$ copies of $d_a$ from $d(r_1,s_1,t_1)$ and $r_2!/(r_2-n)!$ possibilities to pick $n$ copies of $d_{a^{\dag}}$ from $d(r_2,s_2,t_2)$ to form a match of $n$ vertex-vertex matches, but due to indistinguishability one has to divide the product of these numbers of possibilities by $n!$. We thus precisely recover from first principles the standard formula for normal ordering coefficients of the Heisenberg-Weyl rule diagram algebra as presented e.g.\ in~\cite[p.6, eq.~(7)]{blasiak2010combinatorial}.\\

Referring the interested readers to the work of Blasiak et al.\ \cite{blasiak2010combinatorial} for a first account of a diagram algebra construction for the HW algebra (that served in fact as the starting point for the considerations presented in the current paper), let us conclude this subsection by very briefly specializing our general rule diagram algebra Hopf algebra structures to the much simpler case of the HW diagram algebra for illustration (explaining en passent why the HW diagram algebra is in fact such a simple special case). 

\subsubsection{Algebra structure of $\cH\cW$}

Of course, since the HW diagram algebra $\cD_{\cH\cW}$ is clearly a subalgebra of $\cD$, we are guaranteed by virtue of the properties of $\cD$ that this subalgebra is unital and associative. The explicit demonstration of associativity of $\cD_{\cH\cW}$ in the following is purely intended to illustrate in which aspects our construction is analogous to that of~\cite{blasiak2010combinatorial}. Consider thus a generic triple composite of HW rule diagrams, where it w.l.o.g.\ suffices to consider triple composites of normal ordered expressions:
\begin{equation}
\begin{aligned}
d_1*_{\cD}d_2*_{\cD}d_3&\equiv d(r_1,s_1,t_1)*_{\cD} d(r_2,s_2,t_2)*_{\cD} d(r_3,s_3,t_3)\,.
\end{aligned}
\end{equation}
The big simplification that occurs here in comparison to the full rule diagram algebra compositions is that none of the diagrams $d_i$ contains any edges, which entails that all constituents $m_{i,j}$ of a given match $m\in \cM_{1,2,3}$ are simply one-to-one pairings of vertices (i.e.\ no complications due to the delayed edge homomorphism condition arise). For the particular case of the diagrams $d_i$, we analyzed earlier that the only possible nontrivial matches between two diagrams $d_i$ and $d_j$ ($i<j$) consist in pairing copies of $d_a$ in $d_i$ with copies of $d_{a^{\dag}}$ of $d_j$, whence any match $m\in \cM_{1,2,3}$ is simply a set of three binary matches $\{m_{1,2},m_{1,3},m_{2,3}\}$ which consists entirely of one-to-one vertex-vertex pairings, whence we have re-derived quite directly the associativity of the HW rule diagram algebra:
\begin{equation}
\begin{aligned}
&\big(d(r_1,s_1,t_1)*_{\cD} d(r_2,s_2,t_2)\big)*_{\cD} d(r_3,s_3,t_3)\\
&\quad=d(r_1,s_1,t_1)*_{\cD} \big(d(r_2,s_2,t_2)*_{\cD} d(r_3,s_3,t_3)\big)\,.
\end{aligned}
\end{equation}

\subsubsection{Coalgebra structure of $\cH\cW$}

It is precisely in the coalgebra structure of $\cD_{\cH\cW}$ that our rule diagram approach appears to offer some simplifications over the variant presented in~\cite{blasiak2010combinatorial}. The core heuristic of both approaches is that the coproduct should be interpreted as ``storing'' all possibilities to decompose a given object of the diagram algebra. Specializing to the rule diagram algebra $\cD_{\cH\cW}$, we wish to define for the coproduct of a pure basis $d:=\delta(\fd)\in \cD$
\begin{equation*}
\Delta(\delta(\fd)):=\sum_{\stackrel{\text{ways to delete}}{\text{superposition factors from $\fd$}}} \delta\left(\fd_{\text{some factors deleted}}\right)\otimes \delta\left(\fd_{\text{deleted factors}}\right)\,.
\end{equation*}
The counit is the empty rule diagram $d_{\emptyset}$, which is by definition indecomposable, whence
\begin{equation}
\Delta(d_{\emptyset}):= d_{\emptyset}\otimes d_{\emptyset}\,.
\end{equation}
It is then an easy exercise to confirm that a normal ordered expression in $\cD_{\cH\cW}$ aka a linear combination of expressions of the form $d(r,s,t)$ (which coincidentally is a superposition of the three primitive basis diagrams $d_a$, $d_{a^{\dag}}$ and $d_e$) has the coproduct
\begin{equation}
\Delta(d(r,s,t))=\sum_{m=0}^r\sum_{n=0}^s\sum_{\ell=0}^t
\binom{r}{m}\binom{s}{n}\binom{t}{\ell} d(m,n,\ell)\otimes d(r-m,s-n,t-\ell)\,.
\end{equation}
Here, we have used the convention $d(0,0,0)\equiv d_{\emptyset}$, and also the fact that the number of possibilities to choose $n$ copies of indistinguishable objects from a pool of $m$ ($m\geq n$) indistinguishable objects is $\binom{m}{n}$.\\

In comparison, the authors of~\cite{blasiak2010combinatorial} employed a notion of ``spider-like'' diagrams for their diagram algebra, which at least visually would suggest that a given expression of the HW diagram algebra would be connected in some shape or form. We believe that our alternative formulation offers slightly more clarity in this point, emphasizing that the only nontrivial connected subdiagrams are those of the type $d_e$.

\subsubsection{Filtration structure and antipode of $\cD_{\cH\cW}$}

Referring the interested readers to Appendix~\ref{app:FilteredHopf} for the full details on the filtration structure of the rule diagram algebra, the specialization to $\cD_{\cH\cW}$ is particularly simple: each of the elements $d(r,s,t)$ is an element of the filtration components of index at least $r+s+t$,
\begin{equation}
d(r,s,t)\in \cD^n\quad \forall r+s+t\leq n\,,
\end{equation}
whence it follows directly from the explicit formula for composition and coproduct of rule diagrams of the form $d(r,s,t)$ that these operations are compatible with the filtration:
\begin{equation}
\begin{aligned}
d(r_1,s_1,t_1)*_{\cD}d(r_2,s_2,t_2)&\in \cD^{(r_1+r_2+s_1+s_2+t_1+t_2)}\\
\Delta(d(r,s,t))&\in \sum_{p=0}^{r+s+t} \cD^p\otimes \cD^{r+s+t-p}\,.
\end{aligned}
\end{equation}
Finally, by virtue of our general formula~\ref{eq:antipodeRDA} for the antipode of a product of  primitive basis diagrams $d_i:=\delta(\fd_i)\in Prim(\cD)$
\begin{equation*}
S(d_1*\dotsc*d_n)=(-1)^nd_n*\dotsc*d_1\qquad \forall n>0\,,
\end{equation*}
and since the elements $d(r,s,t)$ may be presented as products of the primitive basis diagrams $d_{a^{\dag}}$, $d_a$ and $d_e$, we immediately obtain that
\begin{equation}
S(d(r,s,t))=(-1)^{r+s+t} d_e^{*t}*_{\cD} d_a^{*s}*_{\cD} d_{a^{\dag}}^{*r}\,.
\end{equation}
Combining this result with our explicit formula~\eqref{eq:HWDAnorm} for the normal ordering of HW diagram algebra elements, we obtain the alternative formula
\begin{equation}
S(d(r,s,t))=(-1)^{r+s+t} \sum_{k=0}^{min(r,s)}k!\binom{r}{k}\binom{s}{k} d(r-k,s-k,t)\,,
\end{equation}
thereby reproducing precisely the corresponding results of~\cite{blasiak2010combinatorial}.

\subsection{Algebra structure of $\cR_{\cH\cW}$} 

The algebra structure of $\cR_{\cH\cW}$ is particularly simple, since again the only nontrivial composite of the generators occurs in the composition 
\begin{equation}
r_a* r_{a^{\dag}}=r_a\uplus r_{a^{\dag}}+r_I\,,
\end{equation}
or diagrammatically
\begin{equation}\label{eq:HWnontriv}
\left(\includetikz{HWdA}\right)*\left(\includetikz{HWdAD}\right)
=\bar{\varphi}\left(
\includetikz{HWdA}\;\includetikz{HWdAD}+\includetikz{HWdE}
\right)
=\left(
\includetikz{HWdAD}\right)\uplus\left(\includetikz{HWdA}\right)+r_{\emptyset}\,.
\end{equation}
Here, in the final step we have used the commutativity of the superposition in order to bring the result into standard order. We also note the important special property
\begin{equation}
r_{a^{\dag}}*r_a=r_{a^{\dag}}\uplus r_a\,.
\end{equation}
Therefore, the idea of \emph{normal ordering} as presented for the HW rule diagram algebra carries over to the HW rule algebra \emph{mutatis mutandis}: any element of the HW rule algebra may be expressed in terms of normal ordered expressions of the form
\begin{equation}
r(m,n):= r_{a^{\dag}}^{*\:m}*r_{a}^{*\: n}= r_{a^{\dag}}^{\uplus\:m}\uplus r_{a}^{\uplus\: n}\,,
\end{equation} 
and since we have by definition of the insertion morphism $\psi:\cR_T\rightarrow \cD$ that
\begin{equation}
\psi\big( r(m,n)\big)=d(m,n,0)\,,
\end{equation}
combined with the fact that for any type of rule algebra we have that
\begin{equation}
\bar{\varphi}_T\big(d(m,n,k)\big)=r(m,n)\,,
\end{equation}
we may immediately conclude that
\begin{equation}
\begin{aligned}
&r(m_1,n_1)*r(m_2,n_2)=\bar{\varphi}_T\circ(\psi(r(m_1,n_1))*_{\cD}\psi(r(m_2,n_2)))\\
&\quad =\sum_{p=0}^{min(n_1,m_2)}p!\binom{n_1}{p}\binom{m_2}{p}r(m_1+m_2-p,n_1+n_2-p)\,,
\end{aligned}
\end{equation}
thereby recovering the classical result for the normal ordering coefficients of the composition of two normal ordered expressions in the HW algebra.

\section{Structure theory of the rule algebra}\label{sec:structureTheory}

Insight into the finer structure of the rule algebras $\cR_T$ may be gained via following the standard approach to any algebraic theory: via the study of their \emph{generators} and of the \emph{relations} their compositions via the algebra compositions $*_T$ fulfill. However, unlike well-known examples of countably infinite dimensional algebras such as the multi-type Heisenberg-Weyl algebra (the dimension referring to the number of generators), in the case of the rule algebras we shall see that the relations fulfilled already by the compositions of their generators are by far too complex in order to directly study their presentations in any way of use for applications of the framework. It will thus be quintessential to develop the structure theory of the rule algebras in terms of the study of what we call the \emph{generator subalgebras} (specified via explicit subsets of generators) and the \emph{structural subalgebras} (which we will define as subalgebras of $\cR_T$ that are characterized by the structure of their elements, and which will possess comparatively simpler relations).\\

After introducing the generators of $\cR_T$, we will already be able to identify some well-known algebras as certain subalgebras expressible as generator subalgebras. We will then focus on the structural subalgebras of \emph{graph creators and annihilators} $\cR^{+}_T$, $\cR^{-}_T$, respectively, on their union $\cR^{\pm}_T$, and also on various types of structural subalgebras characterized by their elements having equal left- and right hand side graphs.\\

As one very spectacular special case, we will introduce the notion of a generalization of the multi-type Heisenberg-Weyl algebra in the form of the algebra $\cR_{DPO}^{\pm}$. The study of this particular subalgebra forms the basis of one of the possible approaches to combinatorial physics in the new framework, as will be presented in detail in future work.\\

\textsc{Convention:} For notational convenience, we will from hereon adopt the shorthand notations
\begin{equation}
	*_T\equiv *_{\cR_T}\,,\quad \ntC_T\equiv\ntC_{\cR_T}\,.
\end{equation}

\subsection{The generators and generator subalgebras of the rule algebras}\label{sec:generatorSubalgebras}

The following tentative table of generators $x:=\delta(\frak{x})$ could be seen as a sort of ``periodic table'' of the rule algebras -- each entry is one of the atomic generators of the rule algebras $\cR_T$, according to the idea that any rule algebra element may be realized as a term in the compositions of a number of copies of the generators or of arbitrary (finite) superpositions of the generators (for the loop and edge algebras):
\begin{equation}
\begin{array}{*{2}{rcl|}rcl}
\frak{a}^{\dag} & := & \includetikz{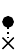} & 
\frak{I} & := & \includetikz{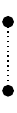} &
\frak{a} & := & \includetikz{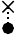}\\
\hline
\frak{l}^{\dag} & := & \includetikz{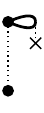}  & 	
\frak{L} & := & \includetikz{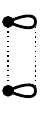} & 
\frak{l} & := & \includetikz{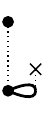}  \\
\hline
\frak{e}_{>}^{\dag} & := & \includetikz{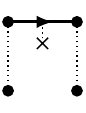}  & 	
\frak{E}_{>} & := & \includetikz{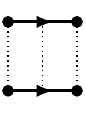}  & 
\frak{e}_{>} & := & \includetikz{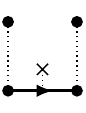}
\end{array}
\end{equation}
However, as we will see in the sequel, it will turn out that the tentative list of generators as presented here is incomplete, see Section~\ref{sec:loopAlgebra}ff.\\

A relatively straightforward to compute part of the relations of the rule algebras consists in the \emph{commutation relations} between the various generators. We will first study some natural subalgebras of the rule algebras generated by subsets of the generators, which we will refer to as \emph{generator subalgebras}. In particular, we will demonstrate to which extent a \emph{presentation} of these subalgebras may be given in the standard form of a free monoid algebra with relations, giving a first hint on the origin of combinatorial complexity in graph rewriting systems.

\subsubsection{The vertex subalgebra $\cR^{\cV}$ aka the Boson algebra}
  
The vertex algebra is generated by the rule algebra elements $x:=\delta(\frak{x})$ ($x\in \{a^{\dag},I,a\}$), with
\begin{equation}
\frak{a}^{\dag} := \includetikz{vC}\,,\quad \frak{I}:=\includetikz{vI}\,,\quad
\frak{a} :=\includetikz{vA}\,.
\end{equation} 
Recalling the convenient shorthand notation
\[
r_1\ntC_T r_2=r_1*_T r_2- r_1\uplus r_2
\]
for the nontrivial parts of the composition of two rule algebra elements $r_1,r_2\in\cR_T$, we obtain the following ``multiplication table'' for the vertex subalgebra $\cR^{\cV}$: 
\begin{equation}\label{eq:vertexAlgebra}
\begin{array}{c|ccc}
\ntC_T & a & I  & a^{\dag}\\
\hline
a^{\dag} & 0 & 0 & 0 \\
I & 0 & I & a^{\dag}\\
a & 0 & a & r_{\emptyset}
\end{array}
\end{equation}
The data in this table does \emph{not} depend on the type $T$ of the underlying rule algebra $\cR_T$, which is why we refer to ``the'' vertex algebra. Moreover, in this particular case the full set of relations that are necessary to describe the vertex algebra is already contained in the \emph{commutation relations}
\begin{equation}
\begin{aligned}\label{eq:vertexAlgebraCRs}
[a,a^{\dag}]&=r_{\emptyset}\\
[a,I]&=a\,, \quad [I,a^{\dag}]=a^{\dag}\,.
\end{aligned}
\end{equation}
All other commutators of the generators are zero. Let us introduce the \emph{normal ordering}
\begin{equation}
\frak{a}^{\dag}\prec \frak{I}\prec \frak{a}\,,
\end{equation}
such that a generic multiple composition of the generators may by repeated use of the commutation relations~\eqref{eq:vertexAlgebraCRs} be expressed in terms of linear combinations of normal-ordered terms of the form
\begin{equation}\label{eq:VAnormalForm}
V(m,n,p):=\left(a^{\dag}\right)^{*m}*I^{*n}*a^{*p}\,.
\end{equation}
In order to prove that any possible element of the rule algebras $\cR_T$ consisting only of discrete graph rewriting rules may indeed be generated by the three (thus far tentative) vertex algebra generators, we need to demonstrate finally that any possible superposition of the a finite, but arbitrary number of copies of the generators may be expressed as linear combinations in the normal ordered terms $V(m,n,p)$. To this end, notice that
\begin{equation}\label{eq:Icomp}
I*(I^{\uplus \: n})=I^{\uplus\: (n+1)}+n I^{\uplus\: n}\,.
\end{equation}
On the other hand, we have that
\begin{equation}
I*I=I^{\uplus \:2}+I\quad \Leftrightarrow\quad I^{\uplus \:2}=I^{*\:2}-I\,,
\end{equation}
whence it is straightforward to prove by induction that
\begin{equation}\label{eq:VAaux}
I^{\uplus\:n}=I*(I-1)*(I-2)*\dotsc*(I-(n-1))\,.
\end{equation}
One may conveniently store the coefficients of $I^{*\:m}$ in the expression for $I^{\uplus\:n}$ according to the above formula by means of the \emph{exponential generating function}
\begin{equation}
EGF_I(\lambda;z):=1+\sum_{n=1}^{\infty}\frac{\lambda^n}{n!}\prod_{k=0}^{n-1}(z-k)
=(1+\lambda)^z\,,
\end{equation}
whence
\begin{equation}
I^{\uplus\:n}=\left[\frac{d^n}{d\lambda^n}(1+\lambda)^z\right]\bigg\vert_{\lambda\to 0}\bigg\vert_{z\to I}\,.
\end{equation}
Because furthermore
\begin{equation}
\left(a^{\dag}\right)^{*\:n}=a^{\dag^{\uplus\:n}}\,,\quad a^{*\:n}=a^{\uplus \:n}\,,
\end{equation}
the explicit formula for expressing superpositions of the generators in terms of their compositions reads
\begin{equation}
\left(a^{\dag}\right)^{\uplus\:m}*\left(I^{\uplus\:n}\right)*a^{\uplus\:p}=\sum_{j=0}^n \alpha_{n}{}^jV(m,j,p)\,,
\end{equation}
where
\begin{equation}
\alpha_n{}^j:=
\frac{1}{j!}\left[\frac{d^j}{dz^j}
	\left[\left(\frac{d^n}{d\lambda^n}(1+\lambda)^z
	\right)\bigg\vert_{\lambda\to 0}\right]\right]\bigg\vert_{z\to 0}\,.
\end{equation}
Combining these relations with~\eqref{eq:VAaux}, we thus obtain a formula for arbitrary normal ordered forms $V(m,n,p)$ in terms of polynomials (under multiplication with $*$) in the generators. In summary, we obtain the following theorem:
\begin{thm}[Structure theorem of the vertex algebra]
The \emph{vertex algebra} $\cR_{\cV}$ is \emph{presented} as
\begin{equation}
\cR_{\cV}:=\frac{\bK.[a^{\dag},I,a]}{\langle [a,a^{\dag}]-r_{\emptyset},
[a,I]-a,[I,a^{\dag}]-a^{\dag}\rangle}\,,
\end{equation}
where the denominator is the ideal generated by the commutation relations~\eqref{eq:vertexAlgebraCRs}, where the notation $\bK.[a^{\dag},I,a]$ denotes the \emph{free monoid algebra} of arbitrary words in the generators (with algebra operation $*$), and where we identify $a^{\dag^0}*I^{0}*a^{0}\equiv r_{\emptyset}$. Fixing the \emph{normal ordering} aka \emph{total order}
\begin{equation}
a^{\dag}\prec I\prec a
\end{equation}
on the set of generators, a generic element of $\cR_{\cV}$ may be expressed in terms of linear combinations of \emph{normal ordered terms}
\begin{equation}
V(m,n,p):=a^{\dag^{*\:m}}*I^{*\:n}*a^{*\:p}\,.
\end{equation}
\end{thm}

Evidently, we recognize upon specialization to the subalgebra of the vertex algebra spanned by $a$ and $a^{\dag}$ besides the commutation relation of the Heisenberg-Weyl algebra presented in the previous section also the standard normal ordering prescription of the HW algebra. Moreover, note that the vertex algebra $\cR_{\cV}$ has precisely the commutation relations of the so-called \emph{Boson algebra} (cf.\ e.g.~\cite{tsohantjis1997boson}), which in the physics literature is typically presented in the form
\begin{equation}
[a,a^{\dag}]=I\,,\quad [N,a]=-a\,,\quad [N,a^{\dag}]=a^{\dag}\,.
\end{equation}
In other words, in the vertex algebra the role of what is typically called the number operator $N:=a^{\dag}a$ is played by the rule algebra element $I$. Interestingly,  upon defining the canonical representation of the rule algebras $\cR_T$ on the space of graph isomorphism classes~\cite{bdg2016,bd2016}, $N$ and $I$ in fact have precisely the same action when acting on discrete graphs only, a feature called \emph{$0$-polymorphicity}. More details will be provided in~\cite{bd2016}.

\subsection{The loop subalgebra $\cR^{\cL}$}
\label{sec:loopAlgebra}
In this subsection, we will consider the algebra generated by elements of the form
\begin{equation}\label{eq:LAgenerators}
\lambda\substack{M\\P\\N}:=\delta\left(
\vcenter{\hbox{\includegraphics[width=0.3\textwidth]{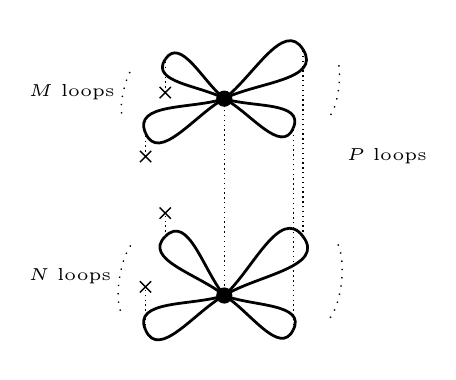}}}\right)\,.
\end{equation}
In other words, the so-called \emph{loop algebra} $\cR_{\cL}$ is the algebra whose atomic elements are vertex preservation rules $I$ with $M$ loop creations, $N$ loop deletions and $P$ loop preservations attached. It is evident from the composition behavior of the vertex preservation rule algebra element $I$ as presented in~\eqref{eq:Icomp} that any element of $\cR_{\cL}$ will be a superposition of the atomic elements of $\cR_{\cL}$, that $\cR_{\cL}$ is closed and whence indeed a subalgebra of the rule algebras $\cR_T$, and that yet again the compositions do not depend on the type $T$ of the respective underlying loop algebra.\\

To study of the loop algebra more closely, let us formulate its tentative generators as $x:=\delta(\frak{x})$ ($x\in\{\ell^{\dag},L,\ell\}$), with
\begin{equation}
\frak{l}^{\dag}:=\includetikz{lC}\,,\quad	
\frak{L}:= \includetikz{lI}\,,\quad
\frak{l}:= \includetikz{lA}\,.
\end{equation}
In terms of the previously introduced notations, we thus have that
\begin{equation}
\ell^{\dag}\equiv \lambda{\substack{1\\0\\0}}\,,\quad L\equiv \lambda\substack{0\\1\\0}\,,\quad
\ell\equiv \lambda\substack{0\\0\\1}\,.
\end{equation}
We thus obtain the ``multiplication table'' for these elements as
\begin{equation}\label{eq:loopAlgMT}
\begin{array}{c|c|c|c}
\ntC & \ell & L & \ell^{\dag}\\
\hline
\ell^{\dag} & 
{\color{blue}\lambda\substack{1\\0\\1}} & 
{\color{blue}\lambda\substack{1\\1\\0}} &
{\color{blue}\lambda\substack{2\\0\\0}}\\
\hline
L & 
{\color{blue}\lambda\substack{0\\1\\1}} &
\left(L+{\color{blue}\lambda\substack{0\\2\\0}}\right) & 
\left(\ell^{\dag}+{\color{blue}\lambda\substack{1\\1\\0}}\right)\\
\hline
\ell & 
{\color{blue}\lambda\substack{0\\0\\2}} & 
\left(\ell+{\color{blue}\lambda\substack{0\\1\\1}}\right) &
\left(I + {\color{blue}\lambda\substack{1\\0\\1}}\right)
\end{array}\,.
\end{equation}
These results entail that the commutation relations close on the set of the three tentative generators $\ell^{\dag}$, $L$ and $\ell$, expanded by the vertex preservation rule $I$ of the vertex algebra, which in turn commutes with all the tentative generators,
\begin{equation}\label{eq:LAelemCRs}
\begin{aligned}
[\ell,\ell^{\dag}]&=I\\
[L,\ell^{\dag}]&=\ell^{\dag}\,,\quad [\ell,L]=\ell\\
[\ell^{\dag},I]&=[L,I]=[\ell,I]=0\,.
\end{aligned}
\end{equation}
It almost appears as if the algebra generated by $\ell^{\dag}$, $L$ and $\ell$ is a realization of a Boson algebra (with unit element $I$). However, as is evident from~\eqref{eq:loopAlgMT}, the loop algebra $\cR_{\cV}$ is much larger than a Boson algebra, because it contains even for just binary compositions of the generators the higher loop terms marked in blue in~\eqref{eq:loopAlgMT}. One possibility to present these terms would be in terms of composites of the tentative generators and their superpositions, as in
\begin{equation}
\begin{aligned}
\lambda\substack{2\\0\\0}&=\ell^{\dag}*\ell^{\dag}-\ell^{\dag}\uplus \ell^{\dag}\\
\lambda\substack{1\\1\\0}&=\ell^{\dag}*L-\ell^{\dag}\uplus L\\
\lambda\substack{1\\0\\1}&=\ell^{\dag}*\ell-\ell^{\dag}\uplus \ell\\
\lambda\substack{0\\1\\1}&=L*\ell - L\uplus \ell\\
\lambda\substack{0\\0\\2}&= \ell*\ell-\ell\uplus \ell\,.
\end{aligned}
\end{equation}
On the other hand, the special property of the atomic tentative generators of $\cR_{\cL}$ is that they are built on the ``backbone'' of a single vertex preservation rule, whence they to a certain extent behave under composition like discrete graph rewriting rules (up to modifications of the loop structure of course). More to the point, we have the rather special properties (for $x\in \{\ell^{\dag},L,\ell,I\}$)
\begin{equation}\label{eq:LAnestedComm}
\begin{aligned}
[x_{i_1},x_{j_1}\uplus \dotsc \uplus x_{j_n}]&=\sum_{k=1}^n [x_{i_1},x_{j_k}] \uplus \biguplus_{\stackrel{p=1}{p\neq k}}^n x_{j_p}\\
[x_{i_1}\uplus x_{i_2},x_{j_1}\uplus \dotsc \uplus x_{j_n}]&=
\sum_{\stackrel{p,q=1}{p\neq q}}^2 \sum_{r=1}^n
[x_{i_p},x_{j_r}]\uplus x_{i_q}\uplus\biguplus_{\stackrel{s=1}{s\neq r}}^n x_{j_s}\\
&\quad +\sum_{\stackrel{p,q=1}{p\neq q}}^2 \sum_{\stackrel{r,s=1}{r\neq s}}^n
[x_{i_p},x_{j_r}]\uplus [x_{i_q},x_{j_s}]\uplus \biguplus_{\stackrel{t=1}{t\neq r,t\neq s}}^n x_{j_t}\,,
\end{aligned}
\end{equation}
with the obvious generalization to higher numbers of superposition factors for the left entry of the commutator. We thus arrive at the slightly peculiar notion of an algebra presented in terms of a set of atomic generators, their arbitrary, but finite superpositions, and in terms of the commutation relations of the generators plus their ``expansions'' as presented in~\eqref{eq:LAnestedComm}:
\begin{thm}[Structure theorem of the loop algebra]
The \emph{loop algebra} $\cR_{\cL}$ is generated by the four atomic rule algebra elements 
\begin{equation*}
\ell^{\dag}\,, \; L\,,\; I\,,\; \ell\,,
\end{equation*}
as well as by arbitrary finite superpositions of these four generators, denoted for brevity by
\begin{equation}
\cL(M,N,P,Q):=\left(\ell^{\dag}\right)^{\uplus\: M}\uplus L^{\uplus\:N}\uplus I^{\uplus\:P}\uplus \ell^{\uplus \: Q}\,,
\end{equation}
with elementary relations~\eqref{eq:LAelemCRs}. These relations are augmented by the nested commutation relations as presented in~\eqref{eq:LAnestedComm}.
\end{thm}
For example, one of these nested commutation relations expressed in the basis elements $\cL(M,N,P,Q)$ reads
\begin{equation}
\begin{aligned}
[x,\cL(M,N,P,Q)]&=\delta_{x,\ell^{\dag}}\left[
-N\cL(M+1,N-1,P,Q)-Q\cL(M,N,P+1,Q-1)
\right]\\
&\quad +\delta_{x,L}
(M-Q)\cL(M,N,P,Q)\\
&\quad \delta_{x,\ell}\left[
M\cL(M-1,N,P+1,Q)+N\cL(M,N-1,P,Q+1)
\right]\,.
\end{aligned}
\end{equation}
One could thus envision to find some sort of generalized normal ordering prescription by defining a total order on the generators $\cL(M,N,P,Q)$. On the other hand, from the preceding chapters it might be alternatively envisioned to define as a basis for this algebra the generic elements
\[
\lambda\substack{M\\N\\P}\,,
\]
employing the various computational relations presented earlier.\\

It is quite remarkable that even for the relatively simple special case of the loop subalgebra $\cR^{\cL}$ we already encounter a very complicated structure, and it will be the topic of future work to determine in applications whether the relations computable via the rule algebraic framework are nonetheless useful to determine the combinatorial properties of such applications. It would depend on the particular applications at hand for the loop algebra to determine such a practical normal form, but there does not appear to exist a natural candidate\footnote{By a ``natural candidate'' for a normal ordered form, we mean a form which has relations just based on the commutation relations of a tractable set of generators, which are sufficient to describe the entire algebra in terms of the free monoid over normal ordered terms modulo the commutation relations.} on generic grounds, in clear contrast to the vertex algebra, where a normal form existed for which all composites are in fact superpositions (up to applying commutation relations).

\subsubsection{Coupling between the vertex and the loop algebra}

A straightforward computation reveals that it is possible to obtain larger subalgebras of $\cR_T$ via performing the union of the generator sets of $\cR^{\cV}$ and $\cR^{\cL}$, which results in the ``multiplication table'' of atomic generators as presented below:
\begin{equation}\label{eq:RVLmt}
\begin{array}{c|c|c|c|c|c|c}
\ntC & a & a^{\dag} & I &\ell & L & \ell^{\dag}\\
\hline
a^{\dag} & 0 & 0 & 0 & 0 & 0 & 0\\
\hline
a & 0 & r_{\emptyset} & a &{\color{blue} (a\ell) }& \delta_{A\in T}\;{\color{blue}(a\ell)} & \delta_{A\in T} \;a\\
\hline
I & 0 & a^{\dag} & I & \ell & L & \ell^{\dag}\\
\hline
\ell^{\dag} & 0 &{\color{blue}(a^{\dag}\ell^{\dag}) }& \ell^{\dag} &
{\color{blue}\lambda\substack{1\\0\\1}} & 
{\color{blue}\lambda\substack{1\\1\\0}} &
{\color{blue}\lambda\substack{2\\0\\0}}\\
\hline
L & 0 & \delta_{B\in T}\; {\color{blue}(a^{\dag}\ell^{\dag}) }& L &
{\color{blue}\lambda\substack{0\\1\\1}} &
\left(L+{\color{blue}\lambda\substack{0\\2\\0}}\right) & 
\left(\ell^{\dag}+{\color{blue}\lambda\substack{1\\1\\0}}\right)\\
\hline
\ell & 0 & \delta_{B\in T}\;a^{\dag} & \ell &
{\color{blue}\lambda\substack{0\\0\\2}} & 
\left(\ell+{\color{blue}\lambda\substack{0\\1\\1}}\right) &
\left(I + {\color{blue}\lambda\substack{1\\0\\1}}\right)
\end{array}\,,
\end{equation}
with
\begin{equation}
(a\ell):=\delta\left(\includetikz{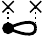}\right)\,,\quad
(a^{\dag}\ell^{\dag}):=\delta\left(\includetikz{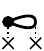}\right)\,.
\end{equation}
As the table reveals, this subalgebra is spanned by atomic elements that consist of one of the vertex algebra elements carrying additional created, annihilated or preserved loops, and of superpositions thereof. Interestingly, this is one of the simplest conceivable subalgebras of $\cR_T$ for which the type plays a role, indicated by 
\[
\delta_{A\in T}\equiv \delta_{T,SPO_A}+\delta_{A,SPO_{AB}}\,,\; \delta_{B\in T}\equiv \delta_{B,SPO_B}+\delta_{B,SPO_{AB}}
\]
in~\eqref{eq:RVLmt}.\\

It would be interesting to determine whether the subalgebras $\cR^{\cV\cL}_T\subsetneq \cR_T$ might have some applications in the realm of statistical physics and chemical reaction systems.

\subsubsection{The edge subalgebra $\cR^{\cE}$}

Consider the subalgebra of $\cR_T$ generated by the elements $x:=\delta(\frak{x})$ ($x\in \{e_{>},E,e^{\dag}_{>},I\}$), with
\begin{equation}
\frak{e}_{>}^{\dag} := \includetikz{eC} \,,\quad 	
\frak{E}_{>}:= \includetikz{eI} \,,\quad 
\frak{e}_{>}:= \includetikz{eA}\,.
\end{equation}
We obtain yet again remarkably simple commutation relations for these tentative atomic generators,
\begin{equation}
\begin{aligned}
[e_{>},e_{>}^{\dag}]&=I^{\uplus\:2}\\
[e_{>},E_{>}]&=e_{>}\,,\quad [E_{>},e^{\dag}_{>}]=e^{\dag}_{>}\\
[e_{>},I]&=[E_{>},I]=[e_{>}^{\dag},I]=0\,,
\end{aligned}
\end{equation}
which is quite similar to the commutation relations of the atomic loop algebra generators. However, computing for example the nontrivial contributions to the composition of the tentative atomic generators $e_{>}$ and $E_{>}$,
\begin{equation}
\begin{aligned}
e_{>}\ntC E_{>}&=
\includetikz{eA}
+\includetikz{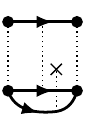}
+\includetikz{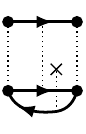}\\
&\quad+
\includetikz{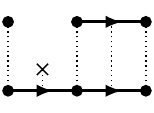}+
\includetikz{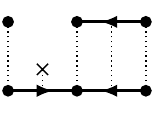}+
\includetikz{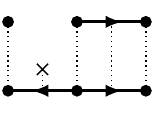}+
\includetikz{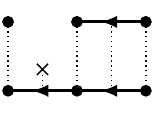}\,,
\end{aligned}
\end{equation}
it becomes immediately evident that the edge algebra is a subalgebra of $\cR_T$ that is generated by many more generators than the simplest atomic ones suggested via the set $\{e_{>},E_{>},e^{\dag}_{>}\}$. In particular, the appearance in arbitrary size and complexity connected irreducible rule diagrams in the higher composites of the tentative generators entail that it is hopeless to find a compact presentation of this algebra in terms of a small set of generators and relations. As in the traditional descriptions of graph rewriting, this is a natural feature of graph rewriting systems, and it will depend in the particular applications at hand whether there are other ways possible to deal with this complexity or not.\\

As a caveat to the general philosophy of combinatorial Hopf algebras, one should thus conclude from this section as presented thus far that while it is certainly feasible for the vertex and loop subalgebras to employ ideas from ordinary combinatorics in describing graph rewriting systems based in these subalgebras, as possible for example via the techniques presented in~\cite{bdg2016}, more in detail work will be necessary to make ``arable'' combinatorial techniques for generic graph rewriting systems.\\

In summary, these facts lie at the heart of the complexity of the study of graph rewriting systems. We will therefore in the following consider viable alternatives to the naive guess for a normal form for specific special and suitably restricted types of ``structural'' subalgebras of $\cR_T$.

\subsection{Structural subalgebras of the rule algebras}

While all rule algebra elements are by definition of the rule algebras irreducible rule diagrams, which are by virtue of the discussion in the previous subsection (linear combinations of) superpositions of atomic basis rule diagrams, the atomic rule diagrams do not form a closed subalgebra for any type of rule algebra, whence thus far we could only speak of one ``structural'' subalgebra of the algebras $\cR_T$, namely trivially $Irr(\cD)$ itself, or of course the generator subalgebras $\cR_{\cV}$, $\cR_{\cL}$ and $\cR_{\cE}$. Here, by ``structural'' subalgebra we mean a subalgebra which may be specified via giving a restriction on the structure of its elements. We will demonstrate in the following that there exist a number of interesting ``structural'' subalgebras that will play an important role in our development of the rule algebraic framework.

\paragraph{Notational convention:} For brevity, in the remainder of the section we will employ the convention that we drop the source and target maps from the specifications of irreducible rule diagrams, as well as the explicit mention of the (for irreducible rule diagrams always empty) match morphism $m$,
\begin{equation}
[I,O,r]\equiv (I,O,r,\emptyset,s_{I\uplus O},t_{I\uplus O})\,.
\end{equation}
The use of this convention should not lead to any confusion, since it is quite apparent by checking the typing (three arguments versus six arguments for the full data convention, square instead of round brackets).

\subsubsection{The structural subalgebras $\cR^{+}$ and $\cR^{-}$}

\begin{defn}
Let for each graph isomorphism class $G\in \cG$ the rule algebra elements $\hat{G}^{\dag}$ and $\hat{G}$ be defined as 
\begin{equation}
\begin{aligned}\label{eq:GGdag}
	\hat{G}^{\dag}:= \delta([\emptyset,G,\emptyset])\in\cR_T\\
	\hat{G}:= \delta([G,\emptyset,\emptyset])\in \cR_T\,.
\end{aligned}
\end{equation}
where the notation is motivated by the fact that
\begin{equation}
	\left(\hat{G}\right)^{\dag}=\hat{G}^{\dag}\,.
\end{equation}
Then the structural subalgebra $\cR^{+}\subsetneq \cR_T$ is defined as the algebra spanned by the elements $\hat{G}^{\dag}$ (with one such element per isomorphism class $G\in\cG$), while the subalgebra $\cR^{-}\subsetneq \cR_T$ is defined as the algebra generated by the elements $\hat{G}$ (with one such element per isomorphism class $G\in\cG$).
\end{defn}
It is straightforward to prove the following proposition:
\begin{prop}\label{prop:structureRPM}
$(i)$ The algebras $\cR^{+}$ and $\cR^{-}$ are \emph{commutative subalgebras} of the rule algebras $\cR_T$ and are independent of the type $T$.
$(ii)$ Let us introduce the convenient \emph{multiindex notations} ($\alpha_G\in \bZ_{\geq 0}$)
\begin{equation}
	\vec{\hat{G}}^{\vec{\alpha}}:=\prod_{G\in\cG}\hat{G}^{\alpha_G}\,,\quad
	\vec{\hat{G}}^{\dag\:\vec{\alpha}}:=\prod_{G\in\cG}\hat{G}^{\dag\:\alpha_G}
\end{equation}
for finite products of elements of $\cR^{+}$ and $\cR^{-}$, respectively (which thus imposes the constraint $\sum_{G\in\cG}\alpha_G<\infty$), where $\cG_c\subsetneq \cG$ is the set of \emph{isomorphism classes of connected finite graphs}. Then the algebras $\cR^{+}$ and $\cR^{-}$ are presentable as \emph{polynomial algebras},
\begin{equation}
\cR^{+}\equiv \bK[\{\hat{G}^{\dag}\mid G\in\cG_c\}]\,,\quad \cR^{-}\equiv \bK[\{\hat{G}\mid G\in\cG_C\}]\,. 
\end{equation}
This entails in particular that any element of $\cR^{+}$ and of $\cR^{-}$, respectively, is of the form
\begin{equation}
	\cR^{+}\ni r=\sum_{\vec{\alpha}}f_{\vec{\alpha}}\vec{\hat{G}}^{\dag\:\vec{\alpha}}\,,\quad \cR^{-}\ni r=\sum_{\vec{\alpha}}g_{\vec{\alpha}}\vec{\hat{G}}^{\vec{\alpha}}\,,
\end{equation}
with $f_{\vec{\alpha}},g_{\vec{\alpha}}\in \bK$, and with the constraints
\begin{equation}
	\sum_{\vec{\alpha}}f_{\vec{\alpha}}<\infty\,,
\quad
\sum_{\vec{\alpha}}g_{\vec{\alpha}}<\infty\,.
\end{equation}
\begin{proof}
	\underline{Ad $(i)$:} It is straightforward to verify that for all $G_{\alpha},G_{\beta}\in \cG$ 
	\begin{equation}
		\begin{aligned}
			\hat{G}_{\alpha}*_{T}\hat{G}_{\beta}&=\hat{G}_{\alpha}\uplus\hat{G}_{\beta}
			\equiv \widehat{G_{\alpha}\uplus G_{\beta}}\\
			\hat{G}_{\alpha}^{\dag}*_{T}\hat{G}_{\beta}^{\dag}&=\hat{G}_{\alpha}^{\dag}\uplus\hat{G}_{\beta}^{\dag}
			\equiv \widehat{G_{\alpha}\uplus G_{\beta}}^{\dag}\,,
		\end{aligned}
	\end{equation}
	thereby proving closedness, type-independence and (via the commutativity of the superposition operation) the commutativity of the subalgebras $\cR^{+}$ and $\cR^{-}$.\\
	
	\underline{Ad $(ii)$:} Each isomorphism class $G\in \cG$ describes \emph{finite} graphs that are not necessarily connected, whence a generic isomorphism class $G\in\cG$ may be presented as
	\begin{equation}
		G=G_{i_1}^{\uplus n_{i_1}}\uplus \dotsc G_{i_G}^{\uplus n_{i_G}}\,,
	\end{equation}
	with $G_{i_j}\in \cG_c$ a finite set of isomorphism classes of \emph{connected graphs}, and with finite multiplicities $n_{i_j}\in \bZ_{\geq 0}$. But the result presented for the proof of claim $(i)$ entails immediately that
	\begin{equation}
		\begin{aligned}
			\hat{G}&=\hat{G}_{i_1}^{*_T\: n_{i_1}}*_T \dotsc \hat{G}_{i_G}^{*_T\: n_{i_G}}\\
			\hat{G}^{\dag}&=\left(\hat{G}_{i_1}^{\dag}\right)^{*_T\: n_{i_1}}*_T \dotsc \left(\hat{G}_{i_G}^{\dag}\right)^{*_T\: n_{i_G}}\,,
		\end{aligned}
	\end{equation}
	from which the claim follows due to multilinearity of the product operation.
\end{proof}
\end{prop}

It is at first sight somewhat surprising to find two subalgebras $\cR^{+}$ and $\cR^{-}$ that are freely presented after all the complications that we encountered with the generator subalgebras. However, upon closer inspection they are merely the analogues of the (rather trivial) subalgebras of the multi-type Heisenberg-Weyl algebra spanned by just the creation or just the annihilation operators, respectively. In the next section, we will study the effect of combining $\cR^{+}$ and $\cR^{-}$ into a joint subalgebra.

\subsubsection{The structural subalgebras $\cR^{\pm}_T$}\label{sec:RAhw}

\begin{defn}\label{def:Rpm}
The subalgebras $\cR^{\pm}_T\subsetneq \cR_T$ are defined as the algebras generated by the rule algebra elements
\[
\{ \hat{G}_{\alpha}\mid G_{\alpha}\in \cG_c\}\cup \{ \hat{G}^{\dag}_{\beta}\mid G_{\beta}^{\dag}\in \cG_c\}\,.
\]
\end{defn}
We may make the particular structure of the subalgebras $\cR^{\pm}_T$ somewhat more explicit via the following Theorem:
\begin{thm}\label{thm:RPMstruct}
Let $\cR^{\pm}_T$ be defined as above. $(i)$ For all isomorphism classes of connected graphs $G_{\alpha},G_{\beta}\in\cG_c$, we have that
\begin{equation}
\begin{aligned}
		\hat{G}_{\alpha}^{\dag}*_{T}\hat{G}_{\beta}^{\dag}&=\hat{G}_{\alpha}^{\dag}\uplus\hat{G}_{\beta}^{\dag}\\
		\hat{G}_{\alpha}*_{T}\hat{G}_{\beta}&=\hat{G}_{\alpha}\uplus\hat{G}_{\beta}\\
		\hat{G}_{\alpha}^{\dag}*_{T}\hat{G}_{\beta}&=\hat{G}_{\alpha}^{\dag}\uplus\hat{G}_{\beta}\\
		\hat{G}_{\alpha}*_{T}\hat{G}_{\beta}^{\dag}&=
		\hat{G}_{\beta}^{\dag}\uplus\hat{G}_{\alpha}
		+\hat{G}_{\alpha}\ntC_{T}\hat{G}_{\beta}^{\dag}\,,
\end{aligned}
\end{equation}	
where as usual (for all $r_1,r_2\in\cR_T$)
\begin{equation}
	r_1\ntC_{T}r_2:=r_1*_T r_2-r_1\uplus r_2
\end{equation}
are the \emph{contributions of compositions along nontrivial matches}.\\

$(ii)$ Let $G,H\in\cG_c$ be isomorphism classes of connected graphs. Then for any nontrivial compositions of $\hat{G}$ with $\hat{H}^{\dag}$ the following \emph{connected commutation relations} hold:
\begin{equation}\label{eq:RPMcomm}
	\begin{aligned}
		\hat{G}\ntC_{DPO}\hat{H}^{\dag}&=\delta_{G\cong H}|Aut(G)|r_{\emptyset}\\
		\hat{G}\ntC_{SPO_A}\hat{H}^{\dag}&=\delta_{G\subseteq H}\sum_m f^{SPO_A}_{GH}{}^m\widehat{H_m}^{\dag}\,,\quad H_m:=H\setminus dom(m)\\
		\hat{G}\ntC_{SPO_{B}}\hat{H}^{\dag}&=\delta_{H\subseteq G}
		\sum_m f^{SPO_B}_{GH}{}^m\widehat{G_m}\,,\quad G_m:=G\setminus im(m)\\
		\hat{G}\ntC_{SPO_{AB}}\hat{H}&=\sum_m f^{SPO_{AB}}_{GH}{}^m\widehat{H_m}^{\dag}\uplus\widehat{G_m}\,,
	\end{aligned}
\end{equation}	
where $|Aut(G)|$ is the size of the automorphism group of the graph $G$, and where the notations $\delta_{G_1\cong G_2}$ and $\delta_{G_1\subseteq G_2}$ entails that $G_1$ must be isomorphic to $G_2$, or a proper subgraph of $G_2$, respectively (or else the contributions are discarded). The constants $f^{T}_{GH}{}^m\in \bZ_{>0}$ are called \emph{structure constants}. The sums range over the nontrivially contributing matches $m$. We furthermore employ the convention that for $G\subseteq H$ the notation $H\setminus G$ stands for the graph that results by deleting from $H$ all vertices of $G$ and any edges that had one or both endpoint vertices in $G$. Note that in the above formulae, depending on the specific rule algebra elements at hand it may be the case that some of the $G_m$ or $H_m$ are not connected, whence the right hand sides of~\eqref{eq:RPMcomm} may not be in the ``normal form'' in terms of formal polynomials in atomic elements of $\cR^{+}$ and $\cR^{-}$. In those cases, one may of course easily invoke the results of Proposition~\ref{prop:structureRPM} to achieve the ``normal form'' if desired.\\

$(iii)$ The algebras $\cR^{\pm}_T$ are \emph{non-commutative subalgebras} of $\cR_T$ for each $T$.
\begin{proof}
\underline{Ad $(i)$:} The proof follows from a straightforward application of the definitions of the elements of $\cR^{+}$ and of $\cR^{-}$.\\
\underline{Ad $(ii)$:}
The specific structure of the nontrivial compositions is \emph{type-dependent}:
\begin{itemize}
\item \underline{Case $T=DPO$:} In the $DPO$-type rule algebra, whenever elements $\hat{G}$ are composed with elements $\hat{H}^{\dag}$ for which $\isoc{G}\neq \isoc{H}$, this entails that the graph $G$ may not be completely matched to $H$. But because we assumed that $G$ and $H$ are connected graphs, this entails that compositions along any nontrivial matches will have ``dangling edges'', whence the claim follows from the fact that $\cF_{DPO}\circ \partial$ discards all such compositions (aka maps them to $0\cdot r_{\emptyset}$. For the case $\isoc{G}= \isoc{H}$, only matches that entirely match $G$ to $H$ will not lead to ``dangling edges'' upon application of $\partial$. But such matches must by definition of matches then be injective partial graph homomorphisms from $G$ to $H$. Since by assumption $\isoc{G}= \isoc{H}$, $G$ and $H$ are both representatives of the same graph isomorphism class, whence the set of matches that lead to nontrivial contributions for the composition in this case coincides with the set of automorphisms of $G$. Finally, upon composition along such a nontrivial match followed by reduction, we always obtain the empty rule diagram, whence the claim for the case $T=DPO$ made in~\eqref{eq:RPMcomm} follows.
\item \underline{Cases $T\neq DPO$:} By the same type of argument, the $SPO$-variants of the compositions tolerate compositions along matches that result upon application of $\partial$ in ``dangling edges''. More precisely, for $SPO_A$ rewriting, ``dangling edges'' resulting from vertex deletions are admissible, but not those resulting from vertex creations. Thus, for this case we only obtain nontrivial contributions to the composition if $G\subseteq H$. If this is the case, any match that should yield a nontrivial contribution to the composition must fully match $G$ into $H$; applying $\partial$ to a composition along such a match $m$, we obtain the irreducible rule pre-diagram $[\emptyset,H-dom(m),\emptyset]\in Irr_{pre}(\cD)$, which potentially contains ``dangling edges'' resulting from deleting those vertices and edges from $H$ that were matched to $G$. Then, application of $\cF_{SPO_A}$ removes those dangling edges, resulting in the proper irreducible rule diagram $[\emptyset,H\setminus dom(m),\emptyset]\in Irr_{pre}(\cD)$. The dual argument leads to the derivation of the $SPO_B$ algebra result, while a combination of both arguments justifies the $SPO_{AB}$ result
\end{itemize}

\underline{Ad $(iii)$:} The claim follows directly from part $(i)$ and $(ii)$ of the theorem.
\end{proof}
\end{thm}

\subsubsection{The subalgebras $\cO$ and $\cO^{\pm}_{DPO}$}

Another interesting type of structural subalgebras may be formulated via considering rule algebra elements with equal left and right hand side graphs:
\begin{defn}
The structural subalgebras $\cO_T\subsetneq \cR_T$ and $\cO^{\pm}_T\subsetneq \cR_T$ are defined (tentatively) to be generated by elements of the form
\begin{equation}
	\begin{aligned}
		\cO_T&\ni \tilde{O}\equiv \delta([O,O,id_O])\\
		\cO^{\pm}_T&\ni \breve{O}\equiv \delta([O,O,\emptyset])\,.
	\end{aligned}
\end{equation}	
\end{defn}

With these definitions, we obtain the following important theorem:
\begin{thm}
$(i)$ We have the following contributions due to compositions along nontrivial matches of connected graphs $O_1$ and $O_2$:
\begin{equation}
	\begin{aligned}
		\tilde{O}_1\ntC_{T}\tilde{O}_2&=\sum_m \delta([O_{1_m2},O_{1_m2},\emptyset])\\
		\\
		\breve{O}_1\ntC_{DPO}\breve{O}_2&:=\delta_{O_1\cong O_2}|Aut(O_1)|\breve{O}_1\\
		\breve{O}_1\ntC_{SPO_A}\breve{O}_2&:=
		\delta_{O_1\subseteq O_2}\sum_m
		\delta([O_2,O_1\uplus(O_2\setminus dom(m)),\emptyset])\\
		\breve{O}_1\ntC_{SPO_B}\breve{O}_2&:=
		\delta_{O_2\subseteq O_1}\sum_m
		\delta([(O_1\setminus im(m))\uplus O_2,O_1,\emptyset])\\
		\breve{O}_1\ntC_{SPO_{AB}}\breve{O}_2&:=
		\sum_m
		\delta([(O_1\setminus im(m))\uplus O_2,O_1\uplus(O_2\setminus dom(m)),\emptyset])\,,
		\end{aligned}
\end{equation}
where the sums run over the nontrivial matches $m$, and where $O_{1_m2}$ is the graph that results from overlapping $O_1$ with $O_2$ along the common subgraph mapped by $m$. Consequently, $\cO_T$ is a proper subalgebra of $\cR_T$ for all types $T$, and so is $\cO^{\pm}_{DPO}$, but for $T\neq DPO$ $\cO^{\pm}_T$ is \emph{not} a proper subalgebra of $\cR_T$ (since it is evidently not closed).

$(ii)$ $\cO_T$ and $\cO^{\pm}_{DPO}$ are \emph{commutative subalgebras of $\cR_T$}.
\begin{proof}
	\underline{Ad $(i)$:} The argument is entirely analogous to the proof of Theorem~\ref{thm:RPMstruct}, whence we omit it here for brevity.\\
	
	\underline{Ad $(ii)$:} Trivially follows from the explicit formulae for the nontrivial contributions (and from commutativity of the superposition operation).
\end{proof}	
\end{thm}

It is clear by definition of the subalgebras that any of their elements is generically a superposition of atomic rule diagrams of the required form. On the other hand, we have generically that superpositions always arise as the contribution along the trivial match of any two rule diagrams, whence one may e.g.\ realize the superposition of two atomic rule diagrams by extracting the respective term from their composition. Therefore, we have the following important result:

\begin{prop}\label{prop:connectedObs}
	The subalgebras $\cO^{\pm}_{DPO}$ and $\cO_T$ are \emph{freely generated by their atomic elements}, whence they may be presented as follows:
	\begin{equation}
		\begin{aligned}
		\cO_T\equiv\bK[\{\delta([O,O,id_O])\in Irr(\cD)\mid O\in\cG_c\}]\\
		\cO^{\pm}_{DPO}\equiv \bK[\{\delta([O,O,\emptyset])\}\in Irr(\cD)\mid O\in \cG_c\}]\,,
		\end{aligned}
	\end{equation}
	where $\cG_c\subsetneq\cG$ again denotes the set of isomorphism classes of \emph{connected} graphs, and where notation $\bK[\dotsc]$ denotes the \emph{algebra of polynomials} in the generators with coefficients in $\bK$ and algebra operation the composition $*_T$.
	\begin{proof}
		This follows directly from
		\begin{equation}
			r_1*_{\cR_T}r_2=r_1\uplus r_2+r_1\ntC_{\cR_T} r_2\,,
		\end{equation}
		whence e.g.\ the superposition $r_1\uplus r_2$ of two atomic rule diagrams $r_1,r_2\in Atom(\cD)$ may be realized as
		\begin{equation}\label{eq:AnsatzConn}
			r_1\uplus r_2=r_1*_{T}r_2-r_1\ntC_{T} r_2\,.
		\end{equation}
		It may then happen in general that $r_1\ntC_{T} r_2$ itself is a superposition of atomic rule diagrams. However, by the definition of $\ntC_{T}$, it is always guaranteed that the potential superposition factors in $r_1\ntC_{T} r_2$ are \emph{smaller} than the original atomic rule diagrams $r_1$ and $r_2$, whence it is guaranteed that the process of replacing superpositions recursively via the Ansatz~\eqref{eq:AnsatzConn} will eventually terminate, whence the claim follows.
	\end{proof}
\end{prop}

\subsection{The structural subalgebra $\cO^{\GRule{}{}}_{DPO}$}

\begin{defn}
	The structural subalgebra $\cO^{\GRule{}{}}_{DPO}\subsetneq \cR_{DPO}$ is the subalgebra generated by the basis rule algebra elements of the form
	\begin{equation}
	\begin{aligned}
		\omega_{O,O'}^r&:=\delta([O,O',r,\emptyset])\in\cR_{DPO}\\
		|V_O|&=|V_{O'}|\,,\; |E_O|:=|E_{O'}|\,,
	\end{aligned}
	\end{equation}
	where $O$ and $O'$ thus are graphs of the same number of vertices and edges, but which are \emph{not} necessarily isomorphic.
	\end{defn}

Remarkably, the basis elements of a certain subalgebra of $\cO^{\GRule{}{}}_{DPO}$, namely those elements for which 
\[
O\cong O' \land r\subseteq id_O\,,
\]
form a commutative subalgebra of $\cO^{\GRule{}{}}_{DPO}$ and play the role of \emph{generalized graph observables} in the framework of the \emph{stochastic mechanics of $DPO$ graph rewriting} -- we refer the reader to~\cite{bdg2016} for the precise details, and content ourselves here with formulating the following proposition:
\begin{prop}
	$(i)$ The algebra $\cO^{\GRule{}{}}_{DPO}$ is \emph{closed} (and thus indeed a subalgebra of $\cR_{DPO}$).
	$(ii)$ The algebra of elements of the form
	\begin{equation}\label{eq:DPOobs}
		\fD_{obs}:=\left\{\omega_{O,O'}^r\in \cO^{\GRule{}{}}_{DPO}\mid O\cong O'\land r\subseteq id_O\right\}
	\end{equation}
	is a commutative subalgebra of $\cO^{\GRule{}{}}_{DPO}$.
	\begin{proof}
	\underline{Ad $(i)$:} Since in the $DPO$-type rewriting the deletion of a given vertex is only possible if also all incident edges are explicitly deleted, it is immediately clear that composition of two elements of  basis diagrams $\cO^{\GRule{}{}}_{DPO}$ along nontrivial matches lead upon reduction where any vertices or edges effectively deleted from the output interface of one of the diagrams had to be deleted by matching them to vertices or edges of the input interface of the other diagram, whence in total we have not changed the property that the resulting reduced diagram has the same number of vertices in input and output interface, and also the same number of edges. In other words, any nontrivial composite is again an element of  $\cO^{\GRule{}{}}_{DPO}$, which by multilinear extension proves the claim.\\
	
	\underline{Ad $(ii)$:} Since the rule diagrams of the form~\eqref{eq:DPOobs}  have by definition the special property that their partial map $r$ is a subset (when described as a relation) of the identity morphism $id_O$ of the input interface, and since moreover $O\cong O'$ for these diagrams, the diagrams are ``vertically symmetric'', i.e.\ invariant under dualization. Therefore, for any two basis diagrams $\fd_1,\fd_2\in \fD_{obs}$, there exists an isomorphism from the set of matches $\cM_{\fd_1}(\fd_2)$ to the set of matches $\cM_{\fd_2}(\fd_1)$ by virtue of the symmetry, thus proving the commutativity property. To prove that the algebra generated from $\fD_{obs}$ is indeed closed, it suffices to observe that due again to the ``vertical symmetry'', for any given composite $\comp{\fd_1}{m_{12}}{\fd_2}$ we have that
	\[
		r_{\comp{\fd_1}{m_{12}}{\fd_2}}\subseteq id_{\cI(\fd_1)}\cup id_{\cI(\fd_2)}
	\]
	as well as ``vertical symmetry'' of the source and target of the composite's partial map, while any vertex or edge deleted from $\cO(\fd_2)$ by matching it to vertices and edges in $\cI(\fd_1)\setminus dom(r_1)$ will be having a counterpart in the output of the reduced composite diagram. This entails that the reduced composite diagram is a member of $\fD_{obs}$, which concludes the proof.
	\end{proof}
\end{prop}

\subsection{Nonexistence of a Hopf algebra structure on $\cR_T$}

Let us conclude the discussion of the general structure of the rule algebras $\cR_T$ with the following elementary observation (the analogue of which for the case of the Heisenberg-Weyl algebras is a well-known fact, cf.\ e.g.\ \cite{blasiak2010combinatorial}):
\begin{prop}
It is not possible to endow the rule algebras $\cR_T$ with the structure of a coalgebra, and consequently in particular not with the structure of a Hopf algebra.
\begin{proof}
	It is enough to demonstrate that already for the special case of rule algebra elements that generate the Heisenberg-Weyl algebra,
	\[
	a:=\delta\left(\includetikz{vA}\right)\,,\quad 
	{a^{\dag}}:=\delta\left(\includetikz{vC}\right)\,,
	\]
	there cannot be defined a consistent counit $\varepsilon$, as one may easily conclude	via a short direct computation:
	\begin{align*}
		\varepsilon(1)&=\varepsilon(a*_{\cR_T}{a^{\dag}}-
		{a^{\dag}}*_{\cR_T}a)\\
		&=\varepsilon(a*_{\cR_T}{a^{\dag}})-
		\varepsilon({a^{\dag}}*_{\cR_T}a)\\
		&=\varepsilon(a)\varepsilon({a^{\dag}})-
		\varepsilon({a^{\dag}})\varepsilon(a)\\
		&=0\quad \neq 1\,.
	\end{align*}%
	Here, we have used that any rule algebra element is also an element of $Irr(\cD)$, which makes the application of the counit map as defined in~\eqref{eq:coUnit} possible. Since on the other hand the Heisenberg-Weyl algebra is a subalgebra of all rule algebras $\cR_T$, this means that one cannot consistently define a counit for any of these algebras, and consequently also no antipode and thus no Hopf algebra structures.
\end{proof}	
\end{prop}

\section{Conclusion and Outlook}

We believe that the well-established research field of graph rewriting will profit profoundly from the novel formulation in terms of rule diagrams as presented in this paper. While it is in principle possible to formulate the rule algebras also in the more traditional category theoretical framework~\cite{bdgs2016}, the formulation in terms of rule diagrams holds a large number of advantages: the framework is entirely intuitive, easy to implement on algorithms, leads to deep insights into the structure and combinatorics of graph rewriting, and allows to make contact with other well-established variants of descriptions of rewriting systems in the mathematical combinatorics, chemistry and statistical physics literature. We are currently finalizing a series of papers that illustrate the great utility of this new approach~\cite{bd2016,behrCP2016}, and have already presented a complete framework for graph rewriting for the case of DPO graph rewriting~\cite{bdg2016}.\\

The three main results of this paper may be summarized as follows: 
\begin{itemize}
\item We find a complete framework to formulate the algebras of graph rewriting for the four types of graph rewriting (two of which are new), based on so-called rule diagrams. The key idea of the construction is to follow the blueprint of diagrammatic combinatorial Hopf algebras to implement a notion of composition of rule diagrams, and four variants of reductions of rule diagrams to normal form in order to implement the four variants of graph rewriting algebras. 
\item Secondly, our framework allows to clearly understand the famous Heisenberg-Weyl algebra of annihilation and creation of indistinguishable particles (of countably infinite types) as a particular special case of all four types of rule algebras, thereby offering immense potential in view of applications: graph rewriting may now be seen as a direct generalization of chemical reactions, and fits perfectly well into the established statistical, combinatorial and probabilistic frameworks! 
\item Thirdly, our work features the first steps towards a \emph{structural analysis} of the rather complex rule algebras, with highlights including the notion of generators of the rule algebras, generator subalgebras and structural subalgebras, which recover the concept as graph observables from the literature as a special type of structural subalgebras, and which feature a sort of direct generalization of the HW algebra in terms of creation and annihilation of graph motives in DPO rewriting. 
\end{itemize}

We conclude this paper by identifying a selection of interesting future directions and applications.

\subsection{Extension to generic types of graphs and graph-like structures}

While we focussed for simplicity on \emph{directed multigraphs} in this paper, it is entirely straightforward to extend the rule algebraic framework in a number of interesting directions: one can handle \emph{undirected graphs} by modifying the definitions of diagrams and their matches (replacing the source and target maps in the definitions with a single endpoint map). One may introduce \emph{colors} aka \emph{types} for both vertices and edges -- the only necessary modification them consists in modifying the definition of matches by adding the extra condition that only vertices and edges of the same colors/types are matchable. It is finally easily conceivable to extend the framework to include hyperedges, surfaces or higher-dimensional structured graph-like objects. 

\subsection{General representation theory of $\cR_T$}

A crucial aspect of the framework is the construction of the analogues of the \emph{canonical representation} of the Heisenberg-Weyl algebra in order to complete the implementation of graph rewriting in the rule-algebraic approach. More precisely, as has been presented in~\cite{bdg2016} for the case of DPO rewriting and will be introduced in~\cite{bd2016} for the other three types of rewriting, defining a representation of $\cR_T$ will amount to defining a specific endomorphism $\rho_T:\cR_T\rightarrow End(\cG)$ from $\cR_T$ to the endomorphisms of a vector space $\cG$, the vector space of graph isomorphism classes. The interpretation of the representations $\rho_T$ might somewhat obscure for a traditional graph rewriting theorist, yet quite evident to mathematical combinatorists: applying the representation $\rho_T(r)$ of a rule algebra element $r\in \cR_T$ to one of the basis vectors $|G\rangle\in \cG$ will result in a linear combination of basis vectors of $\cG$, each representing one of the possible outcomes for applying the graph rewriting rule presented by $r$ onto the input graph $G$, with the prefactor of each basis vector $|G'\rangle\in \cG$ in the output storing the number of possible ways the graph $G'$ can be produced by applying the rule $r$ to the input graph $G$. In this fashion, one might view the rule-algebraic approach to graph rewriting as a sort of \emph{combinatorial approach to concurrency}. 

\subsection{Stochastic mechanics of graph rewriting}

We would like to advertise in particular to the mathematical combinatorics audience that the rule algebras $\cR_T$ offer not only a mathematical structure of interest on its own, but in particular also a very direct bridge to the statistical physics, theoretical computer science and theoretical (bio-) chemistry literature. More precisely, a realization of \emph{continuous time Markov chains}~\cite{norris} based on graph rewriting systems, so-called \emph{stochastic graph rewriting systems}, is easily implementable in the new framework in a form that should be highly transparent to practitioners from theoretical disciplines. The necessary ingredients of what we call the \emph{stochastic mechanics of graph rewriting} include besides the rule algebras $\cR_T$ also their canonical representations and a number of ingredients from the statistical physics and graph rewriting literature. We refer the interested reader to~\cite{bdg2016} for the precise details of DPO stochastic mechanics, and will present the full framework in~\cite{bd2016}. Notably, specializing to discrete graph rewriting (extended to the case of multiple colors), our construction reduces to Doi's second quantization description of \emph{chemical reaction systems}~\cite{doi1976second}, which relies solely on the Heisenberg-Weyl algebra and its canonical representation. This hints at interesting novel types of stochastic dynamical systems akin to chemical reaction systems, where for example a discrete graph rewriting system representing a chemical reaction system is applied to non-discrete input graphs, or where one formulates a sort of ``graphical reaction system'' based on the rule-algebraic analogue of the Heisenberg-Weyl algebra for DPO rewriting as presented in Section~\ref{sec:RAhw}. As already demonstrated in~\cite{bdg2016}, one of the main results of the stochastic mechanics approach will be the streamlined formulation of known results for stochastic graph rewriting systems, as well as novel mathematical results such as the Generalized Fragmentation Theorem for the time evolution of arbitrary moments of graph observables.

\subsection{Towards a generalized combinatorial physics paradigm for graph rewriting systems}

Remarkably, the results presented in Section~\ref{sec:structureTheory} indicate that a direct application of ideas from analytical combinatorics to the study of graph rewriting systems appears to be of rather limited practicality. For example, even in the case of the relatively simple loop subalgebra $\cR^{\cL}$ it is not possible to find a compact normal form for the possible elements of $\cR^{\cL}$, a prerequisite for many constructions in combinatorics. On the other hand, because of the concrete realization of stochastic graph rewriting systems in terms of stochastic mechanics, there would be a huge potential in terms of practical applications for techniques from combinatorics such as umbral calculus, coherent states and generating functions, if only they could be suitably extended. Because of the aforementioned intrinsic complexity of the full rule algebra, it appears that it will in general not be possible to derive exponential generating functions for decompositions of rule diagram subalgebras, at least not via a low number of formal variables -- each possible type of diagram has to be kept track of via its own formal variable, and it will in general not be the case that powers of a given formal variable correspond to an easy to interpret diagram in the class that the given formal variable should encode. It will thus pose a severe technical challenge to construct examples of subalgebras of $\cR_T$ that have a tractable algebro-combinatorial description. However, there exist notable exceptions to the above arguments, one of the most important classes of which consist in a very special variant of graph rewriting systems (site-graph rewriting plus extra structures) on which the programming language \emph{Kappa}~\cite{danos2007rule} is based. This motivates as a future direction the search for \emph{combinatorially tame graph rewriting systems}. These and other ideas will be pursued in~\cite{behrCP2016,bdfk2016}.

\appendix

\section{Background material on diagram Hopf algebras and proofs of the mathematical statements}\label{app:B}

For the interdisciplinary readers' convenience, we briefly recall for each subsection the relevant generic mathematical background materials (which are clearly very well-known to specialists in combinatorics), and also provide the proofs for the results presented in the main text.

\subsection{Algebra structure}\label{app:Algebra}

Let us first recall the relevant generic definitions:
\begin{defn}[Unital associative $\mathbb{K}$-algebras]
 A $\mathbb{K}$-algebra $A\equiv (A,+,\cdot,*)$ is a $\mathbb{K}$-vector space $A$ endowed with a \emph{bilinear multiplication law} aka \emph{composition} $*$, where the multiplication law is a bilinear map
\begin{equation}
*:A\times A\rightarrow A: (v_1,v_2)=(\sum_i\alpha_i e_i,\sum_i \beta_i e_i)\mapsto v_1*v_2:=\sum_{i,j}\alpha_i\beta_j e_i*e_j\,.
\end{equation}
We also define the \emph{multiplication map} (\emph{convention}: $\otimes\equiv \otimes_{\mathbb{K}}$)
\begin{equation}
\mu:A\otimes A\rightarrow A: v_1\otimes v_2\mapsto v_1*v_2
\end{equation}
and the \emph{unit map} 
\begin{equation}
\eta: \mathbb{K}\rightarrow A: \alpha\mapsto \alpha \cdot 1_A\,.
\end{equation}
The special element $1_A$ is the (unique) \emph{unit under composition},
\begin{equation}
1_A*a=a*1_A=a\qquad \forall a\in A\,.
\end{equation}
For a \emph{unital associative algebra}, we require that the diagrams below commute:
\begin{equation}
  \includetikz{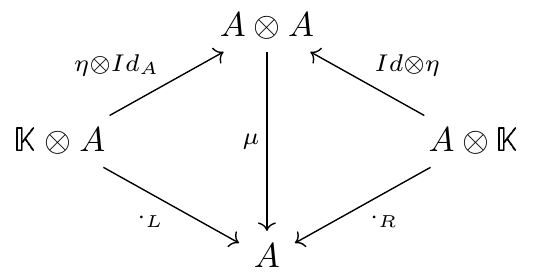}\qquad\qquad
    \includetikz{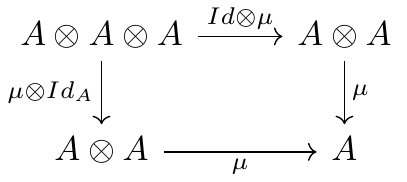}
\end{equation}
\end{defn}

We conclude this subsection by expressing the conditions for having a unital and associative algebra in a useful graphical form, via introducing the so-called \emph{string diagrams}:
\begin{equation}
\begin{aligned}
id_A:A\rightarrow A\quad&\widehat{=}\quad\boxed{\includetikz{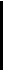}}\,, &\quad
id_{A\otimes A}:A\otimes A\rightarrow A\otimes A\quad&\widehat{=}\quad\boxed{\includetikz{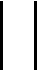}}
\\
\mu:A\otimes A\rightarrow A\quad&\widehat{=}\quad\boxed{\includetikz{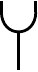}}\,,&\quad
\eta:\bK\rightarrow A\quad&\widehat{=}\quad\boxed{\includetikz{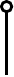}}
\end{aligned}
\end{equation}
Note that we implicitly defined the \emph{``pessimistic arrow of time''}, i.e.\ the order of operations is \emph{from top to bottom}, which appears to be the standard convention in the mathematics literature. With these preparations, the conditions for the unital property reads
\begin{equation}
  \includetikz{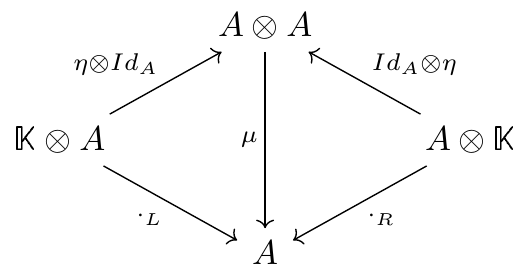}\quad
	\widehat{=}\quad\boxed{\includetikz{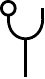}=\includetikz{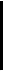}=
\includetikz{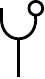}}\,,
\end{equation}
while the associativity property is graphically expressed as
\begin{equation}
  \includetikz{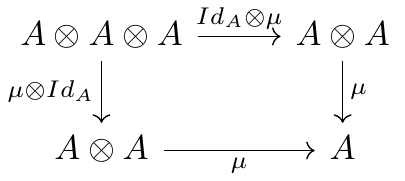}\quad\widehat{=}\quad\boxed{\includetikz{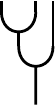}=
\includetikz{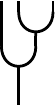}}\,.
\end{equation}
The string diagrams are useful since, as we will see on the next subsection, they allow to express the concept of duality when defining the coalgebra structure next, and moreover will simplify expressing the complicated consistency conditions for the bi- and Hopf algebra structures.

\subsection{Coalgebra structure}\label{app:Coalgebra}

\begin{defn}[Algebra homomorphisms]\label{def:algMorph}
Let $A$ and $B$ be unital associative $\mathbb{K}$-algebras. A $\mathbb{K}$-linear map $\theta:A\rightarrow B$ is called a \emph{$\mathbb{K}$-algebra homomorphism} if the following two diagrams commute:
\begin{equation}
  \includetikz{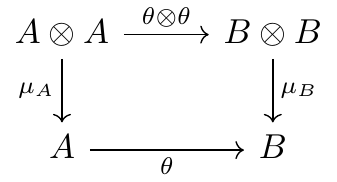}
\quad
	\widehat{=}\quad\boxed{\includetikz{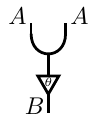}=
\includetikz{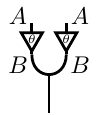}}
\end{equation}
\begin{equation}
  \includetikz{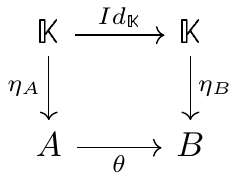}
\quad
	\widehat{=}\quad\boxed{\includetikz{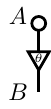}=
\includetikz{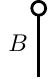}}
\end{equation}
Here, we have introduced the symbol 
\begin{equation}
\includetikz{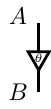}
\end{equation}
for a $\mathbb{K}$-algebra morphism $\theta:A\rightarrow B$.
\end{defn}

\begin{defn}[Tensor product of two algebras]\label{def:algTensorProduct}
Let $A$ and $B$ be unital associative $\mathbb{K}$-algebras. Then\footnote{As in the case of ``the'' tensor product of two vector spaces, there is always a degree of freedom available in choosing the morphisms that make the module properties of the $\bK$-tensor product explicit.} ``the'' \emph{tensor product algebra} $A\otimes B\equiv (A\otimes B,+,\cdot,*)$ is defined via
\begin{equation}
\mu_{A\otimes B}:(A\otimes B)\otimes(A\otimes B)\rightarrow A\otimes B:(a_1\otimes b_1)\otimes(a_2\otimes b_2)\mapsto (a_1*a_2)\otimes(b_1*b_2)\,,
\end{equation} 
or in slightly more detail
\begin{equation}
 \includetikz{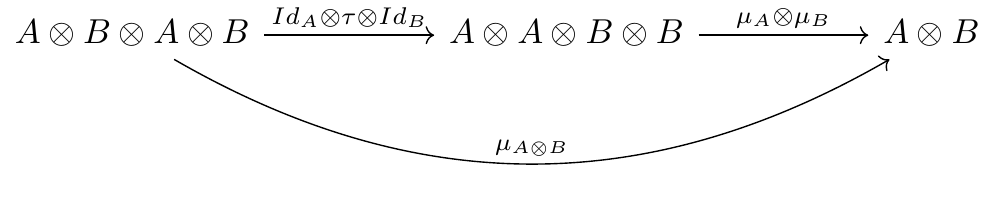}\,.
\end{equation}
The unit map of $A\otimes B$ is defined via
\begin{equation}
  \includetikz{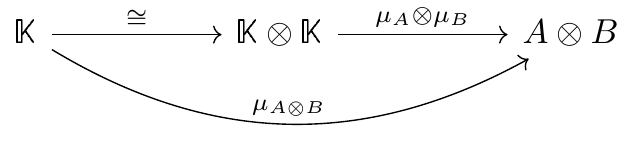}
\end{equation}
and $A\otimes B$ thus defined is a unital associative $\mathbb{K}$-algebra.\\

Evidently, this construction applies in particular also to the case $A=B$.
\end{defn}

It was necessary to introduce these two definitions, since the coalgebra structures we are about to discuss are concepts related to tensor products of the algebra under scrutiny with itself, in the definition of which the concept of homomorphisms of unital associative algebras was necessary. With these preparations, we may now introduce the following definitions:

\begin{defn}[Ccounital coassociative $\mathbb{K}$-coalgebras]\label{def:coalg}
A \emph{counital coassociative $\mathbb{K}$-coalgebra} $C\equiv(C,+,\cdot,\Delta)$ is a $\mathbb{K}$-vector space equipped with two bilinear maps, the \emph{coproduct} (in abstract \emph{Sweedler's notation})
\begin{equation}
\boxed{\includetikz{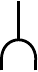}}\quad\widehat{=}\quad \Delta:C\rightarrow C\otimes C: c\mapsto \Delta(c)\equiv \sum_i c^{(1)}_i\otimes c^{(2)}_i\equiv c_{(1)}\otimes c_{(2)}\,,
\end{equation}
and the \emph{counit},
\begin{equation}
\boxed{\includetikz{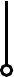}}\quad\widehat{=}\quad \varepsilon:C\rightarrow \mathbb{K}:c\mapsto\begin{cases} \gamma\quad &,\text{ if $c=\gamma\cdot 1_C$}\\
0\quad &,\text{ else}
\end{cases}\,,
\end{equation}
where $1_C\in C$ is the (unique) unit under the coproduct, viz
\begin{equation}
\Delta(1_C)=1_C\otimes 1_C\,.
\end{equation}
In order for $C$ to be co-unital and co-associative, the following diagrams must commute:

\begin{equation}
  \includetikz{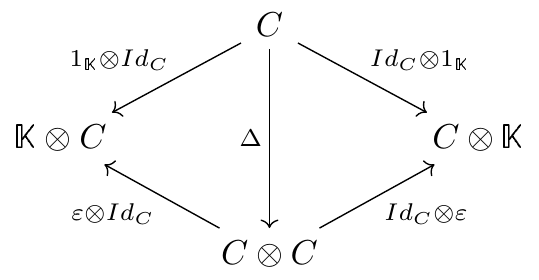}\quad
	\widehat{=}\quad\boxed{\includetikz{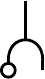}=\includetikz{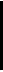}=
\includetikz{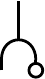}}
\end{equation}
\begin{equation}
  \includetikz{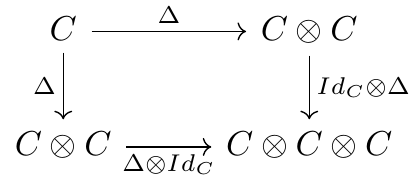}\quad\widehat{=}\quad\boxed{\includetikz{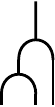}=
\includetikz{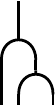}}
\end{equation}
\end{defn}

We present now the proof of the Proposition~\ref{prop:RDAcoalgebra}, which states that the rule diagram algebra may be equipped with the structure of a coassociative, counital coalgebra:
\begin{prop}
The diagram algebra $\cD\equiv(\cD,\Delta,\varepsilon)$ is a \emph{coassociative counital and cocommutative $\bK$-coalgebra}.
\begin{proof}
In order to prove the \emph{counital property}, recall that the counit map $\varepsilon$ maps any diagram but the empty diagram $d_{\emptyset}$ to zero, hence in particular for any basis diagram $\delta(\fd)\in \cD$:
\begin{align*}
\left(\varepsilon\otimes Id_{\cD}\right)\circ \Delta(\delta(\fd))&=\left(\varepsilon\otimes Id_{\cD}\right)\circ\left(
\sum_{(\fd'',\fd')\in \langle \fd \rangle} \delta(\fd'')\otimes \delta(\fd')
\right)\\
&= \sum_{(\fd'',\fd')\in \langle \fd \rangle} \varepsilon(\delta(\fd''))\otimes \delta(\fd')\\
&= 1_{\bK}\otimes \delta(\fd)\,,
\end{align*}%
where $\fd\in\fD$ is an arbitrary rule diagram, and where the notation
\begin{equation*}
(\fd'',\fd')\in \langle \fd \rangle
\end{equation*} 
stands for the set of decompositions of the rule diagram $\fd\in \fD$ (which contains in particular the elements $(\emptyset,\fd)$ and $(\fd,\emptyset)$). An analogous argument holds for $Id_{\cD}\otimes \varepsilon$.\\

The cocommutativity follows simply from the definition of the set of possible decompositions $\langle\fd \rangle$ of any given rule diagram $\fd\in \fD$, since for any given decomposition $(\fd'',\fd')\in\langle \fd\rangle$, the decomposition $(\fd',\fd'')$ is included as well. This then implies that the property of coassociativity follows trivially -- it is immaterial whether one lists the set of all possible double decompositions by further decomposing the  entries $\fd''$ of the $(\fd'',\fd')\in\langle \fd\rangle$ or by further decomposing the entries $\fd'$ of $(\fd'',\fd')\in\langle \fd\rangle$.
\end{proof}
\end{prop}

\subsection{Bialgebra structure}\label{app:Bialgebra}

We first need the following preparations:
\begin{defn}[Tensor product of two coalgebras]
Let $C$ and $D$ be counital coassociative $\mathbb{K}$-coalgebras. Then ``the'' \emph{tensor product coalgebra} $C\otimes D\equiv (C\otimes D,+,\cdot,\Delta)$ is defined via

\begin{equation}
  \includetikz{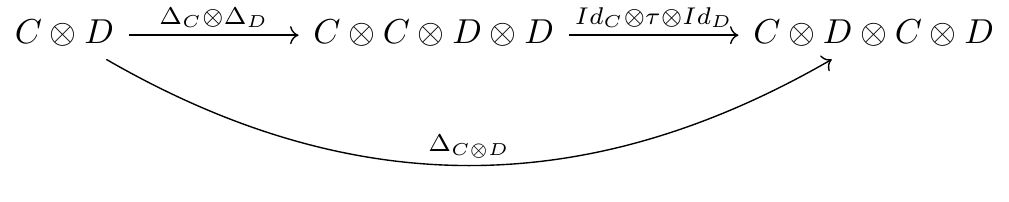}\,.
\end{equation}
The counit map of $C\otimes D$ is defined via
\begin{equation}
  \includetikz{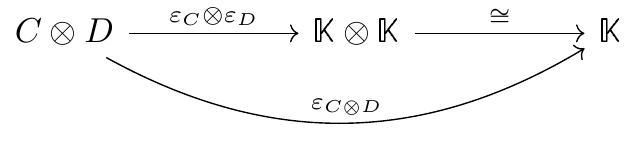}\,.
\end{equation}
and $C\otimes D$ thus defined is a counital coassociative $\mathbb{K}$-coalgebra.\\

Evidently, this construction applies in particular also to the case $C=D$.
\end{defn}

\begin{defn}[The twist map and (co-)commutativity]
Let $A$ be an algebra (or a coalgebra). Then the \emph{twist map} $\tau$ is defined as
\begin{equation}
\boxed{\includetikz{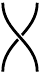}}\quad\widehat{=}\quad \tau:A\otimes A\rightarrow A\otimes A: a\otimes b\mapsto b\otimes a\,.
\end{equation}
An algebra $A$ is said to be \emph{commutative} if the following diagram commutes:
\begin{equation}
  \includetikz{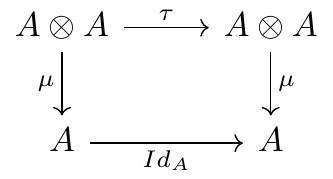}\quad
	\widehat{=}\quad\boxed{\includetikz{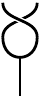}=
\includetikz{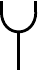}}
\end{equation}
A coalgebra $C$ is said to be \emph{cocommutative} if the following diagram commutes:
\begin{equation}
  \includetikz{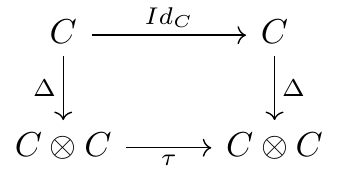}\quad
	\widehat{=}\quad\boxed{\includetikz{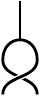}=
\includetikz{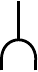}}
\end{equation}
\end{defn}

\begin{defn}[Bialgebras]
A \emph{$\mathbb{K}$-bialgebra} $A$ is a vector space 
\[
A\equiv (A,+,\cdot,\mu,\eta,\Delta,\varepsilon)\,,
\]
where
\begin{itemize}
\item $(A,\mu,\eta)$ is a \emph{unital associative $\mathbb{K}$-algebra}
\item $(A,\Delta,\varepsilon)$ is a \emph{counital coassociative $\mathbb{K}$-coalgebra}
\end{itemize}
and such that either (and hence both) of the following two conditions hold:
\begin{enumerate}
\item[$(i)$] $\Delta$ and $\varepsilon$ are \emph{algebra homomorphisms}
\item[$(ii)$] $\mu$ and $\eta$ are \emph{coalgebra homomorphisms}.
\end{enumerate}
The respective conditions from either viewpoint as well as the relevant string diagrams are presented in Table~\ref{tab:bialg}, where we have introduced the auxiliary definitions
\begin{equation}
\mu_{\bK}:\bK\otimes\bK\xrightarrow{\cong}\bK:(k_1,k_2)\mapsto k_1k_2\,,\quad
\Delta_{\bK}:\bK\xrightarrow{\cong}\bK\otimes\bK: k\mapsto k\cdot(1_{\bK}\otimes 1_{\bK})\,,
\end{equation}
and where the notion of coalgebra homomorphisms is defined via dualizing the string diagrams in Definition~\ref{def:algMorph} of algebra homomorphisms.
\end{defn}

\begin{table}[htbp]
\caption{Bialgebra compatibility conditions, expressed in each of the three equivalent forms: via string diagrams, from the viewpoint of the algebra structure, and from the viewpoint of the coalgebra structure.\vspace{1em}}
\label{tab:bialg}
\centering
\rotatebox{90}{\begin{tabular}{ccc}
\toprule
 string diagrams &algebra & coalgebra \\
 \midrule
 $\boxed{\includetikz{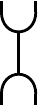}=\includetikz{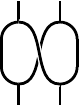}}$ &
$\includetikz{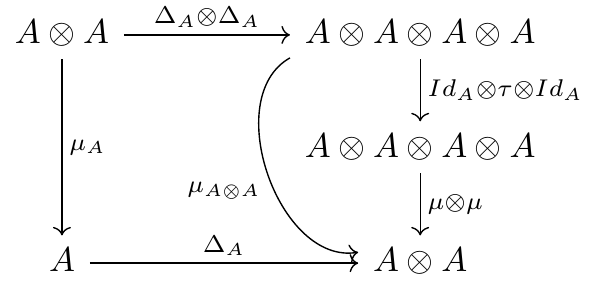}$ &
 $\includetikz{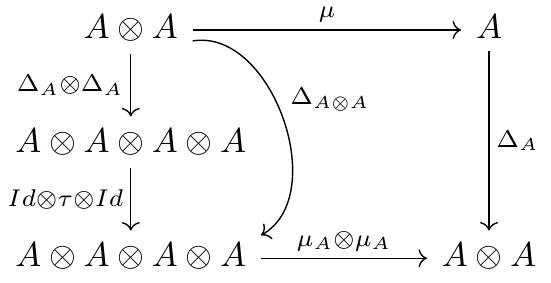}$  \\\midrule
 $\boxed{\includetikz{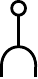}\;=\;\includetikz{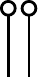}}$
 & $\includetikz{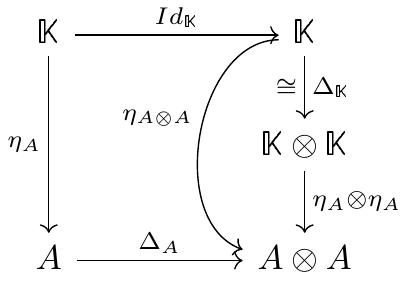}$
 &
 $\includetikz{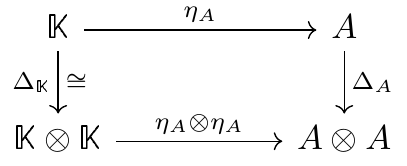}$\\
 \midrule
 $\boxed{\includetikz{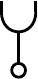}\;=\;\includetikz{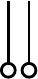}}$
 &
 $\includetikz{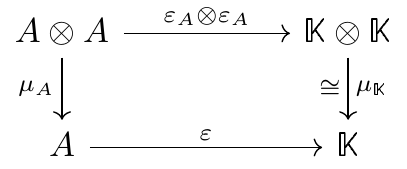}$
 &
 $\includetikz{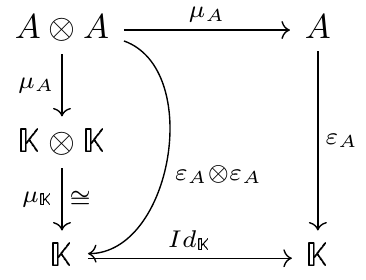}$
 \\\midrule
 $\boxed{\includetikz{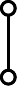}=\quad}$
 &
 $\includetikz{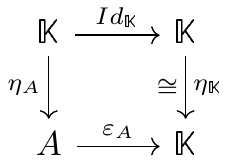}$
 &
 $\includetikz{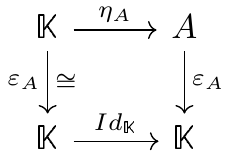}$\\
 \bottomrule
\end{tabular}}
\end{table}

We now present the proof of Theorem~\ref{thm:bialgRD}:
\begin{thm}[Bialgebra structure of $\cD$]
The rule diagram algebra 
\[
\cD\equiv(\cD,\mu,\eta,\Delta,\varepsilon)
\]
is a bialgebra.
\begin{proof}
It is evident that due to linearity and the underlying $\bK$-vector space structure of $\cD$, all compatibility conditions have to be checked only on the basis elements of the vector spaces, i.e.\ on basis diagrams $d\equiv \delta(\fd)\in \cD$ (with $\fd\in \fD$) and on $1_{\bK}$.\\
 
The compatibility of the counit $\varepsilon$ with the multiplication map $\mu$ follows from
\begin{equation}
\begin{aligned}
\varepsilon\circ\mu(\delta(\fd_A)\otimes \delta(\fd_B))&=\varepsilon (\delta(\fd_A*\fd_B))=\begin{cases}
1_{\bK}\quad &\text{, if } \fd_A=\fd_B=\fd_{\emptyset}\\
0\quad &\text{, else}
\end{cases}\\
&\overset{(*)}{=}\begin{cases}
1_{\bK}\otimes1_{\bK}\quad &\text{, if } \fd_A=\fd_B=d_{\emptyset}\\
0\quad &\text{, else}
\end{cases}\\
&=\left(\varepsilon\otimes\varepsilon\right)(\delta(\fd_A)\otimes \delta(\fd_B))\,,
\end{aligned}
\end{equation}
for all $\fd_A,\fd_B\in\fD$. Here, in the step marked $(*)$ we have used the identity 
\begin{equation}
1_{\bK}\otimes 1_{\bK}=1_{\bK}\,,
\end{equation}
which follows from the fact that $\bK$ is a field and from $\otimes\equiv\otimes_{\bK}$.\\

Dually, the consistency of the coproduct with the unit map follows from
\begin{equation}
\begin{aligned}
\Delta\circ \eta(1_{\bK})&=\Delta(d_{\emptyset})=d_{\emptyset}\otimes d_{\emptyset}=\left(\eta\otimes\eta\right)(1_{\bK}\otimes 1_{\bK})\,.
\end{aligned}
\end{equation}

Next, the consistency of the counit map with the unit map follows trivially from
\begin{equation}
\varepsilon\circ\eta(1_{\bK})=\varepsilon(d_{\emptyset})=1_{\bK}\,.
\end{equation}

This leaves to prove the by far less obvious consistency condition between the multiplication and the comultiplication map. We need to show that
\begin{equation}\label{eq:bialgCond4}
\Delta\circ\mu(\fd_A\otimes \fd_B)=(\mu\otimes\mu)\circ(Id_{\cD}\otimes\tau\otimes Id_{\cD})\circ(\Delta\otimes\Delta)(\delta(\fd_A)\otimes \delta(\fd_B))\,.
\end{equation}
Let us begin by recalling from the definition of decompositions of rule diagrams that a generic rule diagram is a \emph{superposition of indecomposable aka primitive rule diagrams}, whence in particular
\begin{equation}\label{eq:bialgAux}
\fd_A=\left[\biguplus_{i=1}^{n_A} \fd(A_i)\right]_{\cong}\,,\quad \fd_B=\left[\biguplus_{j=1}^{n_B}\fd(B_j)\right]_{\cong}\,.
\end{equation}
We may thus sort any possible \emph{match} between $\fd_A$ and $\fd_B$ into the form (upon fixing suitable representatives for $\fd_A$ and $\fd_B$)
\begin{equation}
\cM_{\fd_A}(\fd_B)\ni m_{A,B}\equiv \left\{ m_{A_i,B_j}\mid 1\leq i\leq n_A\,,\; 1\leq j\leq n_B\,,\; i<j\right\}\,.
\end{equation}
If we first compose $\fd_A$ with $\fd_B$, this amounts to 
\begin{equation}
\delta(\fd_A)*\delta(\fd_B)=\sum_{m_{A,B}\in\cM_{\fd_A}(\fd_B)}\delta\left(\comp{\fd_A}{m_{A,B}}{\fd_B}\right)\,.
\end{equation}
Whenever for a particular match $m_{A,B}$ an entry $m_{A_i,B_j}$ is nontrivial (i.e.\ when $m_{A_i,B_j}\neq \emptyset$), this entry ``fuses'' the indecomposable superposition factors $\fd(A_i)$ and $\fd(B_j)$ into an \emph{indecomposable} rule diagram upon performing the composition operation. Hence we have generically that
\begin{equation}\label{eq:biAlgAux2}
\comp{\fd_A}{m_{A,B}}{\fd_B}\equiv \left[\biguplus_{i=1}^{n_{m_{A,B}}} \fd((AB)_i)\right]_{\cong}\,.
\end{equation}
More precisely, some of the $\fd((AB)_i)$ will generically coincide with some of the original superposition factors $\fd(A_i)$ and $\fd(B_j)$ (precisely if these factors had not been involved in the match), while the remaining factors will be composites of some of the factors $\fd(A_i)$ and $\fd(B_j)$ along the nontrivial entries $m_{A_i,B_j}\neq \emptyset$ of the given match $m_{A,B}$. In any event, once we apply the coproduct to such a generic term of the multiplication operation, the set of possible decompositions coincides with the set of all partitions into two unordered sets of the superposition factors in~\eqref{eq:biAlgAux2}.\\

After a moment's thought, we may convince ourselves that listing all possible decompositions of all possible composites of $\fd_A$ with $\fd_B$ is entirely equivalent to the following operation: first, list (independently of one another) all possible decompositions of $\fd_A$ and $\fd_B$ -- these amount as per our previous arguments to all possible partitions into two sets of the superposition factors as described by~\eqref{eq:bialgAux}. Thus we have that
\begin{equation}
\begin{aligned}
&(\Delta\otimes\Delta)(\delta(\fd_A)\otimes \delta(\fd_B))\\
&\quad=
\sum_{(\fd_A{}',\fd_A{}'')\in\langle \fd_A\rangle}
\sum_{(\fd_B{}',\fd_B{}'')\in\langle \fd_B\rangle}
\delta(\fd_A{}')\otimes \delta(\fd_A{}'')\otimes \delta(\fd_B{}')\otimes \delta(\fd_B{}'')\,.
\end{aligned}
\end{equation}
Applying the map $Id\otimes \tau\otimes Id$ (which amounts to swapping the middle two factors) followed by the map $\mu\otimes\mu$, we finally obtain
\begin{equation}
\begin{aligned}
&(\mu\otimes\mu)\circ(Id_{\cD}\otimes\tau\otimes Id_{\cD})\circ(\Delta\otimes\Delta)(\delta(\fd_A)\otimes \delta(\fd_B))\\
&\quad =
\sum_{(\fd_A{}',\fd_A{}'')\in\langle \fd_A\rangle}
\sum_{(\fd_B{}',\fd_B{}'')\in\langle \fd_B\rangle}
(\delta(\fd_A{}')*\delta(\fd_B{}'))\otimes( \delta(\fd_A{}'') * \delta(\fd_B{}''))\,.
\end{aligned}
\end{equation}
Now, recalling that the coproduct is cocommutative, the above contributions in fact amount to tensor products of products of \emph{all possible subsets} of superposition factors of $\fd_A$ and $\fd_B$! For example, the four simplest possible contributions (those for which two of the subsets are empty) read as follows:
\begin{equation}
\begin{aligned}
(\delta(\fd_A)*\delta(\fd_B))\otimes (d_{\emptyset}*d_{\emptyset})&=
(\delta(\fd_A)*\delta(\fd_B))\otimes d_{\emptyset}\\
(d_{\emptyset}*\delta(\fd_B))\otimes (\delta(\fd_{A})*d_{\emptyset})&=
\delta(\fd_B)\otimes \delta(\fd_A)\\
(\delta(\fd_A)*d_{\emptyset})\otimes (d_{\emptyset}*\delta(\fd_{B}))&=
\delta(\fd_A)\otimes \delta(\fd_B)\\
(d_{\emptyset}*d_{\emptyset})\otimes(\delta(\fd_A)*\delta(\fd_B))&=
d_{\emptyset}\otimes (\delta(\fd_A)*\delta(\fd_B))\,.
\end{aligned}
\end{equation}
The first and the last term amount to the trivial part of the coproduct of $\fd_A*\fd_B$, while the second and third term amount to two of the generically possible contributions to the coproduct of
\[
\comp{\fd_A}{\emptyset}{\fd_B}\,.
\]
Generically, all possible terms in
\begin{align*}
&\Delta(\delta(\fd_A)*\delta(\fd_B))\\
&\quad=\sum_{m\in \cM_{\fd_A}(\fd_B)}\sum_{(\fd_{A_m B}{}',\fd_{A_m B}{}'')\in\langle \comp{\fd_A}{m_{A,B}}{\fd_B\rangle}}
\delta(\fd_{A_mB}{}')\otimes \delta(\fd_{A_mB}{}'')
\end{align*}
may be sorted by the superposition factors $\fd(A_i)$ and $\fd(B_j)$ contained in the left factor of the tensor product, which hence amounts to listing all possible compositions of those factors as induced by the matches $m_{A,B}$ together with all possible composites of the remaining superposition factors in the right factor of the tensor product. This completes the proof.
\end{proof}
\end{thm}

\subsection{Taking multiple products and coproducts}\label{app:MultipleProductsAndCoproducts}

Suppose $A\equiv(A,\mu,\eta)$ is an associative unital algebra. Then we saw from the consistency condition for associativity, viz
\[
\mu\circ(Id_A\otimes \mu)=\mu\circ (\mu\otimes Id_A)\,,
\]
that for an associative algebra $A$ it makes sense to define ``the'' triple product of $A$ via choosing one of the possible realizations, which w.l.o.g.\ is customarily chosen as
\begin{equation}
\mu_{A,3}:=\mu\circ (\mu\otimes Id_A):\cA^{\otimes\: n}\rightarrow \cA:a_1\otimes a_2\otimes a_3\mapsto a_1*a_2*a_3\,.
\end{equation}
By induction, associativity implies that it makes sense to choose any of the possible realizations of the higher multiple multiplication maps to represent ``the'' multiple multiplications, which motivates the following definition:
\begin{defn}[$n$-fold multiplication]\label{def:nfoldMult}
Let $A\equiv (A,\mu,\eta)$ be an associative unital algebra. Then the \emph{$n$-fold multiplication maps} are defined as
\begin{equation}
\mu_{A,n}:=\mu\circ (\mu\otimes Id_A)\circ (\mu\otimes Id_A^{\otimes\: 2})\circ\dotsc\circ(\mu\otimes Id_A^{\otimes\: (n-1)})\,,
\end{equation}
which moreover induces the definition of the \emph{$n$-fold products}
\begin{equation}
a_1*a_2*\dotsc*a_n:=\mu_{A\:n}(a_1\otimes\dotsc\otimes  a_n)\,.
\end{equation}
\end{defn}
Dually, for a counital coassociative coalgebra $C\equiv(C,\Delta,\varepsilon)$, coassociativity entails that
\[
(\Delta\otimes Id_C)\circ \Delta=(Id_C\otimes \Delta)\circ \Delta\,,
\]
which by induction motivates the following definition:
\begin{defn}[$n$-fold comultiplication]\label{def:nfoldComult}
Let $C\equiv(C,\Delta,\varepsilon)$ be a coassociative counital coalgebra. Then the \emph{$n$-fold coproduct} is defined as
\begin{equation}
\Delta_{C,n}:=(\Delta\otimes Id_C^{\otimes\: (n-2)})\circ (\Delta\otimes Id_C^{\otimes\: (n-3)})\circ\dotsc \circ (\Delta\otimes Id_C)\circ\Delta\,.
\end{equation}
\end{defn}

\subsection{Filtration and Hopf algebra structure}\label{app:FilteredHopf}

\begin{defn}[Connected filtered bialgebra (cf.~\cite{manchon2008hopf}, ch.~4.2)]\label{def:connFilt}
A bialgebra 
\[
\cA\equiv(\cA,\mu,\eta,\Delta,\varepsilon)
\]
is called \emph{filtered} if there exists an \emph{increasing $\bZ_{\geq 0}$-indexed filtration}
\begin{equation}
\cA^0\subset \cA^1\subset \dotsc \subset \cA^n\subset\dotsc\,,\quad \cup_{n=0}^\infty\cA^n=\cA
\end{equation}
compatible with the bialgebra structure, in the sense that the following equations hold true:
\begin{align}
\cA^p*\cA^q&\subset \cA^{p+q}\\
\Delta(\cA^n)&\subset\sum_{p+q=n}\cA^p\otimes\cA^q=\sum_{p=0}^n\cA^p\otimes\cA^{n-p}\,.
\end{align}%
Moreover, if $dim(\cA^0)=1$, then $\cA$ is called a \emph{connected filtered bialgebra}.
\end{defn}
The usefulness of the concept just introduced lies in a number of very powerful statements available from the mathematics literature for such connected filtered bialgebras:
\begin{prop}[cf.~\cite{manchon2008hopf}, ch.~4.2, Prop.~18]\label{prop:coalg2}
For any $a\in\cA^n$ with $n>0$, we may write the coproduct of $a$ in the form
\begin{equation}
\Delta(a)=1_{\cA}\otimes a+a\otimes 1_{\cA}+\tilde{\Delta}(a)\,,\quad \tilde{\Delta}(a)\in\sum_{\stackrel{p+q=n}{p\neq0,q\neq0}}\cA^p\otimes \cA^q=\sum_{p=1}^{n-1}\cA^p\otimes \cA^{n-p}\,.
\end{equation}
Thus in particular
\begin{equation}
\tilde{\Delta}(a)=0\quad \forall a\in \cA^1\,.
\end{equation}
The map $\tilde{\Delta}$ is coassociative only on $Ker(\varepsilon)$ (unlike $\Delta$, which is coassociative on all of $\cA$). We have moreover that the maps\footnote{Note that our convention for the labelling of the $k$-fold (co-)multiplications differ from those of~\cite{manchon2008hopf} by a shift of the index by $+1$ relative to loc cit.\ -- in our notation, the index indicates how many copies of the algebra are involved, i.e.
\[
\mu_{\cA,k}:\cA^{\otimes\: k}\rightarrow \cA\,,\quad \Delta_{\cA,k}:\cA\rightarrow \cA^{\otimes\: k},\quad \tilde{\Delta}_{\cA,k}:\cA\rightarrow \cA^{\otimes\: k}\,.
\]}
\begin{equation}
\tilde{\Delta}_k:=\left(\tilde{\Delta}\otimes Id^{\otimes (k-2)}\right)\circ\left( \tilde{\Delta}\otimes Id^{\otimes (k-3)}\right)\circ\dotsc\circ\tilde{\Delta}
\end{equation}
send $\cA^n$ into $(\cA^{n-k+1})^{\otimes k}$.
\begin{proof}
The first two statements follow directly from the definitions of $\Delta$ and $\tilde{\Delta}$. To prove the third statement, observe that for any $a\in\cA^n$ ($n>0$)
\begin{equation}
\begin{aligned}
\tilde{\Delta}_3(a)&=(\tilde{\Delta}\otimes Id)\circ\overbrace{\left(\tilde{\Delta}(a)\right)}^{\in \sum_{p=1}^{n-1}\cA^p\otimes \cA^{n-p}}\\
&\overset{(*)}{\subset} \sum_{p=2}^{n-1}\sum_{q=1}^{p-1} \cA^q\otimes  \cA^{p-q}\otimes\cA^{n-p}\subset \left(\cA^{n-2}\right)^{\otimes\: 3}\,,
\end{aligned}
\end{equation}
where in the step marked $(*)$ we have used that $\tilde{\Delta}(a)=0$ for $a\in\cA^1$, which annihilates all contributions for which $p=1$. For $\tilde{\Delta}_4(a)$, we have entirely analogously that
\begin{equation}
\begin{aligned}
\tilde{\Delta}_4(a)&=(\tilde{\Delta}\otimes Id\otimes Id)\circ\overbrace{\left(\tilde{\Delta}_3(a)\right)}^{\in \sum_{p=2}^{n-1}\sum_{q=1}^{p-1}\cA^q\otimes\cA^{p-q}\otimes \cA^{n-p}}\\
&\overset{(*)}{\subset} \sum_{p=3}^{n-1}\sum_{q=2}^{p-1}\sum_{r=1}^{q-1}\cA^r\otimes \cA^{q-r}\otimes  \cA^{p-q}\otimes\cA^{n-p}\subset \left(\cA^{n-3}\right)^{\otimes\: 4}\,,
\end{aligned}
\end{equation}
where again $\tilde{\Delta}$ applied to the first factor of $\cA^q\otimes \cA^{p-q}\otimes\cA^{n-p}$ annihilates all contributions with $q=1$; incidentally, at $p=2$ one only has the contribution $\cA^1\otimes\cA^1\otimes \cA^{n-2}$, hence the sum over $p$ effectively starts at $p=3$. We thus have the induction hypothesis that for all $a\in\cA^n$ with $n>0$
\begin{equation}\label{eq:tildeDeltaA}
\tilde{\Delta}_{m+1}(a)\subset \sum_{p_1=m}^{n-1}\sum_{p_2=m-1}^{p_1-1}\dotsc\sum_{p_m=1}^{p_{m-1}-1}
\cA^{p_m}\otimes\cA^{p_{m-1}-p_m}\otimes \dotsc\otimes \cA^{p_1-p_2}\otimes \cA^{n-p_1}\,.
\end{equation}
For the induction step $m+1\to m+2$, notice that 
\[
\tilde{\Delta}_{m+2}(a)=\left(\tilde{\Delta}\otimes Id^{\otimes\:m}\right)\circ\tilde{\Delta}_{m+1}(a)\,.
\]
Application of $\tilde{\Delta}$ to the first factors of the contributions as presented in~\eqref{eq:tildeDeltaA} annihilates all contributions for which $p_m=1$, hence the nonzero contributions are those for which $p_{m}>1$, which necessitates $p_{m-1}>2$, $p_{m-2}>3$ and so forth, thereby proving the claim.
\end{proof}
\end{prop}
In our later applications, we will make frequent use of the following Corollary:
\begin{cor}\label{corr:tildeDeltaAnn}
Let $\cA$ be a connected filtered bialgebra as before. Then for any $a\in\cA^n$, since 
\begin{equation}
\tilde{\Delta}_k:\cA^n\rightarrow \left(\cA^{n-k+1}\right)^{\otimes\: k}\,,
\end{equation}
we have that
\begin{equation}
\tilde{\Delta}_k(a)=0\quad \forall a\in A^{k-1}\,.
\end{equation}

\end{cor}

Apart from the identity element $1_{\cA}$ (which is unique for a connected filtered bialgebra), the other type of special elements are those for which $\tilde{\Delta}(a)=0$:
\begin{defn}[Primitive elements]\label{def:prim}
Let $\cA$ be a connected filtered bialgebra. Then the elements $a\in\cA$ for which
\begin{equation}
\Delta(a)=1_{\cA}\otimes a+a\otimes 1_{\cA}
\end{equation}
are called \emph{primitive elements} of $\cA$. The subspace of primitive elements of $\cA$ is given by $\cA^1\setminus \cA^0$.
\end{defn}

We now present the proof of Proposition~\ref{prop:connFilt}:
\begin{prop}
The rule diagram algebra $\cD\equiv(\cD,\mu,\eta,\Delta,\varepsilon)$ is a \emph{connected, filtered bialgebra}, with 
\begin{equation}
\cD^0=span_{\bK}(d_{\emptyset})\,,
\end{equation}
and where the subspace $\cD^n$ is in general given by the linear span of all rule diagrams with at most $n$ indecomposable subdiagrams. In particular, the \emph{primitive basis diagrams of $\cD$} are the basis vectors associated to those rule diagrams that are neither empty nor decomposable, whence
\[
Prim(\cD):= (\cD^1\setminus \cD^0)\subsetneq \cD\,.
\]
\begin{proof}
It is evident that due to
\begin{equation}
d_{\emptyset}*d_{\emptyset}=d_{\emptyset}\,,\quad \Delta(d_{\emptyset})=d_{\emptyset}\otimes d_{\emptyset}\,,
\end{equation}
$d_{\emptyset}$ indeed qualifies as an element of the filtered component $\cD^0$. As already presented in the proof of Theorem~\ref{thm:bialgRD}, the composite of two arbitrary rule diagrams, say,  $\fd_A$ with $s_A$ indecomposable subdiagrams and $\fd_B$ with $s_B$ indecomposable subdiagrams, only contains contributions with maximally $s_A+s_B$ subdiagrams, and depending on the possible matches also of a lower number of subdiagrams, which is compatible with the statements
\begin{equation}
\begin{aligned}
\delta(\fd_I)&\equiv \delta\left(\biguplus_{i=1}^{s_I}\fd_I(i)\right)\in \cD^n\quad \forall n\geq s_I\\
\delta(\fd_A)*\delta(\fd_B)&\in \cD^{s_A+s_B}\subset \cD^{s_A+s_B+1}\subset\dotsc\,. 
\end{aligned}
\end{equation}
Dually, given a generic rule diagram $\fd_I\in\fD$ with $s_I$ connected components, it is trivially clear that the coproduct of $\delta(\fd_I)$, being defined as the sum over tensor products of basis vectors associated to the unordered bipartitions of $\fd_I$, naturally has the required structure
\begin{equation}
\Delta(\delta(\fd_I))\in\sum_{\stackrel{p,q=0}{p+q=s_I}}^{s_I} \cD^p\otimes \cD^q\,.
\end{equation}
\end{proof}
\end{prop}

It is a well-known standard result from the mathematics literature that for any connected filtered bialgebra, it is possible to define the following product:
\begin{defn}[Convolution product (cf.~\cite{manchon2008hopf}, Ch. 4.3)]\label{defconv}
Let $\cH$ be a connected filtered bialgebra, and let $\cA$ be an arbitrary $\bK$-algebra (which we call the target algebra). Then the \emph{convolution product} $\star$ on the space $\cL(\cH,\cA)$ of linear maps from $\cH$ to $\cA$ is defined for any maps $\varphi, \psi\in \cL(\cH,\cA)$ as
\begin{equation}
\varphi\star\psi:= \mu_{\cA}\circ (\varphi\otimes\psi)\circ \Delta_{\cH}\,.\quad 
  \includetikz{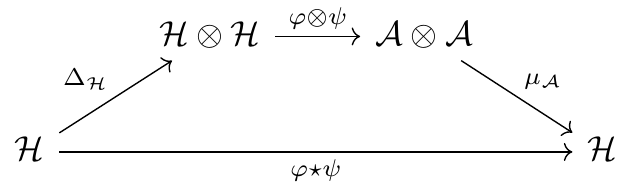}
\end{equation}
\end{defn}

Remarkably, the convolution product may be used to endow the space of linear maps with the structure of a group:
\begin{prop}[cf.~\cite{manchon2008hopf}, Ch.~4.3, Prop.~19]\label{prop:convProd}
Let $\cH$ be a connected filtered bialgebra and $\cA$ an arbitrary $\bK$-algebra as before. Then the linear map
\begin{equation}
e:=\eta_{\cA}\circ \varepsilon_{\cH}: \cH\rightarrow \cA:h\mapsto\begin{cases}
1_{\cA}\quad &\text{, if } h=1_{\cH}\\
0\quad &\text{, else}
\end{cases}
\end{equation}
is a \emph{unit for the convolution product},
\begin{equation}
e\star\psi=\psi\star e=\psi\qquad \forall \psi\in \cL(\cH,\cA)\,.
\end{equation}
Moreover, the set 
\begin{equation}
G(\cA):=\left\{
\varphi\in\cL(\cH,\cA)\mid \varphi(1_{\cH})=1_{\cA}
\right\}
\end{equation}
endowed with the convolution product is a \emph{group}.
\begin{proof}
To prove that $e$ is a unit for the convolution product $*$, it is convenient to introduce \emph{Sweedler's notation}
\begin{equation}
\Delta_{\cH}(h)=\sum_{(h)} h_1\otimes h_2
\end{equation}
for the coproduct of any $h\in\cH$. Then for any $\psi\in\cL(\cH,\cA)$, it follows by direct computation that
\begin{equation}
\begin{aligned}
\mu_{\cA}\circ (e\otimes \psi)\circ\Delta_{\cH}(h)&=\mu_{\cA}\circ (e\otimes \psi)\left(
\sum_{(h)} h_1\otimes h_2
\right)\\
&=\mu_{\cA}\left(
\sum_{(h)} e(h_1)\otimes \psi(h_2)
\right)\\
&\overset{(*)}{=}\psi(h)\,,
\end{aligned}
\end{equation}
where we have used that $e(h_1)=0$ for $h_1\neq 1_{\cH}$. The computation to demonstrate that $\psi*e=\psi$ is entirely analogous.\\

To prove that $G(\cA)$ endowed with the convolution product $*$ is a group, it is first of all evident that $G(\cA)$ is closed under the convolution product -- we have by definition that
\[
\star:\cL(\cH,\cA)\times\cL(\cH,\cA)\rightarrow \cL(\cH,\cA)\,,
\]
hence it suffices to verify that for any two maps $\varphi,\psi\in G(\cA)$, for which by definition of $G(\cA)$ 
\[
\varphi(1_{\cH})=\psi(1_{\cH})=1_{\cA}\,,
\]
we also have that $(\varphi\star \psi)(1_{\cH})=1_{\cA}$:
\begin{align*}
\varphi\star \psi(1_{\cH})&=\mu_{\cA}\circ(\varphi\otimes\psi)\circ\Delta_{\cH}(1_{\cH})\\
&= \mu_{\cA}\circ(\varphi\otimes\psi)(1_{\cH}\otimes1_{\cH})\\
&=\mu_{\cA}(\varphi(1_{\cH})\otimes\psi(1_{\cH}))\\
&=1_{\cA}*1_{\cA}=1_{\cA}\,.
\end{align*}%
The more difficult part of the proof that $G(\cA)\equiv (G(\cA),\star)$ is a group consists in demonstrating that there exist \emph{inverses} for any $\varphi\in G(\cA)$. Using the geometric series
\[
\frac{1}{1-q}=\sum_{k=0}^{\infty}q^k
\]
as a blueprint, we may formally construct\footnote{Note that in analogy to $1/(1-q)=1+q+q^2+\dotsc$, the first term in the formal series is given by the unit, which in this case is the map $e$.} the inverse of a map $\varphi\in G(\cA)$ via (for any $h\in\cH$)
\begin{equation}
\varphi^{\star\:-1}(h)=\left(e-(e-\varphi)\right)^{\star\:-1}(h)=e+\sum_{k=1}^{\infty}(e-\varphi)^{\star\: k}(h)\,.
\end{equation}
This formal series is only well defined if we can demonstrate that it only contains finitely many terms for any given element $h\in\cH$. It is at this point that the requirement of $\cH$ being a connected filtered Hopf algebra comes into play: if $h\in \cH^0$, connectedness of $\cH$ implies that $h=k\cdot 1_{\cH}$ (for some $k\in\bK$), for which we evidently have that $(e-\varphi)(k\cdot1_{\cH})=0$ (since $e(1_{\cH})=1_{\cA}$ and $\varphi(1_{\cH})=1_{\cA}$ for any $\varphi\in G(\cA)$). It immediately follows (since $\varphi\in G(\cA)$ and by definition of $e$) that
\[
(e-\varphi)(1_{\cH})=1_{\cA}-1_{\cA}=0\,,
\]
and thus (using that $G(\cA)$ is closed under composition) also that
\[
(e-\varphi)^{\star\: k}(1_{\cH})=0\quad \forall k\in\bZ_{>0}\,.
\]
Therefore,
\begin{equation}
\varphi^{\star\:-1}(1_{\cH})=e(1_{\cH})+\sum_{k=1}^{\infty}(e-\varphi)^{\star\: k}(1_{\cH})=1_{\cA}\,,
\end{equation}
hence we have proved that $\varphi^{\star\:-1}\in G(\cA)$. It remains to prove that $\varphi^{\star\:-1}$ is well-defined also for an arbitrary element $h\in\cH$. To this end, consider w.l.o.g.\ an element $h\in\cH^n$ for $n>0$ not equal to $1_{\cH}$, i.e.\ an element of $Ker( \varepsilon_{\cH})$ (any other element would be proportional to $1_{\cH}$ by virtue of connectedness of $\cH$). According to Proposition~\ref{prop:coalg2}, such an element has the coproduct
\[
\Delta(h)=1_{\cH}\otimes h+h\otimes 1_{\cH} +\tilde{\Delta}(h)\qquad \forall h\in H^n\cap Ker( \varepsilon_{\cH})\;(n>0)\,.
\] 
A direct computation reveals that 
\begin{equation*}
\left((e-\varphi)\otimes(e-\varphi)\right)(1_{\cH}\otimes h+h\otimes 1_{\cH})=0\,,
\end{equation*}
and hence
\begin{equation}\label{eq:convProdA}
(e-\varphi)^{\otimes\: 2}\circ \Delta_{\cH}(h)=(e-\varphi)^{\otimes\: 2}\circ \tilde{\Delta}_{\cH}(h)\,.
\end{equation}
It is straightforward to verify by induction that (cf.\ Definitions~\ref{def:nfoldMult} and~\ref{def:nfoldComult} for the $k$-fold multiplications $\mu_{\cA,k}$ and $k$-fold comultiplications $\Delta_{\cH,k}$) 
\begin{equation}
(e-\varphi)^{\star\: k}=\mu_{\cA,k}\circ (e-\varphi)^{\otimes\: k}\circ \Delta_{\cH,k}
\overset{\eqref{eq:convProdA}}{=}\mu_{\cA,k}\circ (e-\varphi)^{\otimes\: k}\circ \tilde{\Delta}_{\cH,k}\,,
\end{equation}
where $\tilde{\Delta}_{\cH,k}$ is defined by replacing all occurrences of $\Delta_{\cH}$ in $\Delta_{\cH,k}$ by $\tilde{\Delta}_{\cH}$. But for any $h\in \cH^n$ ($n>0$) we have according to Corollary~\ref{corr:tildeDeltaAnn} that
\[
\tilde{\Delta}_k(h)=0\quad\text{if } n<k\,,
\] 
which entails that 
\begin{equation}\label{eq:convProdInverseExplicit}
\varphi^{\star\:-1}(h)=\left(e-(e-\varphi)\right)^{\star\:-1}(h)=e+\sum_{k=1}^{n}(e-\varphi)^{\star\: k}(h)\qquad \forall h\in\cH^n\quad (n>0)\,.
\end{equation}
In other words, $\varphi^{\star\:-1}(h)$ is given by a \emph{finite} series and thus evidently well-defined, which concludes the proof.
\end{proof}
\end{prop}

The final ingredient is the concept of Hopf algebras:
\begin{defn}[Hopf algebra]\label{def:Hopf}
A \emph{Hopf algebra} $\cH\equiv(\cH,\mu,\eta,\Delta,\varepsilon,S)$ is a $\bK$-bialgebra $\cH$ equipped with a linear map $S:\cH\rightarrow\cH$ called the \emph{antipode} such that the following diagram commutes:
\begin{equation}
  \includetikz{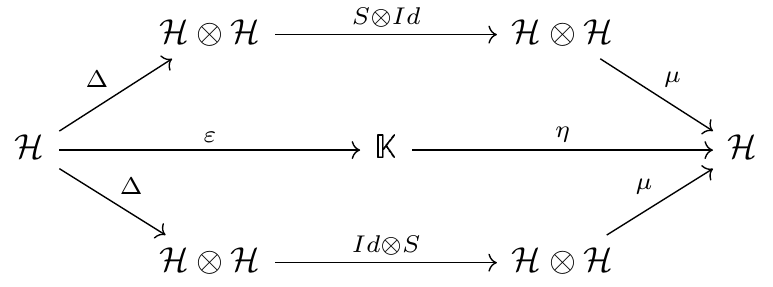}
\end{equation}
In other words, $S$ is a \emph{left and right inverse of the linear map $Id$ under the convolution product $\star$}:
\begin{equation}
S\star Id=e=Id\star S\,.
\end{equation}
\end{defn}

As a direct consequence of Proposition~\ref{prop:convProd}, we have the following very useful Corollary:
\begin{cor}[Any connected filtered bialgebra is a Hopf algebra (cf.\ \cite{manchon2008hopf}, Ch.~4.3, Corr.~5)]\label{corr:bialdCFhopf}
Let $\cA\equiv (\cA,\mu,\eta,\Delta,\varepsilon)$ be a connected filtered $\bK$-bialgebra. Then it is also a Hopf algebra, $\cA\equiv (\cA,\mu,\eta,\Delta,\varepsilon,S)$, with antipode $S$ given according to~\eqref{eq:convProdInverseExplicit} (via the special case $\cA=\cH$ and $\varphi=Id$) by
\begin{equation}\label{}
S(h):=Id^{\star\:-1}(h)=e(h)+\sum_{k=1}^{n}(e-Id)^{\star\: k}(h)\qquad \forall h\in\cH^n\quad (n>0)\,.
\end{equation}
\end{cor}

\end{document}